\documentclass[a4paper,12pt,english]{article}

\usepackage[english]{babel}

\usepackage[T1]{fontenc}          % Paket für die zu verwendendes Schriftart
\usepackage[utf8]{inputenc}       % Eingabe-Zeichenkodierung
\usepackage{lmodern}			  % Schriftart
\usepackage{natbib}
\usepackage{har2nat}
\usepackage{mathtools}
\usepackage{amsfonts}
\usepackage{amsmath}
\usepackage{amssymb}
\usepackage{amsthm}
\usepackage{graphics}
\usepackage{bbm}
\usepackage{pdflscape}
\usepackage{rotating}
\usepackage{booktabs} 			  % \cmidrule,...
\usepackage{adjustbox}
\usepackage[flushleft]{threeparttable}
\usepackage{float}
\usepackage{subcaption}
\usepackage{color}
\usepackage[toc,page]{appendix}
\usepackage{pdfpages}
\usepackage{xspace}
\usepackage{enumitem}
\usepackage[onehalfspacing]{setspace}
\usepackage[paper=a4paper,top=1.25in,bottom=1.25in,left=1.25in,right=1.25in]{geometry}
\usepackage{diagbox}
\usepackage{xr} %cross-ref to external document in the same folder
\newcommand{\mR}{\ensuremath{{\mathbb R}}}
\newcommand{\mZ}{\ensuremath{{\mathbb Z}}}

\newcommand{\mE}{\ensuremath{{\mathbb E}}}
\newcommand{\mP}{\ensuremath{{\mathbb P}}}

\newcommand{\GLS}{\ensuremath{{\tiny \text{GLS}}}}
\newcommand{\ve}{\ensuremath{\varepsilon}}
\newcommand{\ie}{\mbox{i.\,e.}\xspace}

\newcommand{\eg}{\mbox{e.\,g.}\xspace}

\newcommand{\floor}[1]{\lfloor #1 \rfloor}

\newtheorem{remark}{Remark}
\newtheorem{assumption}{Assumption}

\newtheorem{proposition}{Proposition}

\usepackage{authblk}
\usepackage{hyperref}

\author{Karsten Reichold\thanks{I thank Christoph Hanck, Carsten Jentsch and Martin Wagner for helpful comments.\\Correspondence to: Karsten Reichold, Department of Statistics, TU Dortmund University, Vogelpothsweg 87, D--44227 Dortmund, Germany. E-Mail: \href{mailto:reichold@statistik.tu-dortmund.de}{reichold@statistik.tu-dortmund.de}}}

\affil{\small{Department of Statistics, TU Dortmund University, Dortmund, Germany}}

\title{A Residuals-Based Nonparametric Variance Ratio\\ Test for Cointegration}

\date{\today}

\begin{document}
	
\numberwithin{equation}{section}
	
{\singlespacing
\maketitle
}

\begin{abstract} % 141 words
	\onehalfspacing
	 \noindent This paper derives asymptotic theory for \citename{Br02}'s \citeyear[Journal of Econometrics 108, 343--363]{Br02} nonparameteric variance ratio unit root test when applied to regression residuals. The test requires neither the specification of the correlation structure in the data nor the choice of tuning parameters. Compared with popular residuals-based no-cointegration tests, the variance ratio test is less prone to size distortions but has smaller local asymptotic power. However, this paper shows that local asymptotic power properties do not serve as a useful indicator for the power of residuals-based no-cointegration tests in finite samples. In terms of size-corrected power, the variance ratio test performs relatively well and, in particular, does not suffer from power reversal problems detected for, \eg, the frequently used augmented Dickey-Fuller type no-cointegration test. An application to daily prices of cryptocurrencies illustrates the usefulness of the variance ratio test in practice.
	
	
	
\end{abstract}

\newpage

\section{Introduction}\label{sec:Int}
Analyzing the relationship between stochastically trending (economic) time series in a single-equation regression framework entails the risk of obtaining misleading spurious regression results \cite{GrNe74,Ph86}. Practitioners thus rely on statistical tests to assess whether the time series are cointegrated. In this context, it is popular in applications to employ so-called \emph{no-cointegration} tests for the null hypothesis of no cointegration against the alternative of cointegration based on regression residuals estimated by ordinary least squares (OLS).

Alternative approaches include, \eg, single-equation cointegration tests based on conditional error correction models \citeaffixed{KED92,Zi00}{\eg,} and system-based tests \citeaffixed{PhOu90,Jo91,Sh01,Br02,HaPo04,CaSh06}{\eg,}. System-based approaches have the advantage that they do not require the specification of a left-hand side variable and may also allow to test for the number of (linearly independent) cointegrating relations in the system. If, however, there exist reasons for a specific choice of the left-hand side variable, it is convenient and intuitively appealing to analyze the relationship between the variables in a single-equation framework.

Among the most popular residuals-based no-cointegration tests are the parametric augmented Dickey-Fuller \citeaffixed{DiFu79,SaDi84}{ADF,} type test, proposed in \citeasnoun{EnGr87} and asymptotically justified in \citeasnoun{PhOu90}, the semiparametric $\widehat{Z}_{\alpha}$ test \cite{Ph87,PhOu90}, and the parametric MSB test \cite{PeNg96,Pe07}.\footnote{The MSB test and the $\widehat Z_\alpha$ test are popular representatives of the class of $M$ tests \cite{St99,PeNg96,NgPe01} and the class of $Z$ tests \cite{PhPe88}, respectively.} The three tests share the common feature that they require the choice of tuning parameters (\eg, the number of lags in an auxiliary regression and/or kernel and bandwidth choices to estimate a long-run variance parameter) to accommodate the correlation structure in the data. Although these tuning parameter choices allow for asymptotically valid inference, they are likely to have adverse effects on the performance of the tests in finite samples.

In contrast, this paper proposes a nonparametric no-cointegration test, which requires neither the specification of the correlation structure in the data nor the choice of tuning parameters. The test is an extension of \citename{Br02}'s \citeyear{Br02} nonparametric variance ratio unit root test (originally applied to observed univariate time series) to regression residuals.\footnote{\citename{Br02}'s \citeyear{Br02} test, in turn, is a generalization of the unit root test of \citeasnoun{ShSc92}.} The test statistic is easy to compute as it is defined as a (re-scaled) ratio between the sample variances of the regression residuals and their partial sums. Under the null hypothesis, the sample variances converge to random variables whose distributions are scale dependent on the same long-run variance parameter. This makes the limiting null distribution of the test statistic nuisance parameter free without estimating the long-run variance parameter directly. Under the alternative of cointegration, the test statistic converges to zero at rate equal to sample size, which makes the test consistent. In the following, we refer to the test as the (nonparametric) variance ratio (no-cointegration) test.\footnote{Please note that \citeasnoun{We05} has extended \citename{Br02}'s \citeyear{Br02} unit root test to test for cointegration in panel data. However, little is known about the asymptotic and finite sample properties of \citename{Br02}'s \citeyear{Br02} unit root test when applied to regression residuals in a pure time series setting. In particular, the nonparametric variance ratio no-cointegration test is not considered in the two insightful contributions of \citeasnoun{Pe07} and \citeasnoun{PeRo16}.}

The paper derives asymptotic theory for the variance ratio no-cointegration test in a setting that allows for the presence of deterministic time trends both in the regression equation and in the regressors. In the presence of deterministic components, we derive the asymptotic properties of the variance ratio no-cointegration test under both OLS detrending and general least squares (GLS) detrending. Moreover, the paper compares the variance ratio test in terms of local asymptotic power, empirical size and size-corrected power with the ADF, $\widehat Z_\alpha$ and MSB tests in a detailed simulation study. 

We follow, \eg, \citeasnoun{Pe07} and \citeasnoun{PeRo16} and impose a directional restriction on the model, which excludes cointegration between the right-hand side variables. In this case, local asymptotic power of the variance ratio test and its competitors is a function of a single nuisance parameter, $R^2$, which measures the long-run correlation between the regression errors and the regressors (cf. \citename{Pe04}, \citeyear*{Pe04,Pe07}). In addition, we construct a simulation setting that allows to analyze the effects of different short-run dynamics in the data generating process (DGP) on the performance of the tests in finite samples while controlling for effects of $R^2$. This justifies a comparison between local asymptotic power of the tests and their power in finite samples. The results reveal that local asymptotic power properties do not serve as a useful indicator for the performance of residuals-based no-cointegration tests in finite samples and we explain why \citeasnoun[p.\,127]{Pe07} and \citeasnoun[p.\,99]{PeRo16} come to opposing conclusions.

Finally, an empirical illustration applies the variance ratio test and its competitors to daily prices of the four cryptocurrencies with highest market capitalization. Test decisions are heterogeneous across tests, but the variance ratio test, the ADF test, and the ADF test based on a modified information criterion provide reliable evidence for the presence of cointegration between the four cryptocurrencies with the highest market capitalization. The results are in line with those in related literature pointing towards the presence of cointegrating relationships in the cryptocurrency market \citeaffixed{KeZh21,ByGo22}{cf., \eg}.

The paper proceeds as follows: Section~\ref{sec:Setting} introduces the model and its underlying assumptions. Section~\ref{sec:Th} defines the variance ratio no-cointegration test under OLS- and GLS detrending and derives its asymptotic properties. Section~\ref{sec:FiniteSample} analyzes the performance of the variance ratio test in finite samples and Section~\ref{sec:Illu} contains the empirical illustration. Section~\ref{sec:Conc} summarizes and concludes. All proofs are relegated to the Appendix, which also contains additional asymptotic and finite sample results. Supplementary Material, available on the author's homepage, provides further finite sample results.

Throughout, $\floor{x}$ denotes the integer part of a real number $x$, $1_m$ denotes the $m$-dimensional vector of ones and $I_m$ denotes the ($m\times m$)-dimensional identity matrix. The symbols $\overset{p}{\longrightarrow}$ and $\overset{w}{\longrightarrow}$ signify convergence in probability and weak convergence, respectively, as the sample size $T\rightarrow\infty$.

\section{The Model and Assumptions}\label{sec:Setting}
We consider the model
\begin{align}
	x_t &= \mu + x_{t-1} + v_t = x_0 + \mu t + \sum_{s=1}^t v_s\label{eq:x}\\
	y_t &= d_t'\tau + x_t'\beta + u_t\label{eq:y}\\
	u_t &=\rho u_{t-1} + \xi_t\label{eq:u},
\end{align}
for $t=1,\ldots,T$, where $\{y_t\}_{t\in\mZ}$ is a scalar integrated process, $\{x_t\}_{t\in\mZ}$ is a $m$-dimensional vector of integrated processes with potentially non-zero deterministic drift $\mu\in\mR^m$, and $u_0=O_\mP(1)$.\footnote{For the asymptotic results in this paper it in fact suffices to assume $T^{-1/2}u_0 = o_\mP(1)$. Section~\ref{sec:FiniteSample_u0} analyzes the impact of a large initial value $u_0$ of order $T^{1/2}$ on the finite sample performance of the variance ratio test.} The $p$-dimensional vector $d_t$ contains the deterministic components included in the model. Under the null hypothesis $\rho=1$ the error $\{u_t\}_{t\in\mZ}$ is an integrated process, \ie, there exists no cointegrating relation between $\{y_t\}_{t\in\mZ}$ and $\{x_t\}_{t\in\mZ}$. Under the alternative $\vert\rho\vert<1$, however, the error process $\{u_t\}_{t\in\mZ}$ is stationary and $\{y_t\}_{t\in\mZ}$ and $\{x_t\}_{t\in\mZ}$ are cointegrated with (normalized) cointegrating vector $[1,-\beta']'$, $\beta\neq 0$.
\begin{assumption}\label{ass:FCLTw}%cf. Phillips and Hansen (1990, p. 101)
	Let $\{w_t\}_{t\in\mZ}\coloneqq \{[\xi_t,v_t']'\}_{t\in\mZ}$ be a strictly stationary and ergodic process with $\mE(w_t)=0$, finite covariance matrix $\mE\left(w_tw_t'\right)>0$ and continuous spectral density matrix $f_{ww}(\lambda)$ on $(-\pi,\pi]$. Moreover, $\{w_t\}_{t\in\mZ}$ fulfills a functional central limit theorem of the form
	\begin{align}\label{eq:FCLTw}
		T^{-1/2}\sum_{t=1}^{\floor{rT}} w_t \overset{w}{\longrightarrow} B(r)= \Omega^{1/2} W(r), \quad 0\leq r\leq 1,
	\end{align}
	where $W(r)=[W_{\xi\cdot v}(r),W_v(r)']'$ is an $(1+m)$-dimensional vector of independent standard Brownian motions and
	\begin{align}
		\Omega = \begin{bmatrix}
			\Omega_{\xi\xi}&\Omega_{\xi v}\\
			\Omega_{v\xi}&\Omega_{vv}
		\end{bmatrix} \coloneqq 2\pi f_{ww}(0) >0 
	%=\sum_{j=-\infty}^\infty \mE\left(w_t w_{t-j}'\right) = 
	\end{align}
	denotes the long-run covariance matrix of $\{w_t\}_{t\in\mZ}$.
\end{assumption}
In particular, positive definiteness of $\Omega_{vv}$ rules out cointegration among the elements of $\{x_t\}_{t\in\mZ}$. To express asymptotic results, it is convenient to work with
\begin{align}
	\Omega^{1/2} =\begin{bmatrix}
		\Omega_{\xi \cdot v}^{1/2} & \Omega_{\xi v}(\Omega_{vv}^{-1/2})'\\
		0 & \Omega_{vv}^{1/2}
	\end{bmatrix},
\end{align}
where $\Omega_{\xi\cdot v}\coloneqq\Omega_{\xi \xi}-\Omega_{\xi v}\Omega_{vv}^{-1}\Omega_{v\xi}$, such that $\Omega^{1/2}(\Omega^{1/2})'=\Omega$. For later usage, we partition $B(r)=[B_\xi(r),B_v(r)']'$ analogously to the partitioning of $W(r)$.

The asymptotic results depend on the specification of the deterministic components in~\eqref{eq:x} and~\eqref{eq:y}. This paper considers three different cases: no deterministics (D0), intercept only (D1) and intercept and linear trend (D2). In case D0, we set $x_0=\mu=0$ and remove $d_t\tau$ from~\eqref{eq:y}. In case D1, we allow for $x_0=O_\mP(1)$ and set $\mu=0$ and $d_t=1$. Finally, in case D2, we allow for $x_0=O_\mP(1)$ and a deterministic trend in $x_t$ and set $d_t=[1,t]'$.\footnote{\citeasnoun{PeRo16} also consider the case with a deterministic trend in the regressors but only a constant (rather than a constant and a linear trend) in the regression. Following \citeasnoun{Pe07}, we abstain from considering this case as it implies that the limiting null distributions of the test statistics depend on the drift parameter $\mu$ being exactly zero or not -- and thus leaves the opportunity of choosing wrong critical values in applications. Considering more general deterministic components, \eg, polynomial time trends, in~\eqref{eq:y} and/or~\eqref{eq:x} is also possible but beyond the scope of this paper.}

\section{Asymptotic Theory}\label{sec:Th}
\subsection{Preliminary Data Detrending}\label{sec:Th-Detrending}
If deterministic components are included in~\eqref{eq:y}, it is standard practice in applications to first detrend $z_t\coloneqq [y_t,x_t']'$ according to the choice of $d_t$ using ordinary least squares (OLS). In this case, the OLS detrended time series are defined as
\begin{align}
	\widetilde z_t' = [\widetilde y_t,\widetilde x_t']\coloneqq z_t' - d_t' \left(\sum_{s=1}^T d_sd_s'\right)^{-1} \sum_{s=1}^T d_s z_s'.
\end{align}
In case D1 and D2 the definition reduces to $\widetilde z_t = z_t - T^{-1}\sum_{s=1}^T z_s$ and
\begin{align}
	\widetilde z_t = z_t - \frac{4T-6t+2}{T-1}\frac{1}{T}\sum_{s=1}^T z_s - \frac{-6T+12t-6}{(T-1)(T+1)}\sum_{s=1}^T\left(\frac{s}{T}\right)z_s,
\end{align}
respectively. For notational brevity, we define $\widetilde z_t = [\widetilde y_t,\widetilde x_t']'\coloneqq z_t$ in case D0. With these definitions in place, it follows from~\eqref{eq:y} that
\begin{align}\label{eq:ytilde}
	\widetilde y_t = \widetilde x_t' \beta + \widetilde u_t,
\end{align}
where $\widetilde u_t$ is defined analogously to $\widetilde z_t$. The OLS residuals in~\eqref{eq:ytilde} are given by
\begin{align}\label{eq:uhat}
	\widehat u_t&\coloneqq \widetilde y_t - \widetilde x_t'\widehat \beta = \widetilde u_t - \widetilde x_t'\left(\widehat \beta - \beta\right) = \widetilde u_t - \widetilde x_t'\left(\sum_{s=1}^T \widetilde x_s \widetilde x_s'\right)^{-1}\sum_{s=1}^T \widetilde x_s \widetilde u_s,
\end{align}
where $\widehat \beta$ denotes the OLS estimator of $\beta$ in~\eqref{eq:ytilde}.

To capture the asymptotic effects of detrending, define for a potentially multivariate integrable stochastic process $P(r)$, $0\leq r\leq 1$, the detrended process $\widetilde P(r) = P(r)-\int_0^1 P(s)ds$ in case D1 and $\widetilde P(r) = P(r) - (4-6r)\int_0^1 P(s)ds-(12r-6)\int_0^1 s P(s)ds$ in case D2. In case D0, we set $\widetilde P(r)\coloneqq P(r)$.
\begin{remark}\label{rem:GLSDetrending}
	\citeasnoun{PeRo16} suggest to use GLS detrended data rather than OLS detrended data in cases D1 and D2 to increase local asymptotic power of residuals-based no-cointegration tests. For $\bar \rho\coloneqq 1+\bar c/T$, with some constant $\bar c\leq 0$, define $z_t^{\bar \rho} \coloneqq z_t-\bar \rho z_{t-1}$ and $d_t^{\bar \rho} \coloneqq d_t-\bar \rho d_{t-1}$, $t=2,\ldots,T$. The GLS detrended variables are constructed as $\widetilde y_t^{(\GLS)} \coloneqq y_t - d_t'\widehat \tau^*$ and $\widetilde x_t^{(\GLS)} \coloneqq x_t - \widehat \Psi_x^{*\prime}d_t$, where $[\widehat\tau^*,\widehat\Psi_x^*]\coloneqq (D^{\bar \rho\prime}D^{\bar \rho})^{-1}D^{\bar \rho\prime}Z^{\bar \rho}$, with $D^{\bar \rho} \coloneqq [d_1,d_2^{\bar \rho},\ldots,d_t^{\bar \rho}]'$ and $Z^{\bar \rho}\coloneqq [z_1,z_2^{\bar \rho},\ldots,z_T^{\bar \rho}]'$.\footnote{Including $d_1$ and $z_1$ in the definitions of $D^{\bar \rho}$ and $Z^{\bar \rho}$, respectively, is important, see, \eg, \citeasnoun{BrTa03}.} The corresponding test statistics are then based on the OLS residuals $\widehat u_t^{(\GLS)}$ in the regression $\widetilde y_t^{(\GLS)} = \widetilde x_t^{(\GLS)\prime}\beta + \widetilde u_t^{(\GLS)}$, where $\widetilde u_t^{(\GLS)} \coloneqq u_t - d_t'(\widehat\tau^*-\tau) + \widehat\beta'\widehat\Psi_x^{*\prime}d_t$.
\end{remark}
In what follows, the main text focuses in cases D1 and D2 on OLS detrended data, whereas a series of remarks is dedicated to the use of GLS detrended data.

\subsection{The Variance Ratio Test}\label{sec:Th-Test}
Typically, residuals-based no-cointegration tests require tuning parameter choices (\eg, the number of lags in an auxiliary regression and/or kernel and bandwidth choices to estimate a long-run variance parameter) to accommodate the correlation structure in the data. Various simulation results indicate that these tuning parameter choices can affect the finite sample performance of no-cointegration tests considerably. This is particularly unfavorable in situations where different tuning parameter choices lead to different test decisions.

To obtain a tuning parameter free no-cointegration test, we apply the nonparametric variance ratio test of \citeasnoun{Br02}, originally proposed to test for a unit root in an observed univariate time series, to the OLS residuals $\widehat u_t$ defined in~\eqref{eq:uhat}, \ie, based on OLS detrended data. The variance ratio test statistic is thus defined as
\begin{align}\label{eq:PuSN}
	\text{VR}\coloneqq \frac{T^{-2} \sum_{t=1}^T \left(\sum_{s=1}^t \widehat{u}_s \right)^2}{\sum_{t=1}^T \widehat{u}_t^2}= \frac{\widehat{\eta}_T}{T^{-2}\sum_{t=1}^T \widehat{u}_t^2},
\end{align}
where $\widehat{\eta}_T\coloneqq T^{-4} \sum_{t=1}^T \left(\sum_{s=1}^t \widehat{u}_s \right)^2$. 

In contrast to, \eg, the MSB and $\widehat Z_\alpha$ test statistics, the variance ratio test statistic does not depend on a consistent estimator of the long-run variance parameter $\Omega_{\xi\cdot v}$. Instead, it depends on a random variable $\widehat{\eta}_T$, whose limiting null distribution is, as we shall see in the proof of Proposition~\ref{prop:VR_H0}, scale dependent on $\Omega_{\xi\cdot v}$. In this sense, the variance ratio test statistic may be interpreted as a self-normalized version of the MSB test statistic.\footnote{This self-normalizing feature of the variance ratio test fits well to the (bootstrap-assisted) self-normalized testing approach of \citeasnoun{ReJe22} for hypotheses on the cointegrating vector.} Scale dependence of the limiting null distribution of $\eta_T$ on $\Omega_{\xi\cdot v}$ is key to obtain a nuisance parameter free limiting distribution of the variance ratio test statistic under the null hypothesis of no cointegration.
\begin{proposition}\label{prop:VR_H0} Let $\{x_t\}_{t\in\mZ}$, $\{y_t\}_{t\in\mZ}$, and $\{u_t\}_{t\in\mZ}$ be generated by~\eqref{eq:x},~\eqref{eq:y}, and~\eqref{eq:u}, respectively, and let $\{w_t\}_{t\in\mZ}$ satisfy Assumption~\ref{ass:FCLTw}. Then it holds under the null hypothesis of no cointegration ($\rho=1$) in cases D0, D1, and D2 that
\begin{align}
	\text{VR} \overset{w}{\longrightarrow} \frac{\int_0^1\left(\int_0^r \widetilde W_{\xi\cdot v}^+(s) ds\right)^2dr}{\int_0^1 \left( \widetilde W_{\xi\cdot v}^+(r)\right)^2dr},
\end{align}
where 
\begin{align}
	\widetilde W_{\xi\cdot v}^+(r) \coloneqq \widetilde W_{\xi\cdot v}(r) - \widetilde W_v(r)'\left(\int_0^1 \widetilde W_v(s)\widetilde W_v(s)'ds\right)^{-1}\int_0^1 \widetilde W_v(s)\widetilde W_{\xi\cdot v}(s)ds.
\end{align}
\end{proposition}
The limiting null distribution of the variance ratio test statistic is nonstandard but free of any nuisance parameters. It only depends on the dimension $m$ of $x_t$, through the dimension of $W_v(r)$, and on the deterministic component in~\eqref{eq:y}, as the detrended processes appear in the limit.\footnote{The limiting null distribution derived in Proposition~\ref{prop:VR_H0} differs from the limiting null distribution derived in \citeasnoun[Prop.~3]{Br02}, reflecting the application of the variance ratio test to regression residuals rather than to an observed (potentially detrended) univariate time series.}

We proceed with analyzing the behavior of the variance ratio test statistic under the alternative of cointegration.
\begin{proposition}\label{prop:VR_H1} Let $\{x_t\}_{t\in\mZ}$, $\{y_t\}_{t\in\mZ}$, and $\{u_t\}_{t\in\mZ}$ be generated by~\eqref{eq:x},~\eqref{eq:y}, and~\eqref{eq:u}, respectively, and let $\{w_t\}_{t\in\mZ}$ satisfy Assumption~\ref{ass:FCLTw}. Then it holds under the alternative of cointegration ($\vert\rho\vert<1$) in cases D0, D1, and D2 that $\text{VR}=O_\mP(T^{-1})$.
\end{proposition}
Proposition~\ref{prop:VR_H1} shows that in the presence of cointegration, the variance ratio test statistic converges to zero at rate equal to sample size. Hence, the variance ratio test is a left-tailed test, rejecting the null hypothesis of no cointegration in favor of the alternative of cointegration for small (\ie, close to zero) realizations of $\text{VR}$. Table~\ref{tab:critvalsVR} in Appendix~\ref{app:Critvals} provides corresponding asymptotic critical values in cases D0, D1, and D2 for $m=1,\ldots,5$.

\begin{remark}
	The variance ratio test applied to the residuals from the regression with GLS detrended data is defined as $\text{VR}^{(\GLS)}\coloneqq T^{-2} \sum_{t=1}^T \left(\sum_{s=1}^t \widehat{u}_s^{(\GLS)} \right)^2/\sum_{t=1}^T \left(\widehat{u}_t^{(\GLS)}\right)^2$, with $\widehat u_t^{(\GLS)}$ as defined in Remark~\ref{rem:GLSDetrending}. Remarks~\ref{rem:GLSLAP} and~\ref{rem:cbar} in Section~\ref{sec:Th-LAPower} derive the limiting distribution of $\text{VR}^{(\GLS)}$ both under the null hypothesis of no cointegration and under local alternatives and also provide some guidance on the choice of $\bar{c}$.
\end{remark}

\subsection{Local Asymptotic Power}\label{sec:Th-LAPower}
To analyzes the performance of the variance ratio test statistic under local alternatives, we set $\rho=\rho_T = 1+c/T$, with $c\leq0$, in~\eqref{eq:u}. For $c=0$, the regression errors $\{u_t\}_{t\in\mZ}$ are integrated of order one, \ie, there is no cointegration between $\{y_t\}_{t\in\mZ}$ and $\{x_t\}_{t\in\mZ}$. Under the local alternative $c<0$, the error process $\{u_t\}_{t\in\mZ}$ is a near unit root process such that $T^{-1/2}u_{\floor{rT}} \overset{w}{\longrightarrow} \Omega_{\xi \cdot v}^{1/2}J_{\xi\cdot v}^c(r)$, $0\leq r\leq 1$, where
\begin{align}
	J_{\xi\cdot v}^c(r) \coloneqq \int_0^r e^{(r-s)c}d\left(W_{\xi \cdot v}(s)+\sqrt{R^2/(1-R^2)}\overline W_v(s)\right),
\end{align}
is an Ornstein-Uhlenbeck process, with $R^2\coloneqq \Omega_{\xi \xi}^{-1}\Omega_{\xi v}\Omega_{vv}^{-1}\Omega_{v\xi} = 1-\Omega_{\xi \cdot v}/\Omega_{\xi \xi}$ and $\overline W_v(r)\coloneqq m^{-1/2}1_m'W_v(r)$ \citeaffixed[Lemma~5.1]{PeRo16}{cf., \eg,}. The coefficient $R^2$ lies between zero and one and measures the squared long-run correlation between the regression errors $\{u_t\}_{t\in\mZ}$ and the regressors $\{x_t\}_{t\in\mZ}$. \citename{Pe04} \citeyear{Pe04,Pe07} shows analytically that $R^2$ is the only nuisance parameter affecting local asymptotic power of the ADF, MSB, and $\widehat{Z}_{\alpha}$ tests. The following propositions shows that $R^2$ is also the only nuisance parameter affecting local asymptotic power of the variance ratio test.
\begin{proposition}\label{prop:VR_Local} Let $\{x_t\}_{t\in\mZ}$ and $\{y_t\}_{t\in\mZ}$ be generated by~\eqref{eq:x} and~\eqref{eq:y}, respectively and let $\{u_t\}_{t\in\mZ}$ be generated by~\eqref{eq:u} with $\rho=\rho_T = 1+c/T$, where $c\leq 0$. Let $\{w_t\}_{t\in\mZ}$ satisfy Assumption~\ref{ass:FCLTw}. Then it holds for the variance ratio test statistic in cases D0, D1, and D2 that
	\begin{align}
		\text{VR} \overset{w}{\longrightarrow} \mathcal{G}_{{\scriptstyle\text{VR}},c} \coloneqq \frac{\int_0^1\left(\int_0^r \widetilde J_{\xi\cdot v}^{c,+}(s) ds\right)^2dr}{\int_0^1 \left( \widetilde J_{\xi\cdot v}^{c,+}(r)\right)^2dr},
	\end{align}
	where
	\begin{align}
		\widetilde J_{\xi\cdot v}^{c,+}(r) \coloneqq \widetilde J_{\xi\cdot v}^c(r) - \widetilde W_v(r)'\left(\int_0^1 \widetilde W_v(s)\widetilde W_v(s)'ds\right)^{-1}\int_0^1 \widetilde W_v(s)\widetilde J_{\xi\cdot v}^c(s)ds.
	\end{align}
\end{proposition}
Proposition~\ref{prop:VR_Local} shows that the limiting distribution of the variance ratio test statistic under local alternatives depends on the location parameter $c$, the number of regressors $m$ and on the nuisance parameter $R^2$. Local asymptotic power of the variance ratio test at the nominal level $\alpha$ is given by the probability that $\mathcal{G}_{{\scriptstyle\text{VR}},c}$ is smaller than the $\alpha$-quantile of the limiting null distribution of $\text{VR}$. For $c=0$, $\mathcal{G}_{{\scriptstyle\text{VR}},c}$ coincides with the limiting null distribution given in Proposition~\ref{prop:VR_H0}, \ie, local asymptotic power at $c=0$ is equal to $\alpha$. To ease comparisons between the limiting distribution of the variance ratio test and the limiting distributions of the tests considered in \citeasnoun{Pe07} under local alternatives, rewrite $\mathcal{G}_{{\scriptstyle\text{VR}},c}=\left(\widetilde \kappa'\widetilde C_c \widetilde\kappa_c\right)/\left(\widetilde \kappa'\widetilde A_c \widetilde\kappa_c\right)$, where $\widetilde{C}_c \coloneqq \int_0^1 \left(\int_0^r \widetilde{W}_c(s)ds\right)\left(\int_0^r \widetilde{W}_c(s)ds\right)'dr$, $\widetilde{A}_c \coloneqq \int_0^1 \widetilde W_c(r) \widetilde W_c(r)'dr$, $\widetilde \kappa_c'=[1,-\int_0^1 \widetilde W_v(r)'\widetilde J_{\xi\cdot v}^c(r)dr\left(\int_0^1 \widetilde W_v(r)\widetilde W_v(r)'dr\right)^{-1}]$ and $\widetilde{W}_c(r) \coloneqq [\widetilde J_{\xi\cdot v}^c(r),\widetilde W_v(r)']'$.\footnote{In this notation, the limiting distributions of the ADF, $\widehat Z_\alpha$ and MSB statistics under local alternatives are given by $c \frac{\left(\tilde{\kappa}_c'\widetilde A_c \tilde{\kappa}_c\right)^{1/2}}{\left(\tilde{\kappa}_c'D\tilde{\kappa}_c\right)^{1/2}} + \frac{\tilde{\kappa}_c'\widetilde B_c \tilde{\kappa}_c}{\left(\tilde{\kappa}_c'D\tilde{\kappa}_c\right)^{1/2}\left(\tilde{\kappa}_c'\widetilde{A}_c\tilde{\kappa}_c\right)^{1/2}}$, $c + \frac{\tilde{\kappa}_c'\widetilde B_c \tilde{\kappa}_c}{\tilde{\kappa}_c'\widetilde A_c \tilde{\kappa}_c}$, and $\frac{\left(\tilde{\kappa}_c'\widetilde A_c \tilde{\kappa}_c\right)^{1/2}}{\left(\tilde{\kappa}_c'D\tilde{\kappa}_c\right)^{1/2}}$, respectively, with $\widetilde{B}_c \coloneqq \int_0^1 \widetilde{W}_c(r)d\left((W_{\xi \cdot v}(s)+\sqrt{R^2/(1-R^2)}\overline W_v(s)\right)$, $D\coloneqq \begin{bmatrix}
		1+\bar{\delta}'\bar{\delta} &\bar{\delta}'\\
		\bar{\delta}& I_m
\end{bmatrix}$ and $\bar \delta'\coloneqq \Omega_{\xi \cdot v}^{-1/2}\Omega_{\xi v}(\Omega_{vv}^{-1/2})'$, see \citeasnoun[Theorem~1]{Pe07}, where we corrected a typo in the limiting distribution of $\widehat{Z}_{\alpha}$. By definition, it holds that $\bar{\delta}'\bar{\delta}=R^2/(1-R^2)$.}

\begin{figure}[!ht]
\begin{center}
	\begin{subfigure}{0.33\textwidth}
		\centering
		\caption*{D0\\$R^2=0$}
		\vspace{-1ex}
		\includegraphics[trim={2cm 0.2cm 2cm 0.5cm},width=0.98\textwidth,clip]{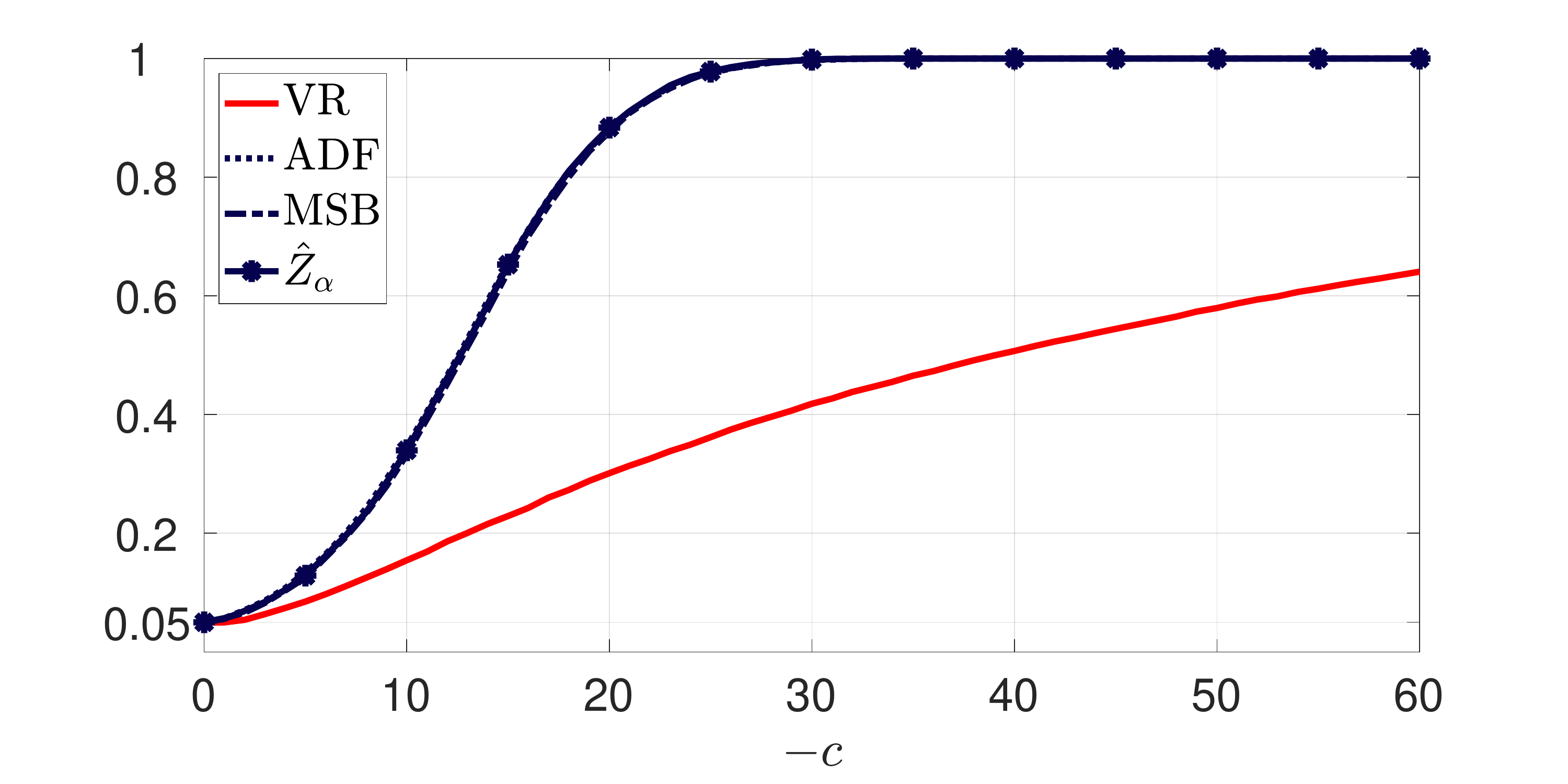}
	\end{subfigure}\begin{subfigure}{0.33\textwidth}
		\centering
		\caption*{D1\\$R^2=0$}
		\vspace{-1ex}
		\includegraphics[trim={2cm 0.2cm 2cm 0.5cm},width=0.98\textwidth,clip]{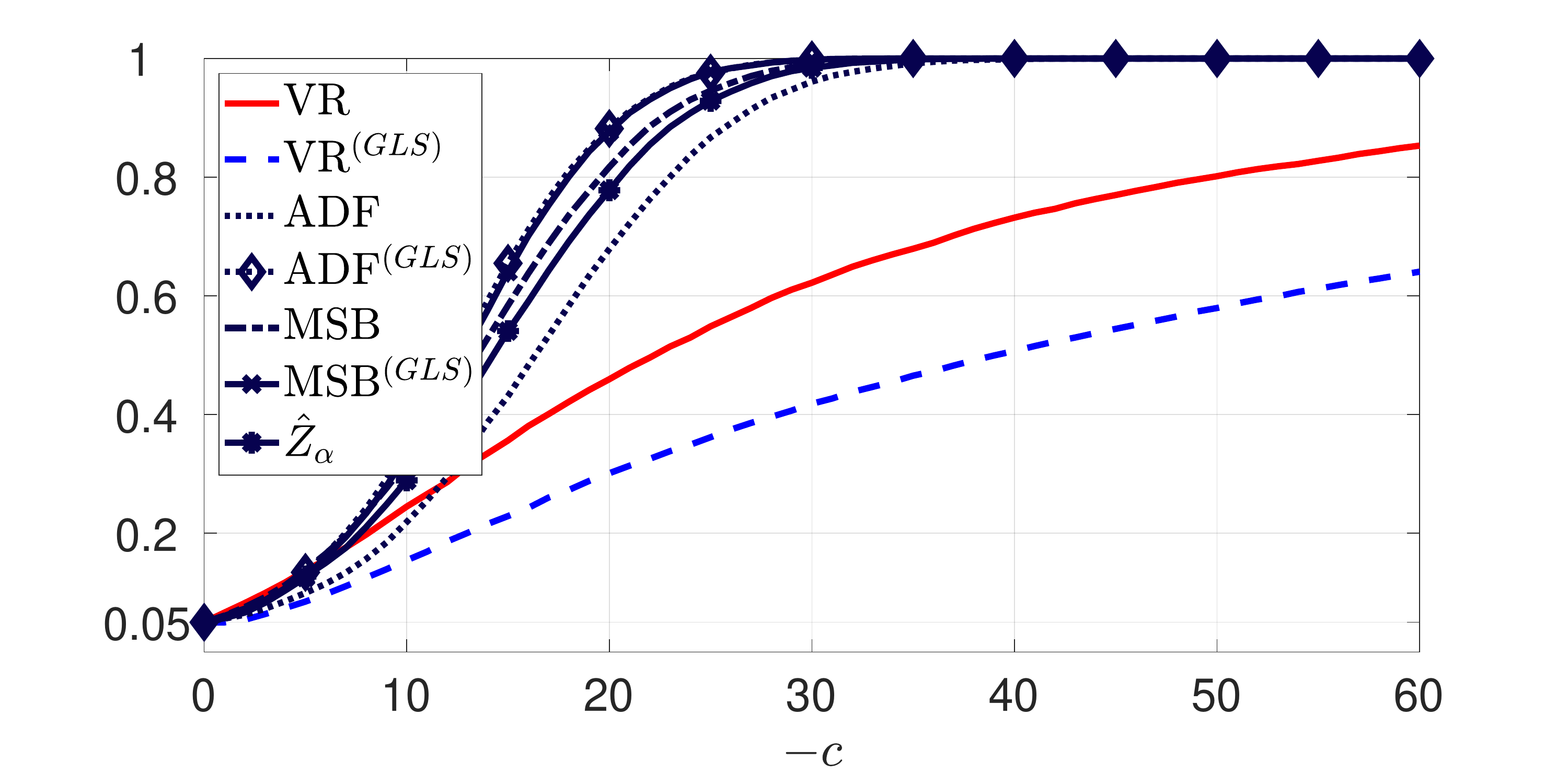}
	\end{subfigure}\begin{subfigure}{0.33\textwidth}
		\centering
		\caption*{D2\\$R^2=0$}
		\vspace{-1ex}
		\includegraphics[trim={2cm 0.2cm 2cm 0.5cm},width=0.98\textwidth,clip]{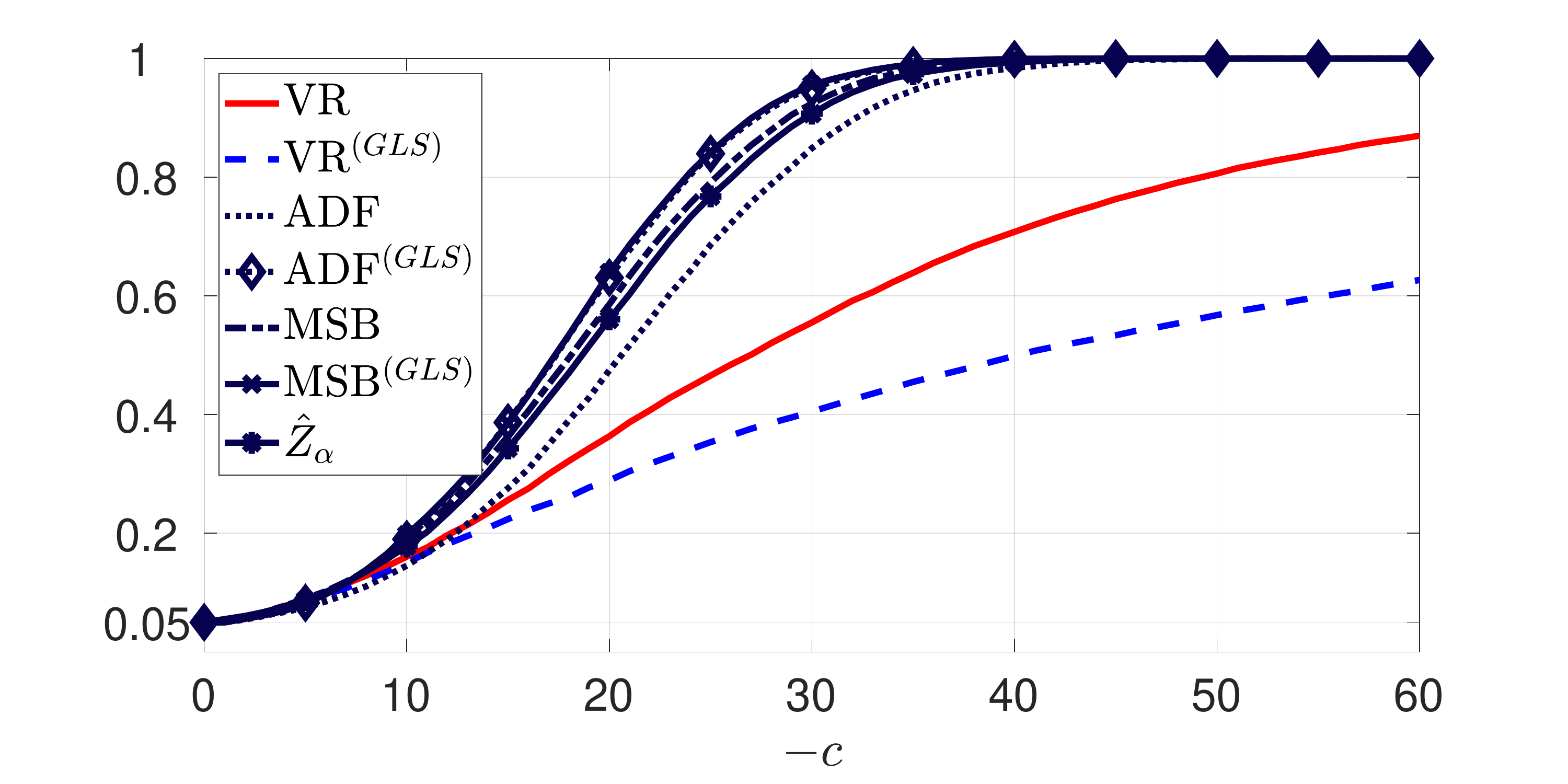}
	\end{subfigure}
	
	\vspace{1ex}
	
	\begin{subfigure}{0.33\textwidth}
		\centering
		\caption*{$R^2=0.4$}
		\vspace{-1ex}
		\includegraphics[trim={2cm 0.2cm 2cm 0.5cm},width=0.98\textwidth,clip]{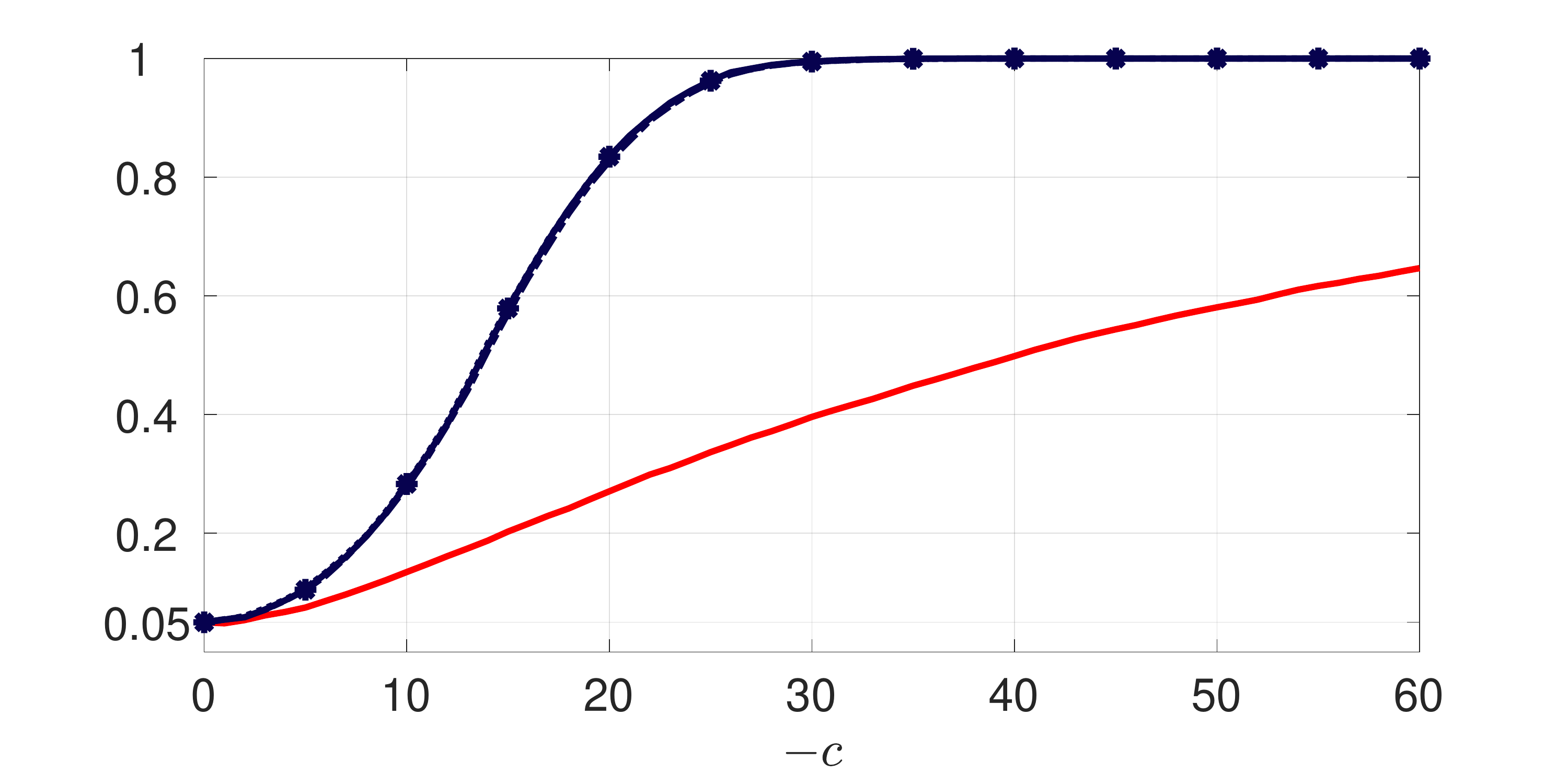}
	\end{subfigure}\begin{subfigure}{0.33\textwidth}
		\centering
		\caption*{$R^2=0.4$}
		\vspace{-1ex}
		\includegraphics[trim={2cm 0.2cm 2cm 0.5cm},width=0.98\textwidth,clip]{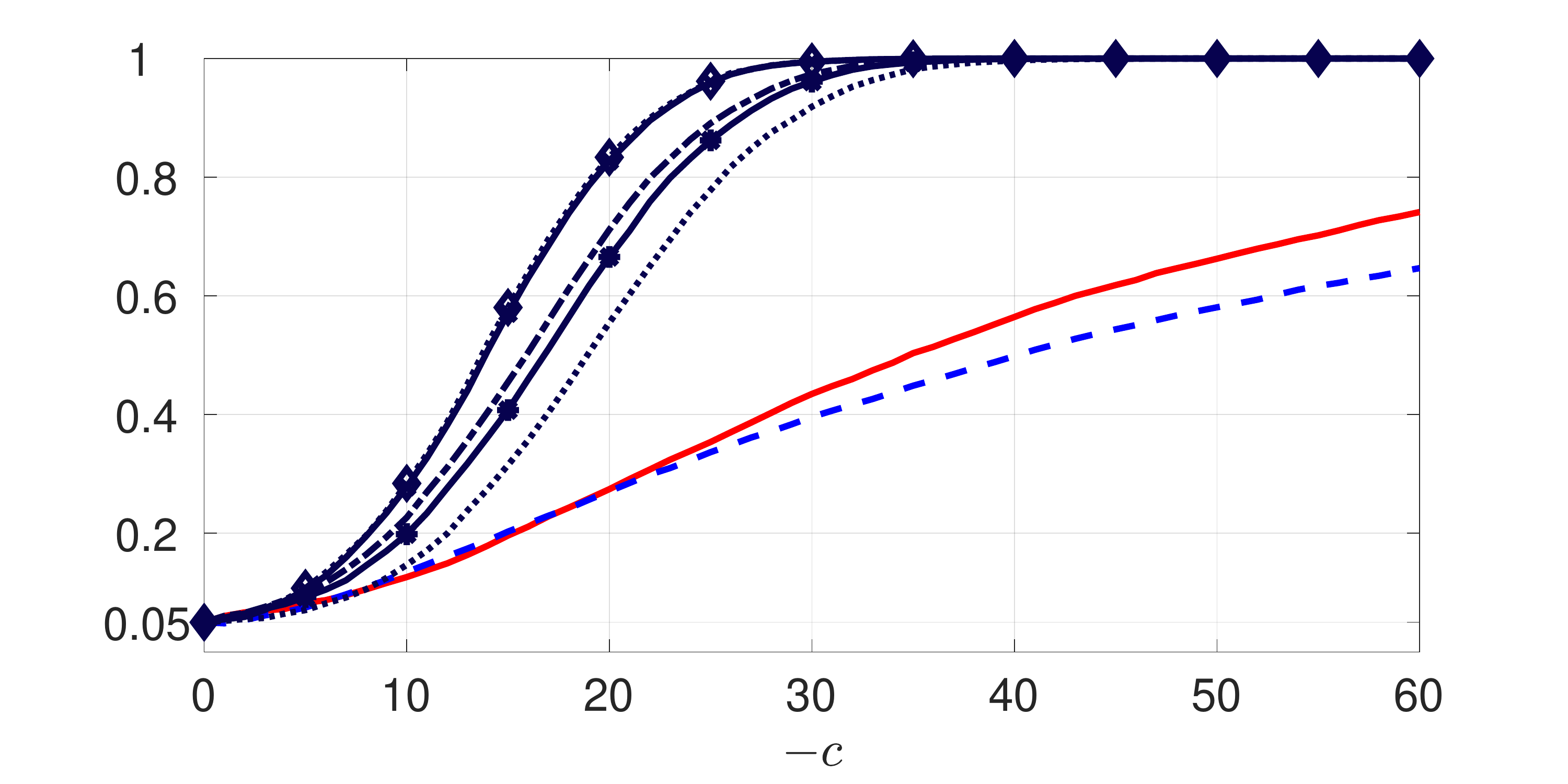}
	\end{subfigure}\begin{subfigure}{0.33\textwidth}
	\centering
	\caption*{$R^2=0.4$}
	\vspace{-1ex}
	\includegraphics[trim={2cm 0.2cm 2cm 0.5cm},width=0.98\textwidth,clip]{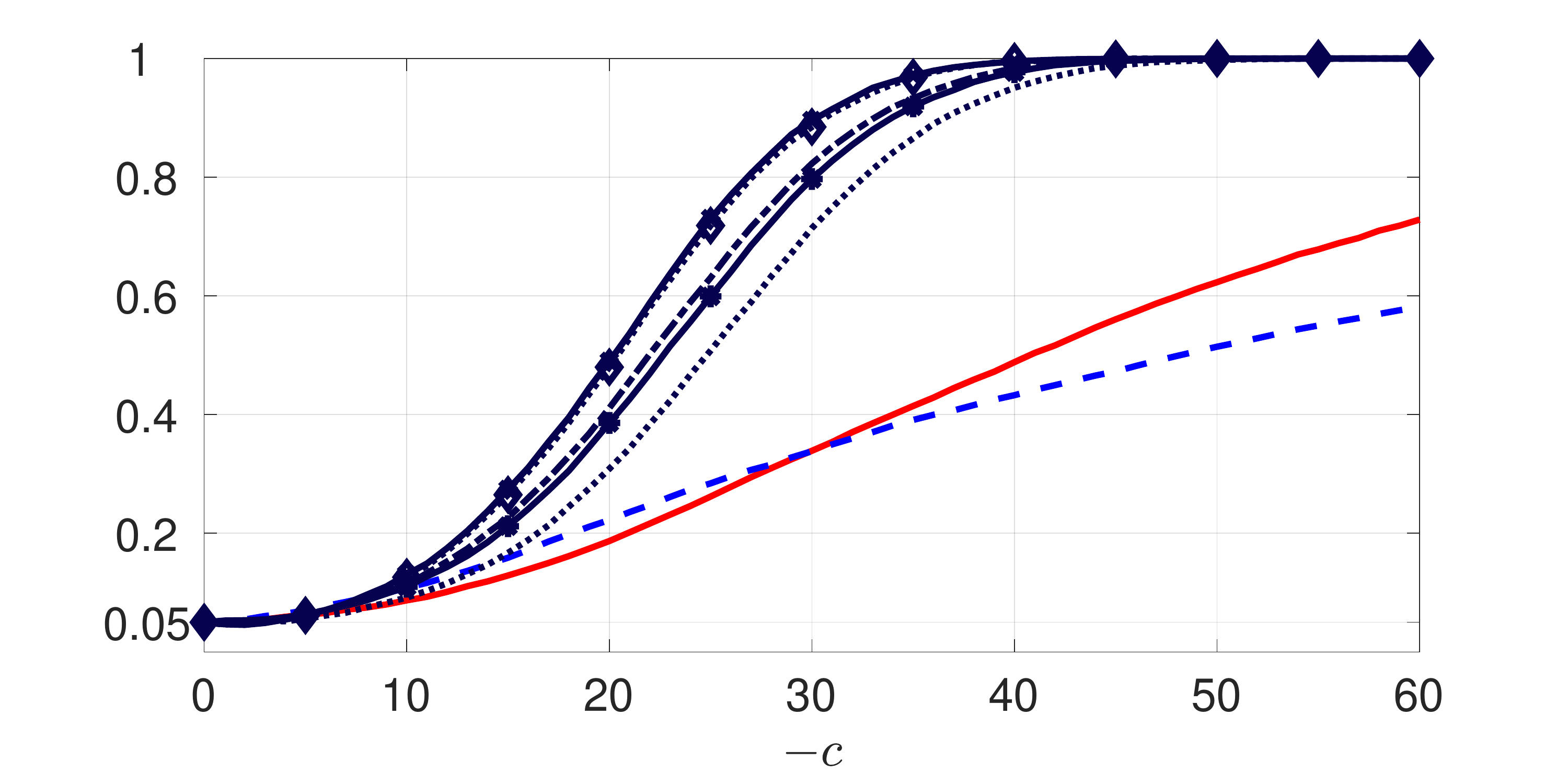}
\end{subfigure}

	\vspace{1ex}
	
	\begin{subfigure}{0.33\textwidth}
		\centering
		\caption*{$R^2=0.8$}
		\vspace{-1ex}
		\includegraphics[trim={2cm 0.2cm 2cm 0.5cm},width=0.98\textwidth,clip]{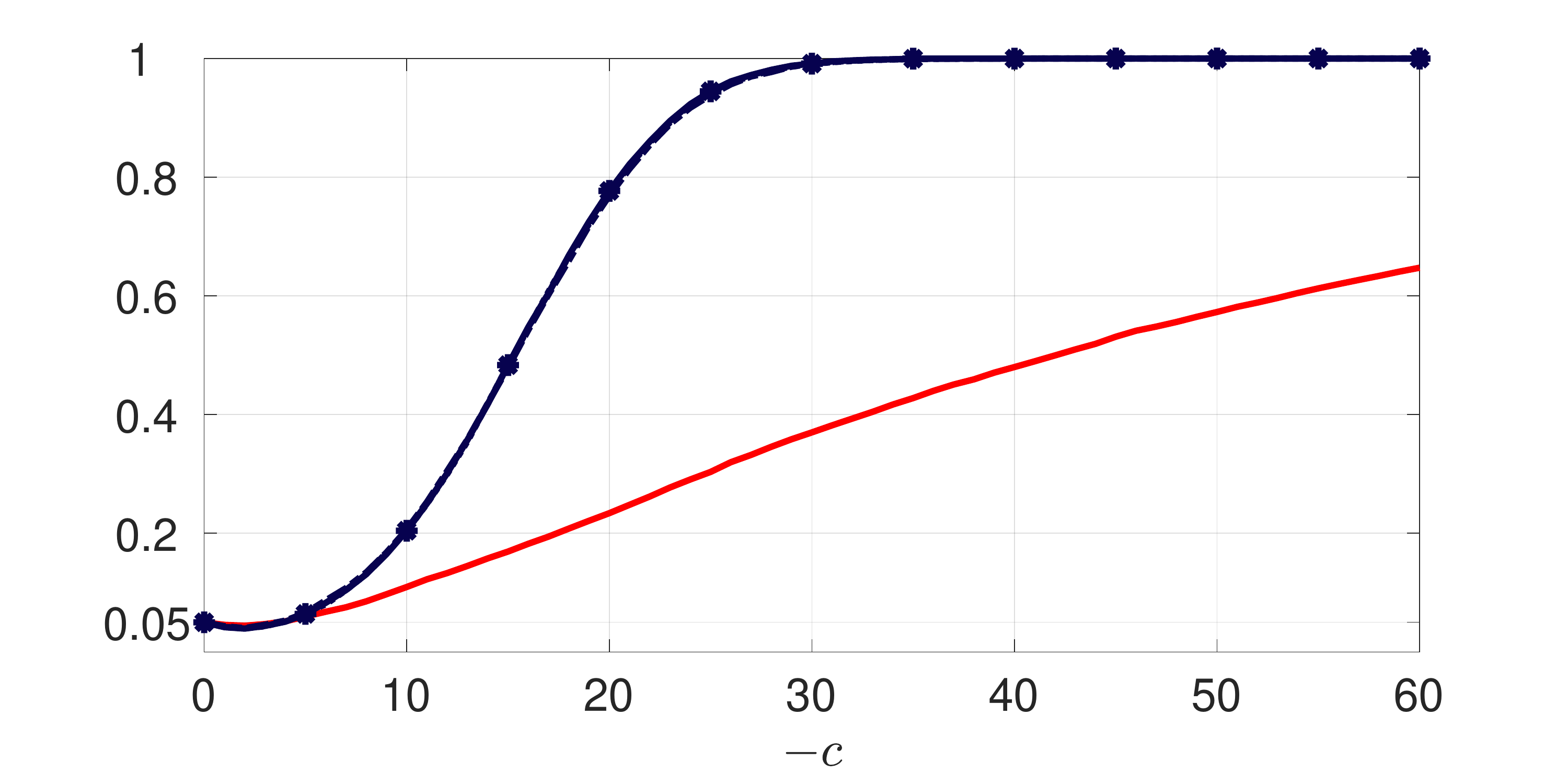}
	\end{subfigure}\begin{subfigure}{0.33\textwidth}
		\centering
		\caption*{$R^2=0.8$}
		\vspace{-1ex}
		\includegraphics[trim={2cm 0.2cm 2cm 0.5cm},width=0.98\textwidth,clip]{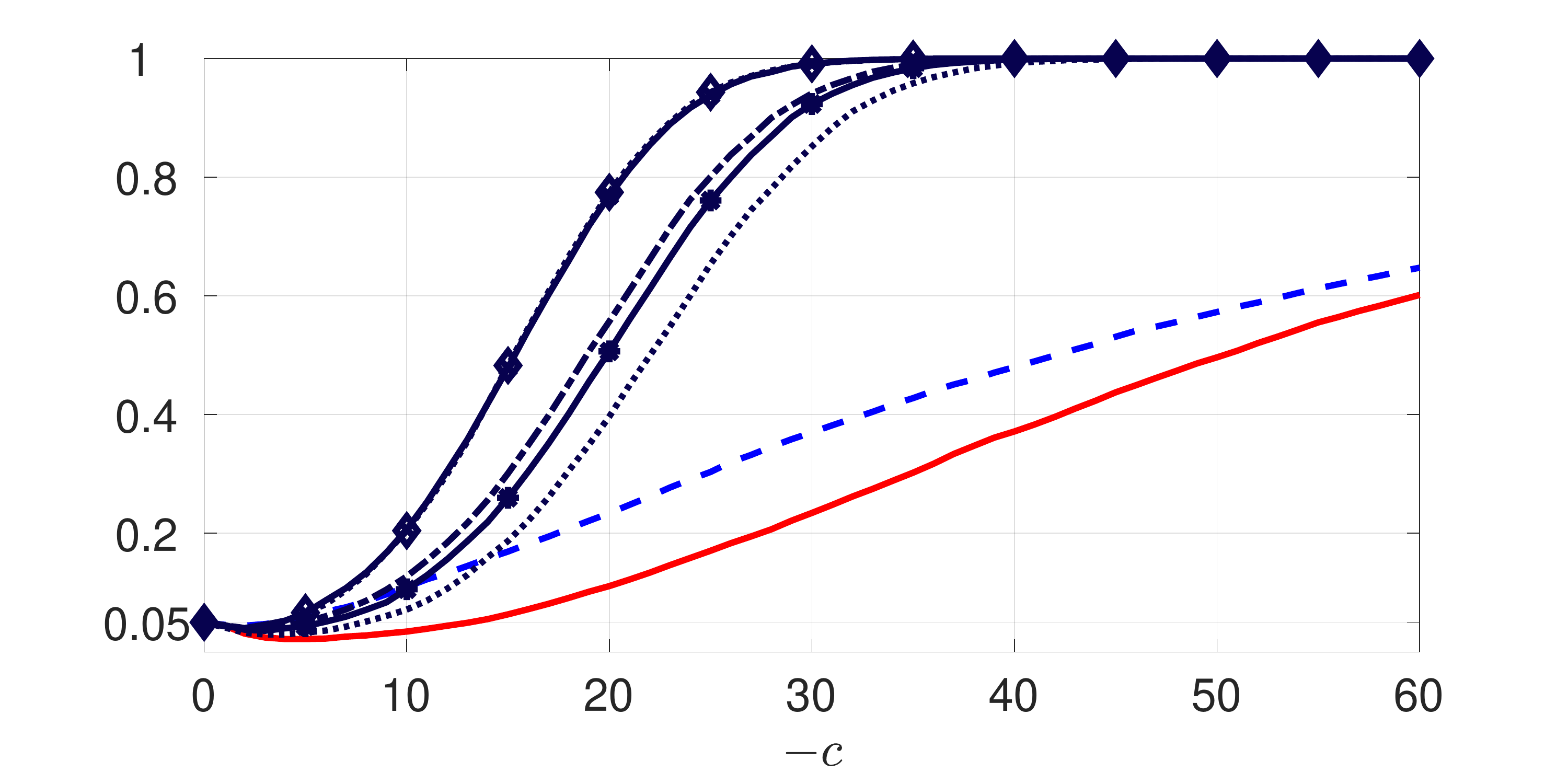}
	\end{subfigure}\begin{subfigure}{0.33\textwidth}
		\centering
		\caption*{$R^2=0.8$}
		\vspace{-1ex}
		\includegraphics[trim={2cm 0.2cm 2cm 0.5cm},width=0.98\textwidth,clip]{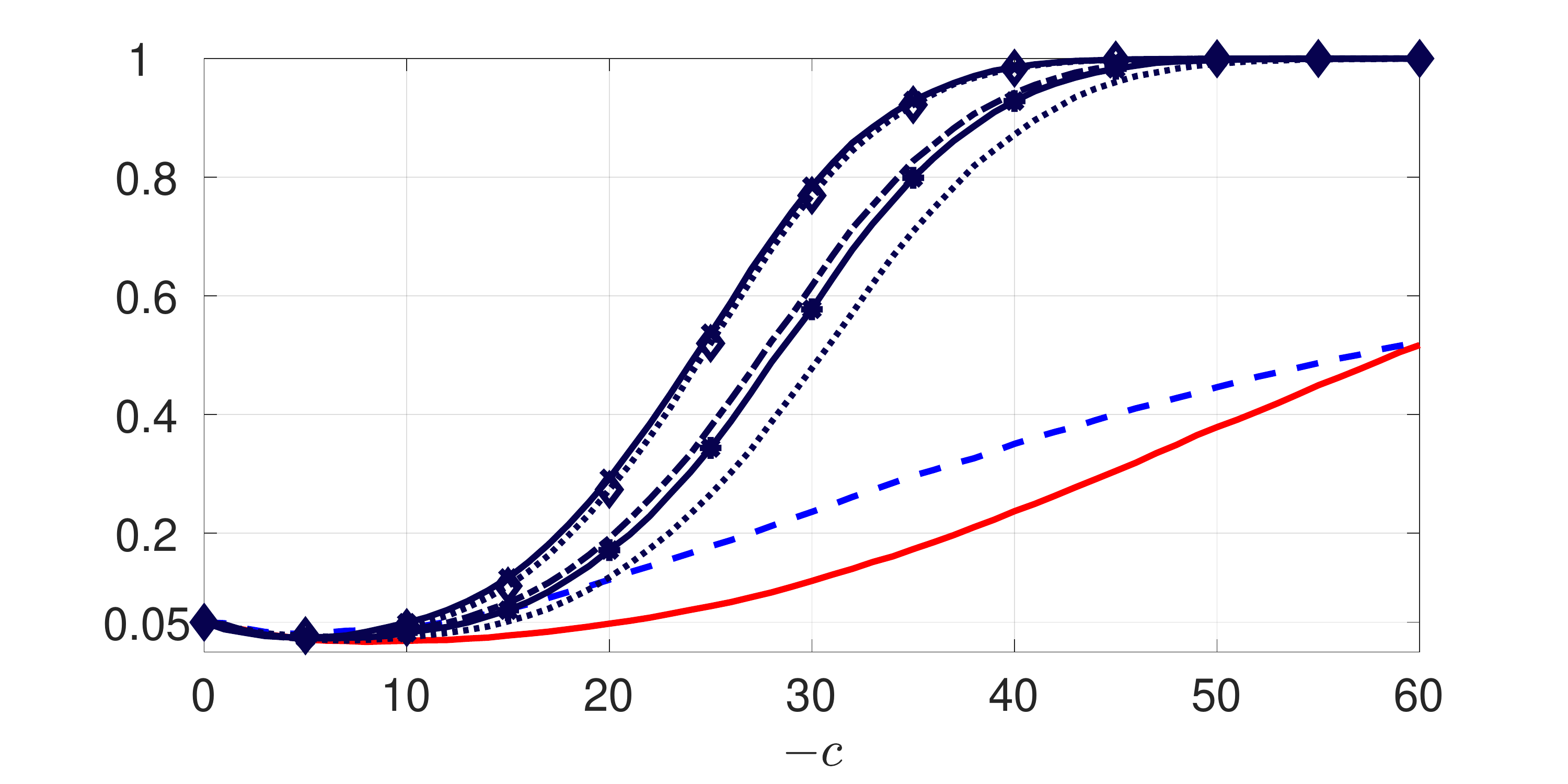}
\end{subfigure}
\end{center}
\vspace{-2ex}
\caption{Asymptotic power of the tests at the nominal $5\%$ level for $\text{$\text{H}_0$:}\ \rho = 1$ under the local alternative $\rho = \rho_T = 1+c/T$ in cases D0 (first column), D1 (second column), and D2 (third column) for $m=1$.\\
{\footnotesize Note: The results are based on $10{,}000$ Monte Carlo replications and standard Brownian motions are approximated by normalized partial sums of $10{,}000$ i.i.d.\ standard normal random variables.}}
\label{fig:local_power_m_1}
\end{figure}

\begin{remark}\label{rem:GLSLAP}
	In case of GLS detrending, it follows from \citeasnoun[Lemma~5.3]{PeRo16} and similar arguments as used in \citeasnoun[Proof of Theorem~5.2]{PeRo16} and in the proof of Proposition~\ref{prop:VR_H0} that the variance ratio test statistic converges under the local alternative $\rho=\rho_T = 1+c/T$, where $c\leq 0$, to
	\begin{align}
		\text{VR}^{(\GLS)} \overset{w}{\longrightarrow} \mathcal{G}_{\scriptstyle\text{VR},c}^{(\GLS)}(\bar c) \coloneqq \frac{\kappa_c^{(\GLS)\prime} C_c^{(\GLS)} \kappa_c^{(\GLS)}}{\kappa_c^{(\GLS)\prime} A_c^{(\GLS)} \kappa_c^{(\GLS)}},
	\end{align}
	with $A_c^{(\GLS)}$, $C_c^{(\GLS)}$ and $\kappa_c^{(\GLS)}$ defined analogously to $\widetilde A_c$, $\widetilde C_c$ and $\widetilde \kappa_c$, respectively: $\widetilde J_{\xi\cdot v}^c(r)$ and $\widetilde W_v(r)$ have to be replaced by $J_{\xi\cdot v}^c(r)$ and $W_v(r)$, respectively, in case D1 and with $J_{\xi\cdot v}^c(r) - \left(\lambda J_{\xi\cdot v}^c(1) + 3(1-\lambda)\int_0^1 s J_{\xi\cdot v}^c(s)ds\right)r$ and $W_v(r) - \left(\lambda W_v(1) + 3(1-\lambda)\int_0^1 s W_v(s)ds\right)r$, respectively, in case D2, where $\lambda \coloneqq (1-\bar c)/(1-\bar c + \bar{c}^2/3)$ and $\bar{c}$ as chosen in Remark~\ref{rem:GLSDetrending}. In case D1, $\mathcal{G}_{\scriptstyle\text{VR},c}^{(\GLS)}(\bar c)$ does not depend on $\bar c$ and coincides with $\mathcal{G}_{{\scriptstyle\text{VR}},c}$ as defined in Proposition~\ref{prop:VR_Local} in case D0. The limiting null distribution of the variance ratio test based on GLS detrending is given by $\mathcal{G}_{\scriptstyle\text{VR},0}^{(\GLS)}(\bar c)$.
\end{remark}
\begin{remark}\label{rem:cbar}
	To provide some guidance on the choice of $\bar c$, we follow \citeasnoun[p.\,93]{PeRo16} and choose $\bar{c}$ such that $P\left(\mathcal{G}_{\scriptstyle\text{VR},\bar c}^{(\GLS)}(\bar c)<q_{0.05}(\bar c)\right) = 0.5$ when $R^2=0.4$, where $q_\alpha(\bar c)$ satisfies $P\left(\mathcal{G}_{\scriptstyle\text{VR},0}^{(\GLS)}(\bar c)<q_\alpha(\bar c)\right) = 0.05$.\footnote{See also the discussion in \citeasnoun[p.\,1528]{Ni09}.} Table~\ref{tab:cbar} in Appendix~\ref{app:Critvals} displays the values of $\bar{c}$ in cases D1 and D2 for $m=1,\ldots,5$. Given these values of $\bar c$, Table~\ref{tab:critvalsVR} in Appendix~\ref{app:Critvals} tabulates the corresponding critical values $q_\alpha(\bar c)$ for the variance ratio test based on GLS detrended data.
\end{remark}
Figure~\ref{fig:local_power_m_1} illustrates the local asymptotic power curve of the variance ratio test at the nominal $5\%$ level (in cases D1 and D2 based on OLS detrended and GLS detrended data) for $m=1$ and $R^2\in\{0,0.4,0.8\}$. In case of GLS detrending, $\bar{c}$ is chosen as suggested in Table~\ref{tab:cbar} in Appendix~\ref{app:Critvals} -- with results being qualitatively similar for other choices of $\bar{c}$. The figure also displays the local asymptotic power curves of the ADF test, the $\widehat{Z}_{\alpha}$ test, and the MSB test based on OLS detrended data derived in~\citeasnoun{Pe07} and the local asymptotic power curve of the ADF and MSB tests based on GLS detrended data derived in \citeasnoun{PeRo16}, with the value of $\bar{c}$ as suggested in \citeasnoun[Table~1]{PeRo16}.\footnote{\citeasnoun{PeRo16} derive, among others, the limiting distributions of the ADF, $\widehat{Z}_{\alpha}$, and MSB tests based on GLS detrended data under local alternatives. However, the analytical expressions of the limiting distributions of the ADF test and the $\widehat{Z}_{\alpha}$ test (the $\widehat{Z}_{\hat\rho}$ test in their notation) in Theorem~5.2 contain typos. The square root should be removed from the numerator in the second term of the limiting distribution of the ADF test and from both the numerator and the denominator in the second term of the limiting distribution of the $\widehat{Z}_{\alpha}$ test \citeaffixed[Theorem~1]{Pe07}{cf.}. The results in Figure~\ref{fig:local_power_m_1} are based on the corrected version of the limiting distribution of the ADF test.} 

As expected, local asymptotic power of all tests decreases when $R^2$ increases, with the power loss being least pronounced in case D0 and most pronounced in case D2. Figure~\ref{fig:local_power_m_1} further reveals that local asymptotic power of the ADF test (in the presence of deterministics based on OLS detrended data), the $\widehat{Z}_{\alpha}$ test, and the MSB test is very similar throughout and considerably larger than local asymptotic power of the variance ratio test, irrespective of whether the variance ratio test is based on OLS detrended data or GLS detrended data.\footnote{\citeasnoun{Ho14} analyzes local asymptotic power of \citename{Br02}'s \citeyear{Br02} unit root test when applied to an observed (potentially OLS detrended) univariate time series and finds similar power losses.}

In cases D1 and D2, GLS detrending increases local asymptotic power of the variance ratio test for $R^2=0.8$, but this power improvement seems to be negligible when compared with local asymptotic power of the other tests considered here. On the contrary, for $R^2=0$ local asymptotic power of the variance ratio test based on GLS detrended data is much smaller than local asymptotic power of its OLS-based counterpart. For $R^2=0.4$, GLS detrending seems to lead to minor power improvements for small values of $c$, but as $c$ moves further away from zero GLS detrending becomes disadvantageous. We conclude that in terms of local asymptotic power, the variance ratio test does not benefit from GLS detrending, which is in line with the findings in \citeasnoun{BrTa03} and \citeasnoun{Ni09} for \citename{Br02}'s \citeyear{Br02} unit root test applied to observed univariate time series. In contrast, GLS detrending improves local asymptotic power of the ADF and MSB tests considerably throughout, which is in line with the findings in \citeasnoun{PeRo16}. Finally, note that results for other values of $m$ are qualitatively similar, with local asymptotic power being lower overall for larger values of $m$.

The results clearly demonstrate that the ADF, $\widehat{Z}_{\alpha}$, and MSB tests outperform the variance ratio test in terms of local asymptotic power. These power losses might not come as a surprise, given the self-normalizing property of the variance ratio test and the fact that self-normalized tests are well known for having smaller local asymptotic power than their (semi-)parametric counterparts \citeaffixed{Sh15}{cf., \eg,}. However, local asymptotic power analyses are unable to reveal finite sample effects of both different short-run dynamics in the DGP and tuning parameter choices. The next section complements the local asymptotic power analysis with a careful assessment of the performance of the tests in finite samples.

\section{Finite Sample Performance}\label{sec:FiniteSample}
We generate data according to \eqref{eq:x} and \eqref{eq:y} with $m=1$ regressor, \ie,
\begin{align} 
	x_{t} &= \mu + x_{t-1} + v_{t} = x_{0} + \mu t + \sum_{s=1}^t v_{s},\label{eq:FSx}\\
	y_t &= d_t'\tau + x_{t}\beta + u_t,\label{eq:FSy}
\end{align}
for $t=1,\ldots,T$. The regression errors are generated as $u_t =\rho u_{t-1} + \xi_t$ using the following five different short-run dynamics:
\begin{align*}
	\xi_t =
	\begin{cases}
		\ve_t & \text{(IID)}\\
		\phi \xi_{t-1} + \ve_t & \text{(AR)}\\
		\ve_t - \theta\ve_{t-1} & \text{(MA)}\\
		\phi \xi_{t-1} + \ve_t - \theta\ve_{t-1} & \text{(ARMA)}\\
		\sqrt{h_t}\ve_t, \quad h_t = (1-a_1-a_2) + a_1\xi_{t-1}^2 + a_2 h_{t-1} & \text{(GARCH)},
	\end{cases}
\end{align*}
for $t=-99,\ldots,0,1,\ldots,T$.\footnote{We set $u_{-100}=\xi_{-100}=0$ and $h_{-100}=1$. The period $t=-99,\ldots,0$ serves as a burn-in period to eliminate these starting-value effects. Section~\ref{sec:FiniteSample_u0} analyzes the effect of a large initial value $u_0$ on the performance of the tests in finite samples.} The vectors $[\ve_t,v_t]'$, are i.i.d.~across $t$ and follow a zero-mean bivariate normal distribution with covariance matrix $\Sigma\coloneqq \begin{bmatrix}1&\sigma_{\ve v}\\\sigma_{\ve v}&1\end{bmatrix}$. To ensure stationarity of $\xi_t$, the parameter values are restricted to $\vert \phi \vert,\vert \theta \vert < 1$ and $a_1+a_2 <1$, with $a_1,a_2\geq 0$. In the IID and GARCH cases, the long-run covariance matrix of $[\xi_t,v_t]'$ is given by $\Omega_{\text{IID}}=\Omega_{\text{GARCH}}=\Sigma$, whereas in the ARMA case it is given by $\Omega_{\text{ARMA}} \coloneqq \begin{bmatrix}
	\frac{(1-\theta)^2}{(1-\phi)^2}& \frac{(1-\theta)\sigma_{\ve v}}{1-\phi}\\
	\frac{(1-\theta)\sigma_{\ve v}}{1-\phi}& 1
\end{bmatrix}$. In the AR and MA cases, the long-run covariance matrices $\Omega_{\text{AR}}$ and $\Omega_{\text{MA}}$ can be directly deduced from $\Omega_{\text{ARMA}}$ by setting $\theta$ and $\phi$, respectively, equal to zero. Although the long-run covariance matrix of $[\xi_t,v_t]'$ differs across types of short-run dynamics, the only nuisance parameter affecting local asymptotic power of the residuals-based no-cointegration tests, \ie, $R^2$ (compare the discussion in Section~\ref{sec:Th-LAPower}), is equal to $\sigma_{\ve v}^2$ in each case. The DGP thus allows to analyze the effects of different short-run dynamics in $\xi_t$ on the performance of residuals-based no-cointegration tests in finite samples while controlling for effects of $R^2$.\footnote{Controlling for effects of $R^2$ allows to compare the finite sample results in this section with the local asymptotic power results obtained in Section~\ref{sec:Th-LAPower}.} In contrast, the local asymptotic power analysis in Section~\ref{sec:Th-LAPower} only allows to assess the effect of $R^2$.
\begin{remark}\label{rem:DGPs}
	\citeasnoun[p.\,127]{Pe07} and \citeasnoun[p.\,99]{PeRo16} state that local asymptotic power results serve as a useful indicator for the performance of residuals-based no-cointegration tests in finite samples. However, \citeasnoun{Pe07} only considers a DGP similar to our IID case, while \citeasnoun{PeRo16} allow the errors to be serially correlated -- using a vector autoregression model of order one to generate $[\xi_t,v_t]'$ -- but restrict the diagonal and off-diagonal elements of the long-run covariance matrix of $[\xi_t,v_t]'$ to one and a constant $r$, respectively, such that $R^2=r^2$. \citeasnoun[p.\,99]{PeRo16} correctly point out that holding the long-run covariance matrix fixed implies that changes in the autoregressive parameters are compensated by changes in the covariance matrix of $[\xi_t,v_t]'$. Consequently, changes in the autoregressive parameters do not affect the performance of the tests in their DGP. In contrast, we shall see below that the DGP considered in this paper is able to detect severe effects of different short-run dynamics on the performance of residuals-based no-cointegration tests. The finite sample results in this section thus differ considerably from the local asymptotic power results in Section~\ref{sec:Th-LAPower}.
\end{remark}
Parameters are chosen as follows: In all cases, we set $\beta=1$. Moreover, we set $\phi,\theta\in\{0.3,0.6,0.9\}$ in the AR and MA models, $(\phi,\theta)\in\{(0.3,0.6),(0.3,0.3),(0.6,0.3)\}$ in the ARMA model, and $(a_1,a_2)\in\{(0.05,0.94),(0.01,0.98)\}$ in the GARCH model. In addition, we set $x_{0}=0$ and $\mu=0$ in case D0, $x_{0}=1$, $\mu=0$ and $\tau=1$ in case D1, and $x_{0}=1$, $\mu=1$ and $\tau=[1,1]'$ in case D2. We present results for $R^2\in\{0,0.4,0.8\}$ by choosing $\sigma_{\ve v}\geq0$ accordingly and $T\in\{100,250\}$. All results are based on $5{,}000$ Monte Carlo replications and all tests are carried out at the nominal $5\%$ level.

We compare the variance ratio test with the ADF test, the MSB test, and the $\widehat Z_\alpha$ test in terms of empirical size and size-corrected power. In cases D1 and D2, the tests are based on OLS detrended data. In addition, we also consider the variance ratio test and the ADF test based on GLS detrended data, indicated by the superscript ``(GLS)'', where $\bar{c}$ is chosen as suggested in Table~\ref{tab:cbar} in Appendix~\ref{app:Critvals} and in Table~1 of \citeasnoun{PeRo16}, respectively. 

In contrast to the variance ratio test, the ADF, MSB, and $\widehat Z_\alpha$ tests require tuning parameter choices. The number of lags for the ADF test and the MSB test is selected using AIC. For the ADF test and its GLS-version, we also analyze the results based on a modified AIC (MAIC) criterion proposed in \citeasnoun{NgPe01} for unit root testing in an observed univariate time series, taking into account further modifications suggested in \citeasnoun{PeQu07}.\footnote{\citeasnoun{PeQu07} suggest to construct the ADF statistic based on GLS detrended data and to determine the number of lags in the auxiliary regression based on OLS detrended data.} We also analyze the performance of the GLS-version of the MSB test in combination with the MAIC. The use of MAIC is indicated by the superscript ``$*$''. For both information criteria, the minimal number of lags is zero and the maximal number of lags is restricted to not exceed $\floor{12(T/100)^{1/4}}$, the upper bound suggested in \citeasnoun{PeQu07}.\footnote{The Supplementary Material provides results based on BIC and MBIC, with MBIC defined analogously to MAIC (compare Section~\ref{app:Conv-Tests-IC} in the Appendix). The tests based on MBIC perform similar to those based on MAIC. In contrast, the tests based on BIC are more prone to severe size distortions than those based on AIC, but often reveal some power advantages compared to their AIC based counterparts. These differences in size-corrected power, however, do not alter the overall picture emerging from the discussion in Section~\ref{sec:FiniteSample_Power}. In particular, the tests based on both BIC and MBIC suffer from similar power reversal problems as observed for the tests based on AIC and MAIC in Section~\ref{sec:FiniteSample_Power}.} The $\widehat Z_\alpha$ test is based on a kernel estimator of a long-run variance parameter. We present results for the quadratic spectral (QS) kernel together with the corresponding data-dependent bandwidth-selection rule of~\citeasnoun{An91}.\footnote{Using the Bartlett kernel instead of the QS kernel often leads to slightly larger size distortions.} For convenience, Appendix~\ref{app:Conv-Tests} describes the construction of the ADF, MSB, and $\widehat Z_\alpha$ tests in more detail.\footnote{When the ADF and MSB tests are used in applications, it seems to be more popular to perform OLS detrending rather than GLS detrending and employing the AIC or BIC rather than the MAIC or MBIC. The following results thus also allow to assess whether practitioners should stick to these ``default'' choices or not.}

\subsection{Empirical Size}\label{sec:FiniteSample_Size}
Table~\ref{tab:RejectionRates-m1-T100-D1D2} presents empirical sizes of the tests under the null hypothesis of no cointegration ($\rho=1$) in cases D1 and D2 for $T=100$. The results directly reveal that the performance of the tests depends heavily on the short-run dynamics in $\xi_t$. In particular, size distortions in the MA and ARMA cases can be more severe than in the IID, AR, and GARCH cases. Moreover, relative to the IID case with $R^2=0$, changing the short-run dynamics in $\xi_t$ can have much larger adverse effects on the performance of the tests than increasing $R^2$ to 0.8. 

Focusing on the performance of the different tests in detail reveals that the variance ratio test is less size-distorted than the ADF and MSB tests. The variance ratio test often also outperforms the $\widehat Z_\alpha$ test, especially in the MA case with $\theta<0.9$ and in the ARMA(0.3,0.6) case. GLS detrending worsens the performance of the variance ratio test, especially in case D2, and of the ADF test, but the GLS detrended version of the ADF test in combination with the MAIC performs relatively well. Similarly, the GLS detrended version of the MSB test in combination with the MAIC has much smaller size distortions than the MSB test, but it tends to be very conservative. 

Comparing the ADF test with the $\text{ADF}^*$ test reveals that using the MAIC criterion is also advantageous under OLS detrending. In most cases, the variance ratio test and the $\text{ADF}^*$ test perform similarly, but the $\text{ADF}^*$ test has some performance advantages over the variance ratio test in the MA case with $\theta >0.3$ and in the ARMA(0.3,0.6) case. These performance advantages are the more pronounced the larger $R^2$. 

Increasing the sample size reduces the size distortions of the tests, especially for $\text{VR}^{(\GLS)}$ in case D2, and makes the $\text{MSB}^{(\GLS)*}$ less conservative, compare Table~\ref{tab:RejectionRates-m1-T250-D1D2} in Appendix~\ref{app:addresults}, which presents the results for $T=250$. Beyond that, the results are qualitatively similar to those for $T=100$. Overall, the findings also hold in case D0, although the variance ratio test seems to be rather conservative in this case even for $T=250$, compare Table~\ref{tab:RejectionRates-m1-D0} in Appendix~\ref{app:addresults}. 

\begin{table}[!ht]%14.07.2022
\centering
\adjustbox{max width=\textwidth}{\begin{threeparttable}
\caption{Empirical sizes of the tests in cases D1 and D2 for $T=100$.}
\label{tab:RejectionRates-m1-T100-D1D2}
\begin{tabular}{clcccccccccccc}
	\toprule[1pt]\midrule[0.3pt]
	\multicolumn{2}{c}{}&\multicolumn{1}{c}{}&\multicolumn{3}{c}{AR}&\multicolumn{3}{c}{MA}&\multicolumn{3}{c}{ARMA}&\multicolumn{2}{c}{GARCH}\\
	\cmidrule(lr){4-6}
	\cmidrule(lr){7-9}
	\cmidrule(lr){10-12}
	\cmidrule(lr){13-14}
	$R^2$&Test & IID & 0.3 & 0.6 & 0.9 & 0.3 & 0.6 & 0.9 & (0.3,0.6) & (0.3,0.3) & (0.6,0.3) & (0.05,0.94) & (0.01,0.98) \\\midrule
	\multicolumn{14}{l}{Deterministic specification D1}\\
	\midrule
	0&$\text{VR}$ & 0.05 & 0.04 & 0.03 & 0.01 & 0.07 & 0.16 & 0.72 & 0.10 & 0.05 & 0.03 & 0.05 & 0.05 \\
	&$\text{VR}^{(\GLS)}$ & 0.12 & 0.11 & 0.08 & 0.03 & 0.15 & 0.26 & 0.82 & 0.19 & 0.12 & 0.09 & 0.12 & 0.12 \\
	&ADF & 0.08 & 0.07 & 0.07 & 0.05 & 0.13 & 0.23 & 0.83 & 0.24 & 0.08 & 0.06 & 0.09 & 0.08 \\
	&$\text{ADF}^{(\GLS)}$ & 0.11 & 0.10 & 0.09 & 0.06 & 0.17 & 0.26 & 0.70 & 0.28 & 0.11 & 0.09 & 0.11 & 0.11 \\
	&$\text{ADF}^*$ & 0.04 & 0.02 & 0.02 & 0.02 & 0.05 & 0.07 & 0.34 & 0.08 & 0.04 & 0.01 & 0.04 & 0.04 \\
	&$\text{ADF}^{(\GLS)*}$ & 0.06 & 0.03 & 0.03 & 0.01 & 0.08 & 0.10 & 0.37 & 0.12 & 0.06 & 0.02 & 0.06 & 0.06 \\
	&MSB & 0.11 & 0.14 & 0.14 & 0.15 & 0.14 & 0.21 & 0.77 & 0.22 & 0.11 & 0.14 & 0.12 & 0.11 \\
	&$\text{MSB}^{(\GLS)*}$ & 0.02 & 0.02 & 0.03 & 0.03 & 0.04 & 0.05 & 0.27 & 0.08 & 0.02 & 0.01 & 0.03 & 0.02 \\
	&$\widehat{Z}_{\alpha}$ & 0.05 & 0.02 & 0.01 & 0.01 & 0.19 & 0.72 & 1.00 & 0.33 & 0.05 & 0.01 & 0.05 & 0.05 \\\midrule
	0.4&$\text{VR}$ & 0.05 & 0.04 & 0.03 & 0.01 & 0.07 & 0.19 & 0.78 & 0.11 & 0.05 & 0.03 & 0.05 & 0.05 \\
	&$\text{VR}^{(\GLS)}$ & 0.12 & 0.11 & 0.09 & 0.05 & 0.16 & 0.29 & 0.85 & 0.20 & 0.12 & 0.09 & 0.12 & 0.12 \\
	&ADF & 0.08 & 0.08 & 0.08 & 0.05 & 0.14 & 0.28 & 0.87 & 0.26 & 0.08 & 0.07 & 0.09 & 0.08 \\
	&$\text{ADF}^{(\GLS)}$ & 0.11 & 0.11 & 0.10 & 0.07 & 0.17 & 0.29 & 0.75 & 0.28 & 0.11 & 0.09 & 0.11 & 0.11 \\
	&$\text{ADF}^*$ & 0.04 & 0.02 & 0.03 & 0.02 & 0.05 & 0.07 & 0.38 & 0.08 & 0.04 & 0.02 & 0.04 & 0.04 \\
	&$\text{ADF}^{(\GLS)*}$ & 0.06 & 0.03 & 0.03 & 0.02 & 0.08 & 0.10 & 0.42 & 0.12 & 0.06 & 0.03 & 0.06 & 0.06 \\
	&MSB & 0.12 & 0.14 & 0.14 & 0.14 & 0.14 & 0.24 & 0.82 & 0.25 & 0.12 & 0.14 & 0.12 & 0.12 \\
	&$\text{MSB}^{(\GLS)*}$ & 0.03 & 0.02 & 0.03 & 0.03 & 0.04 & 0.05 & 0.30 & 0.08 & 0.03 & 0.02 & 0.03 & 0.03 \\
	&$\widehat{Z}_{\alpha}$ & 0.06 & 0.03 & 0.03 & 0.02 & 0.21 & 0.79 & 1.00 & 0.39 & 0.06 & 0.02 & 0.06 & 0.06 \\\midrule
	0.8&$\text{VR}$ & 0.05 & 0.05 & 0.06 & 0.01 & 0.09 & 0.31 & 0.87 & 0.15 & 0.05 & 0.05 & 0.05 & 0.05 \\
	&$\text{VR}^{(\GLS)}$ & 0.12 & 0.12 & 0.14 & 0.06 & 0.18 & 0.43 & 0.93 & 0.26 & 0.12 & 0.12 & 0.12 & 0.12 \\
	&ADF & 0.08 & 0.10 & 0.16 & 0.06 & 0.15 & 0.41 & 0.92 & 0.33 & 0.08 & 0.11 & 0.10 & 0.08 \\
	&$\text{ADF}^{(\GLS)}$ & 0.11 & 0.13 & 0.20 & 0.10 & 0.19 & 0.40 & 0.79 & 0.35 & 0.11 & 0.15 & 0.12 & 0.11 \\
	&$\text{ADF}^*$ & 0.04 & 0.04 & 0.07 & 0.03 & 0.05 & 0.10 & 0.45 & 0.10 & 0.04 & 0.05 & 0.04 & 0.04 \\
	&$\text{ADF}^{(\GLS)*}$ & 0.06 & 0.06 & 0.11 & 0.05 & 0.08 & 0.14 & 0.48 & 0.14 & 0.06 & 0.07 & 0.06 & 0.06 \\
	&MSB & 0.11 & 0.13 & 0.16 & 0.12 & 0.16 & 0.35 & 0.89 & 0.32 & 0.11 & 0.14 & 0.12 & 0.12 \\
	&$\text{MSB}^{(\GLS)*}$ & 0.03 & 0.03 & 0.06 & 0.03 & 0.04 & 0.06 & 0.36 & 0.09 & 0.03 & 0.04 & 0.03 & 0.03 \\
	&$\widehat{Z}_{\alpha}$ & 0.06 & 0.07 & 0.14 & 0.04 & 0.36 & 0.95 & 1.00 & 0.59 & 0.06 & 0.08 & 0.06 & 0.06 \\
	\midrule
	\multicolumn{14}{l}{Deterministic specification D2}\\
	\midrule
	0&$\text{VR}$ & 0.05 & 0.04 & 0.02 & 0.00 & 0.09 & 0.26 & 0.93 & 0.14 & 0.05 & 0.03 & 0.06 & 0.05 \\
	&$\text{VR}^{(\GLS)}$ & 0.30 & 0.26 & 0.20 & 0.07 & 0.36 & 0.59 & 0.99 & 0.45 & 0.30 & 0.22 & 0.30 & 0.30 \\
	&ADF & 0.10 & 0.10 & 0.08 & 0.05 & 0.19 & 0.35 & 0.91 & 0.35 & 0.10 & 0.08 & 0.10 & 0.10 \\
	&$\text{ADF}^{(\GLS)}$ & 0.14 & 0.13 & 0.12 & 0.07 & 0.23 & 0.37 & 0.85 & 0.39 & 0.14 & 0.11 & 0.14 & 0.14 \\
	&$\text{ADF}^*$ & 0.03 & 0.01 & 0.02 & 0.01 & 0.06 & 0.09 & 0.51 & 0.11 & 0.03 & 0.01 & 0.04 & 0.03 \\
	&$\text{ADF}^{(\GLS)*}$ & 0.06 & 0.02 & 0.03 & 0.01 & 0.08 & 0.12 & 0.54 & 0.15 & 0.06 & 0.01 & 0.06 & 0.06 \\
	&MSB & 0.15 & 0.20 & 0.22 & 0.25 & 0.19 & 0.28 & 0.85 & 0.27 & 0.15 & 0.22 & 0.17 & 0.15 \\
	&$\text{MSB}^{(\GLS)*}$ & 0.01 & 0.01 & 0.03 & 0.03 & 0.04 & 0.06 & 0.42 & 0.09 & 0.01 & 0.01 & 0.02 & 0.01 \\
	&$\widehat{Z}_{\alpha}$ & 0.04 & 0.01 & 0.00 & 0.00 & 0.24 & 0.87 & 1.00 & 0.43 & 0.04 & 0.00 & 0.05 & 0.04 \\\midrule
	0.4&$\text{VR}$ & 0.05 & 0.04 & 0.03 & 0.00 & 0.09 & 0.30 & 0.94 & 0.15 & 0.05 & 0.03 & 0.05 & 0.05 \\
	&$\text{VR}^{(\GLS)}$ & 0.30 & 0.26 & 0.22 & 0.08 & 0.38 & 0.64 & 0.99 & 0.47 & 0.30 & 0.23 & 0.30 & 0.30 \\
	&ADF & 0.11 & 0.10 & 0.09 & 0.05 & 0.20 & 0.40 & 0.92 & 0.38 & 0.11 & 0.09 & 0.11 & 0.11 \\
	&$\text{ADF}^{(\GLS)}$ & 0.14 & 0.13 & 0.11 & 0.08 & 0.23 & 0.41 & 0.86 & 0.41 & 0.14 & 0.11 & 0.14 & 0.14 \\
	&$\text{ADF}^*$ & 0.04 & 0.01 & 0.02 & 0.01 & 0.06 & 0.10 & 0.48 & 0.11 & 0.04 & 0.01 & 0.04 & 0.04 \\
	&$\text{ADF}^{(\GLS)*}$ & 0.06 & 0.02 & 0.03 & 0.01 & 0.09 & 0.14 & 0.51 & 0.16 & 0.06 & 0.01 & 0.06 & 0.06 \\
	&MSB & 0.16 & 0.19 & 0.20 & 0.23 & 0.19 & 0.33 & 0.86 & 0.30 & 0.16 & 0.21 & 0.17 & 0.16 \\
	&$\text{MSB}^{(\GLS)*}$ & 0.02 & 0.01 & 0.02 & 0.03 & 0.04 & 0.07 & 0.39 & 0.10 & 0.02 & 0.01 & 0.02 & 0.02 \\
	&$\widehat{Z}_{\alpha}$ & 0.04 & 0.01 & 0.01 & 0.00 & 0.28 & 0.93 & 1.00 & 0.48 & 0.04 & 0.01 & 0.05 & 0.05 \\\midrule
	0.8&$\text{VR}$ & 0.05 & 0.05 & 0.06 & 0.00 & 0.11 & 0.48 & 0.98 & 0.20 & 0.05 & 0.04 & 0.05 & 0.05 \\
	&$\text{VR}^{(\GLS)}$ & 0.29 & 0.29 & 0.32 & 0.09 & 0.43 & 0.81 & 1.00 & 0.56 & 0.29 & 0.28 & 0.30 & 0.29 \\
	&ADF & 0.11 & 0.11 & 0.15 & 0.04 & 0.22 & 0.57 & 0.93 & 0.47 & 0.11 & 0.11 & 0.11 & 0.11 \\
	&$\text{ADF}^{(\GLS)}$ & 0.13 & 0.15 & 0.20 & 0.06 & 0.25 & 0.55 & 0.86 & 0.48 & 0.13 & 0.15 & 0.14 & 0.14 \\
	&$\text{ADF}^*$ & 0.04 & 0.03 & 0.05 & 0.01 & 0.06 & 0.14 & 0.45 & 0.14 & 0.04 & 0.03 & 0.04 & 0.04 \\
	&$\text{ADF}^{(\GLS)*}$ & 0.06 & 0.05 & 0.08 & 0.01 & 0.09 & 0.18 & 0.48 & 0.18 & 0.06 & 0.05 & 0.06 & 0.06 \\
	&MSB & 0.16 & 0.17 & 0.17 & 0.17 & 0.21 & 0.48 & 0.88 & 0.41 & 0.16 & 0.17 & 0.17 & 0.16 \\
	&$\text{MSB}^{(\GLS)*}$ & 0.02 & 0.02 & 0.03 & 0.01 & 0.04 & 0.10 & 0.36 & 0.12 & 0.02 & 0.01 & 0.02 & 0.02 \\
	&$\widehat{Z}_{\alpha}$ & 0.05 & 0.05 & 0.09 & 0.00 & 0.46 & 1.00 & 1.00 & 0.70 & 0.05 & 0.05 & 0.05 & 0.05 \\
	\midrule[0.3pt]\bottomrule[1pt]
\end{tabular}
\begin{tablenotes}
	\item Note: The superscripts ``(GLS)'' and ``$*$'' indicate GLS detrending instead of OLS detrending and the use of MAIC instead of AIC, respectively.
\end{tablenotes}
\end{threeparttable}}
\end{table}

\subsection{Size-Corrected Power}\label{sec:FiniteSample_Power}
To analyze the finite sample properties of the tests under deviations from the null hypothesis, we generate data under the alternatives $\rho = \rho_T = 1+c/T$ using an equidistant grid of 21 points over the interval $[0,60]$ for the values of $-c$.\footnote{Note that the alternatives move closer to the null hypothesis as the sample size increases. Hence, for (very) large sample sizes, the size-corrected power results should match the local asymptotic power results in Figure~\ref{fig:local_power_m_1}, irrespective of the short-run dynamics in $\xi_t$.} To account for the large performance differences between the tests in terms of empirical sizes under the null hypothesis ($c=0$), the analysis focuses on size-corrected (empirical) power. To this end, test decisions are based on case-specific empirical critical values obtained from simulations under the null hypothesis rather than on asymptotic critical values. All size-corrected power curves thus start at the nominal $5\%$ level.

\begin{sidewaysfigure}[!ht]
%\begin{figure}[!ht]
\begin{center}
\begin{subfigure}{0.25\textheight}
	\centering
	\caption*{IID}
	\vspace{-1ex}
	\includegraphics[trim={2cm 0.2cm 2cm 0.5cm},width=0.98\textwidth,clip]{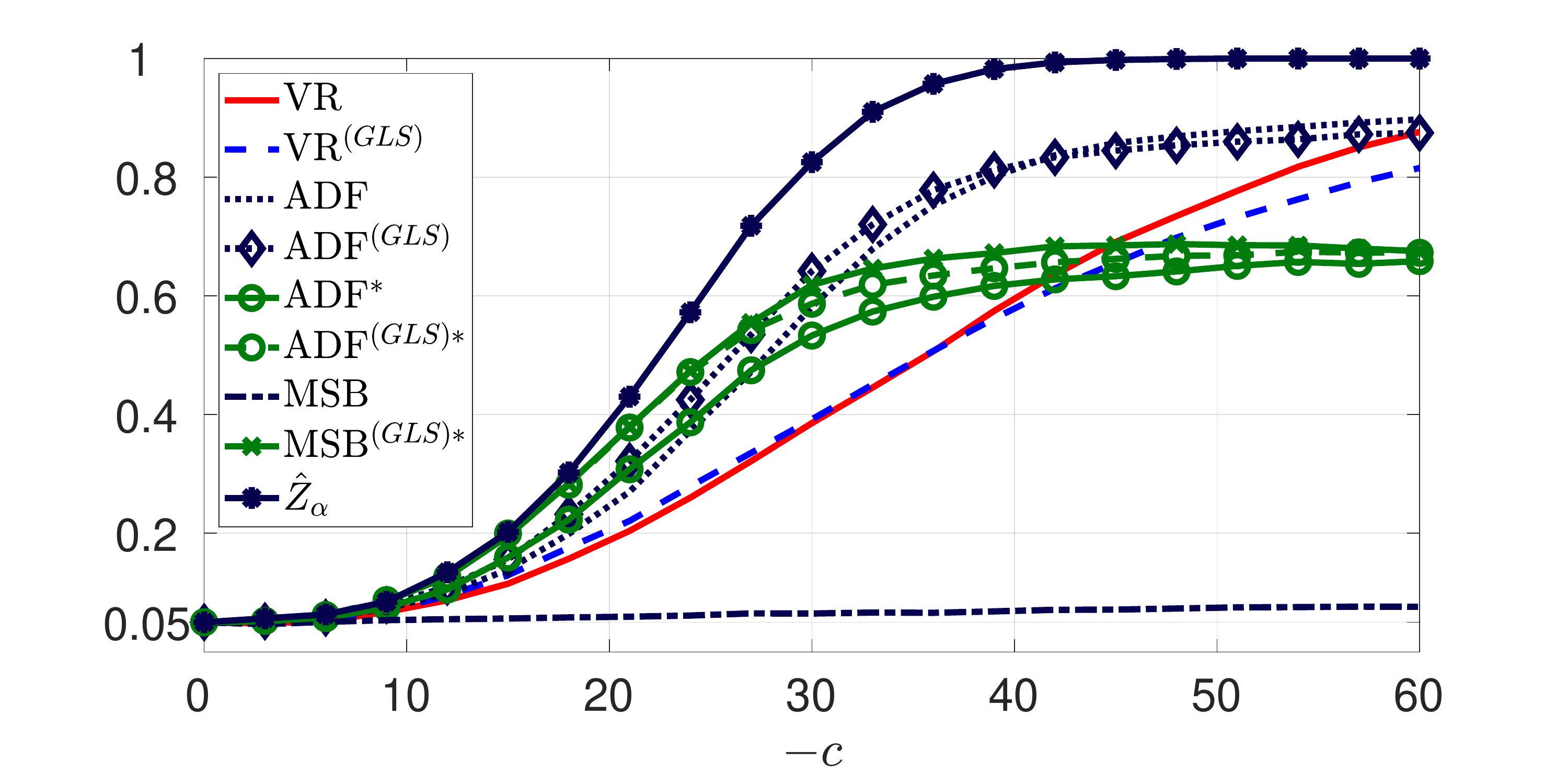}
\end{subfigure}\begin{subfigure}{0.25\textheight}
	\centering
	\caption*{AR, $\phi=0.3$}
	\vspace{-1ex}
	\includegraphics[trim={2cm 0.2cm 2cm 0.5cm},width=0.98\textwidth,clip]{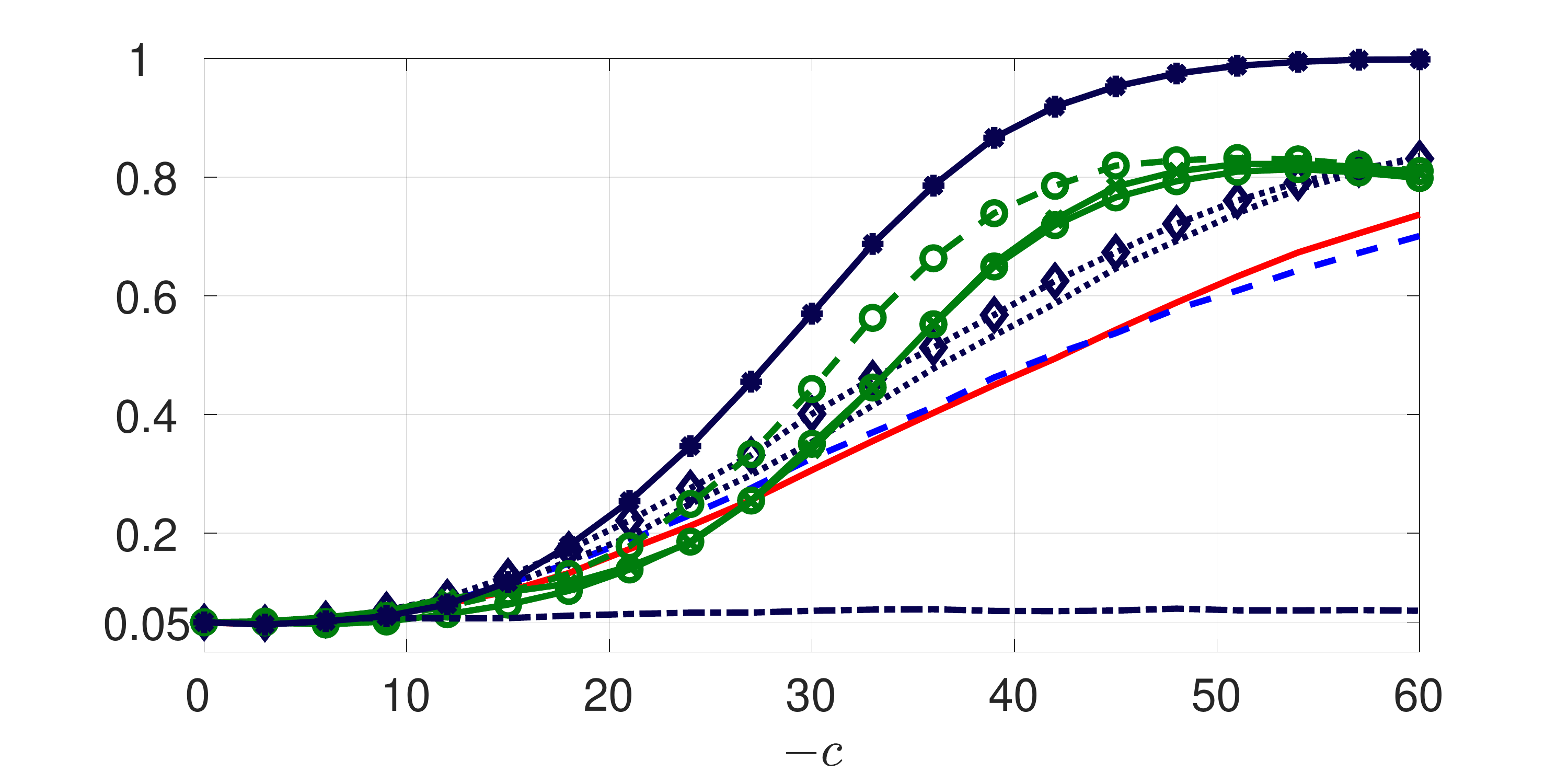}
\end{subfigure}\begin{subfigure}{0.25\textheight}
\centering
\caption*{AR, $\phi=0.6$}
\vspace{-1ex}
\includegraphics[trim={2cm 0.2cm 2cm 0.5cm},width=0.98\textwidth,clip]{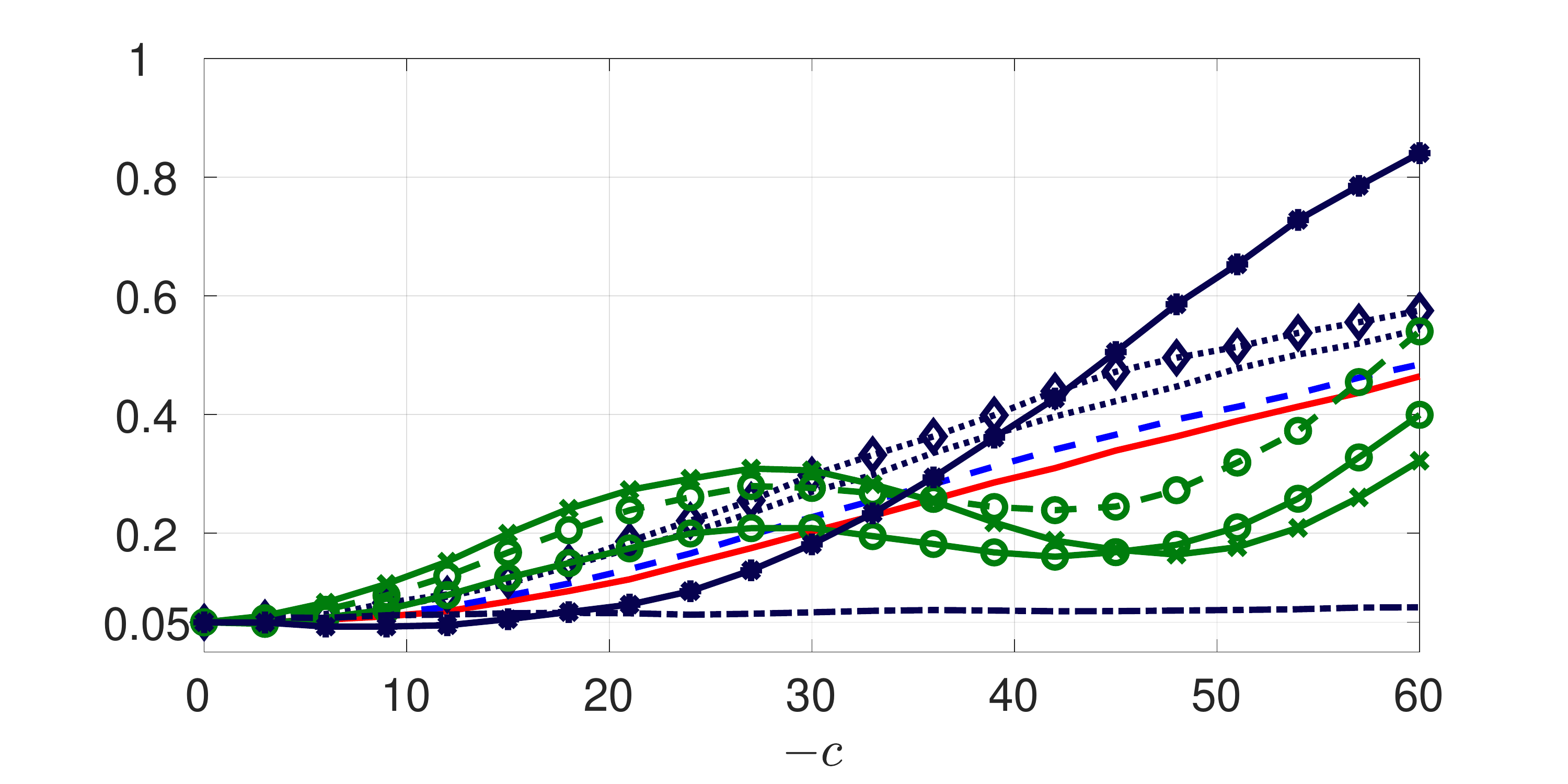}
\end{subfigure}\begin{subfigure}{0.25\textheight}
	\centering
	\caption*{AR, $\phi=0.9$}
	\vspace{-1ex}
	\includegraphics[trim={2cm 0.2cm 2cm 0.5cm},width=0.98\textwidth,clip]{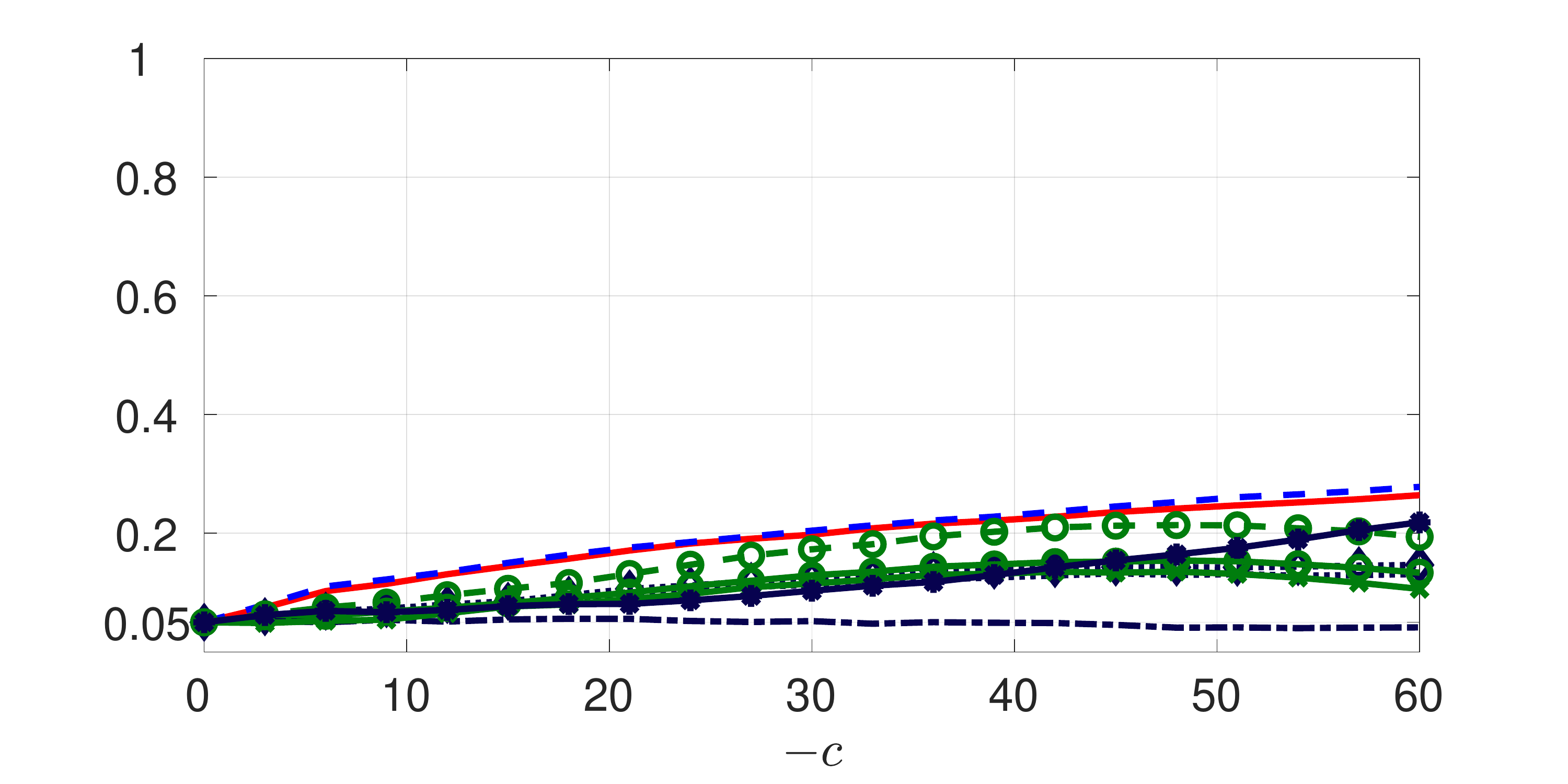}
\end{subfigure}

\vspace{1ex}

\begin{subfigure}{0.25\textheight}
	\centering
	\caption*{MA, $\theta=0.3$}
	\vspace{-1ex}
	\includegraphics[trim={2cm 0.2cm 2cm 0.5cm},width=0.98\textwidth,clip]{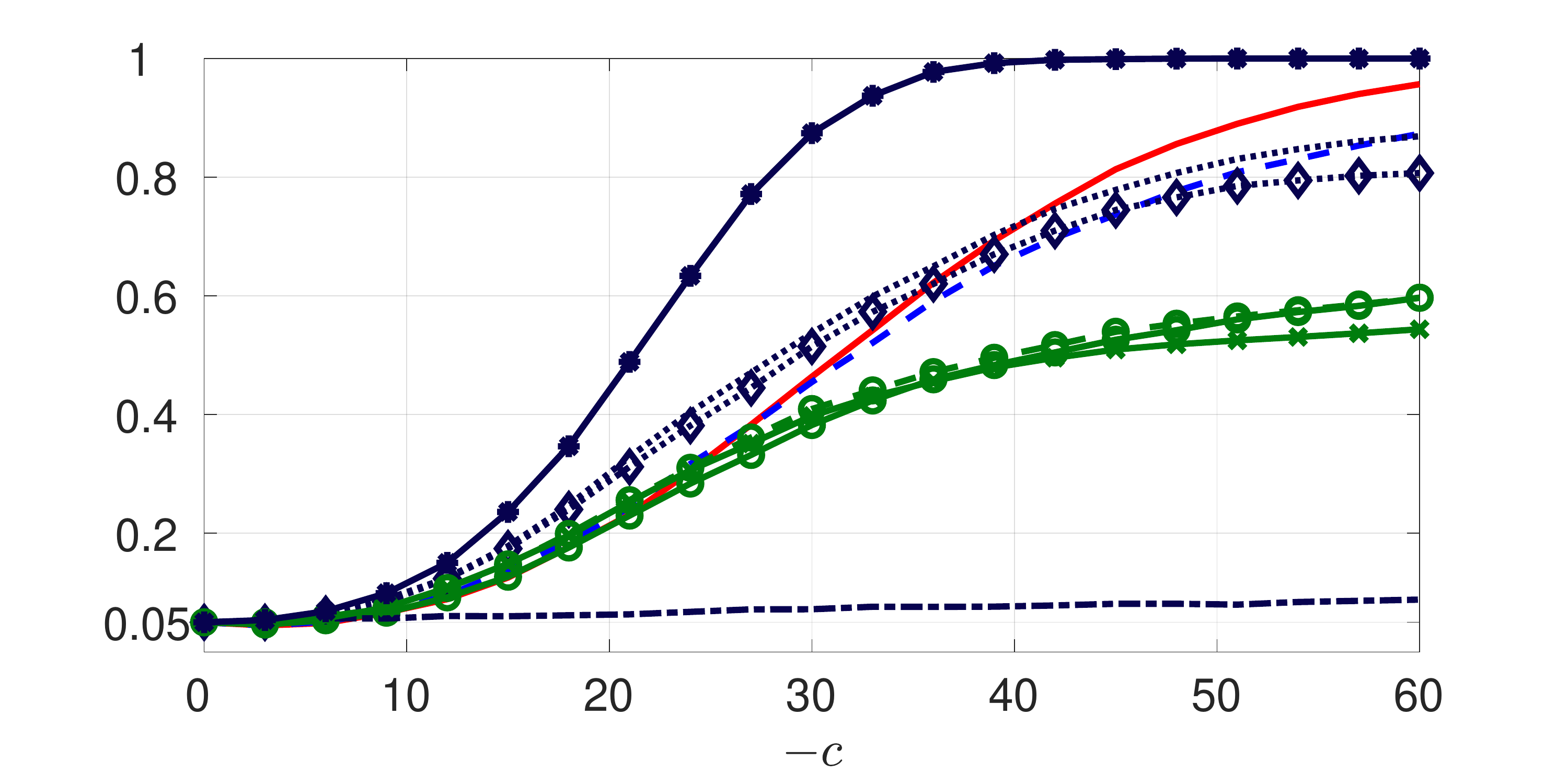}
\end{subfigure}\begin{subfigure}{0.25\textheight}
\centering
\caption*{MA, $\theta=0.6$}
\vspace{-1ex}
\includegraphics[trim={2cm 0.2cm 2cm 0.5cm},width=0.98\textwidth,clip]{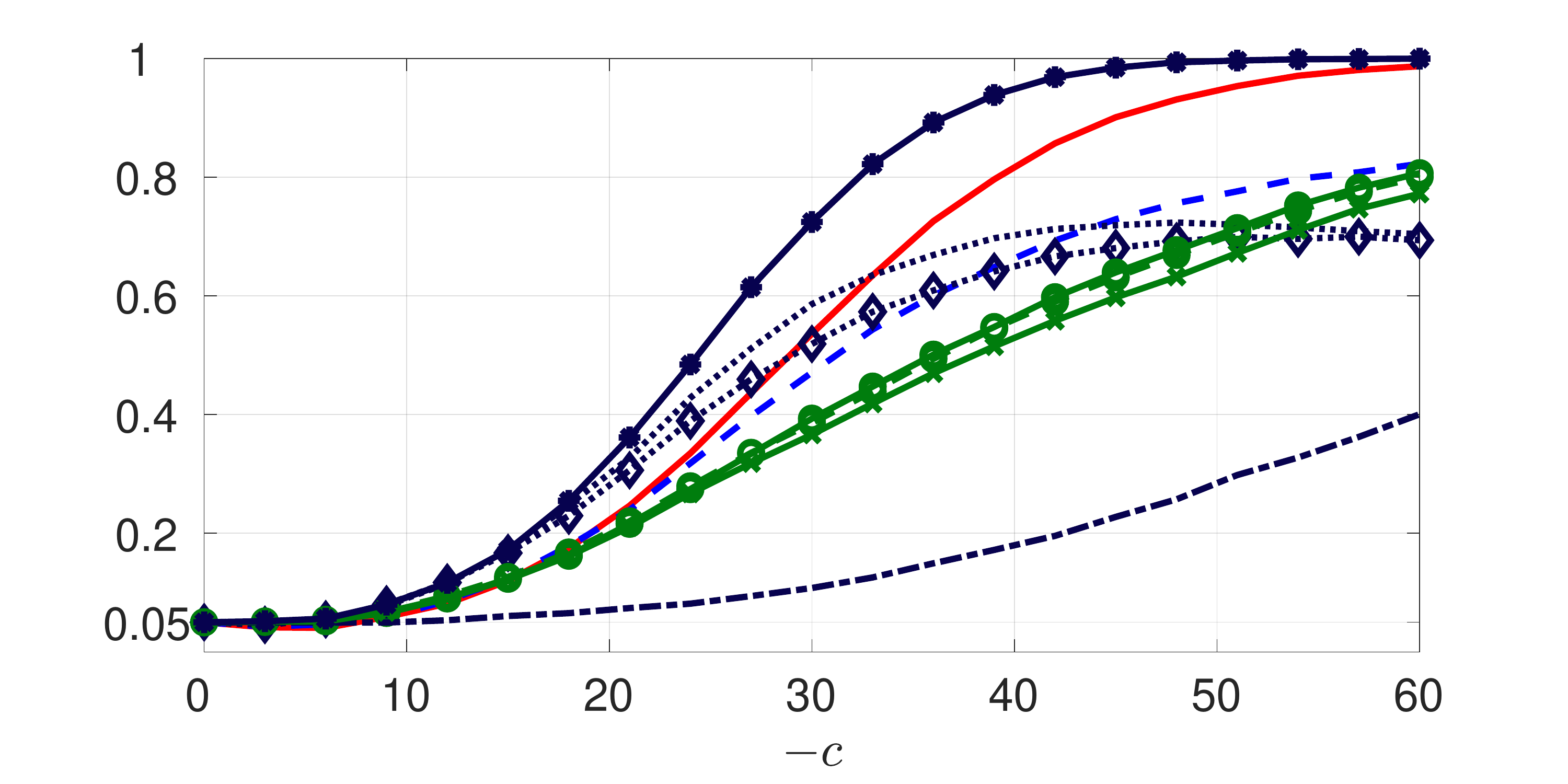}
\end{subfigure}\begin{subfigure}{0.25\textheight}
	\centering
	\caption*{MA, $\theta=0.9$}
	\vspace{-1ex}
	\includegraphics[trim={2cm 0.2cm 2cm 0.5cm},width=0.98\textwidth,clip]{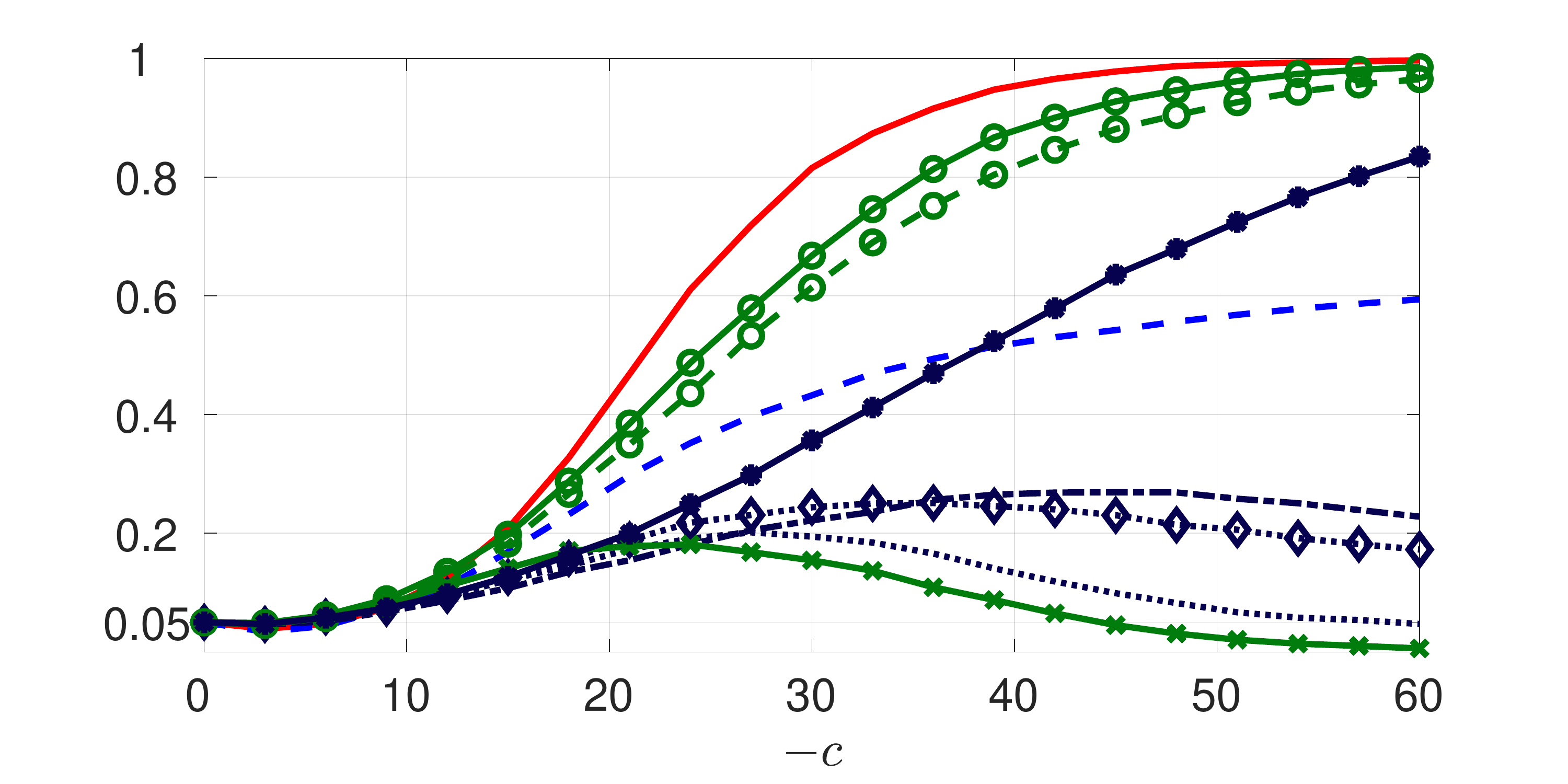}
\end{subfigure}\begin{subfigure}{0.25\textheight}
	\centering
	\caption*{ARMA, $\phi=0.3,\theta=0.6$}
	\vspace{-1ex}
	\includegraphics[trim={2cm 0.2cm 2cm 0.5cm},width=0.98\textwidth,clip]{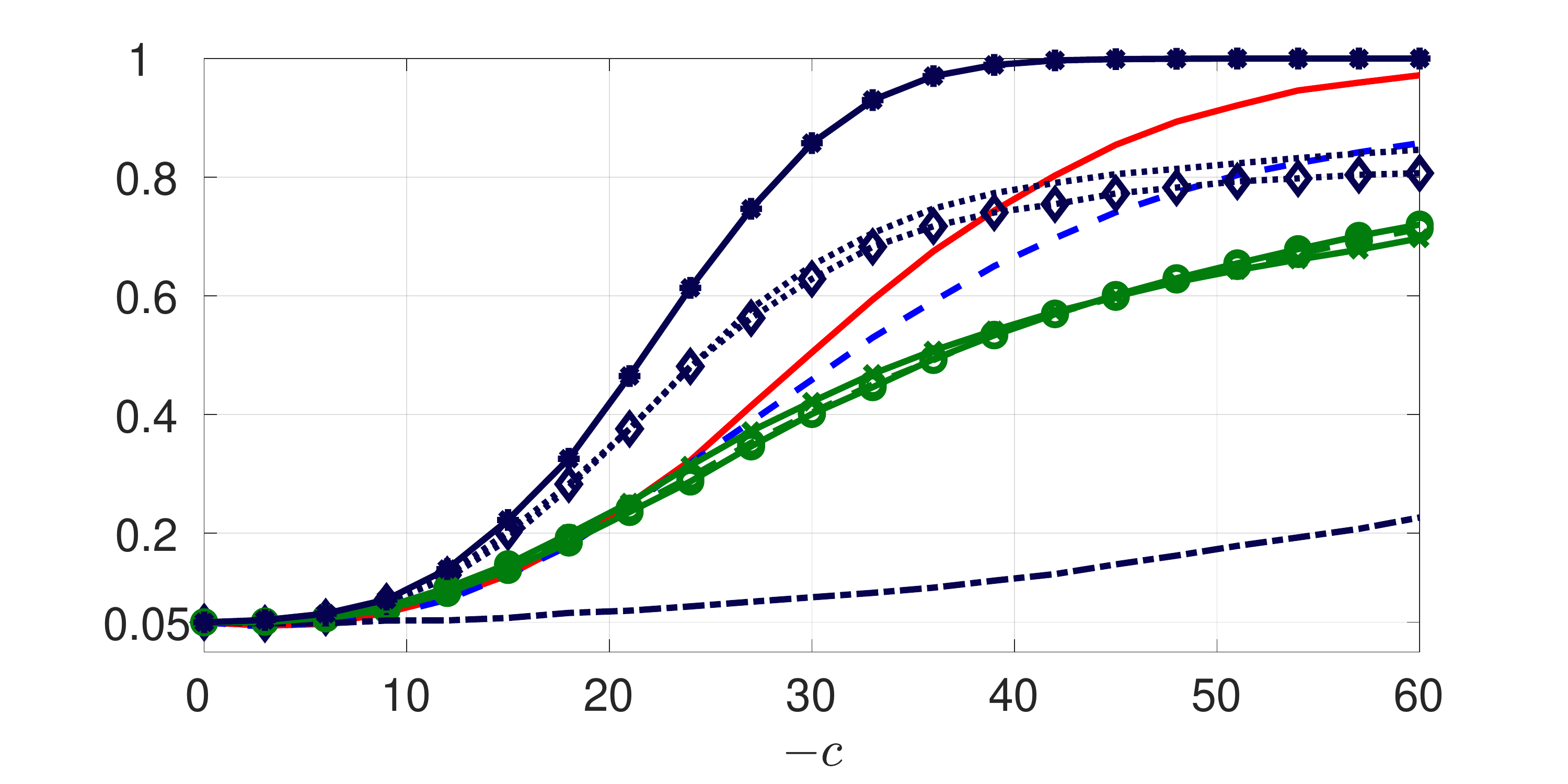}
\end{subfigure}

\vspace{1ex}

\begin{subfigure}{0.25\textheight}
\centering
\caption*{ARMA, $\phi=0.3,\theta=0.3$}
\vspace{-1ex}
\includegraphics[trim={2cm 0.2cm 2cm 0.5cm},width=0.98\textwidth,clip]{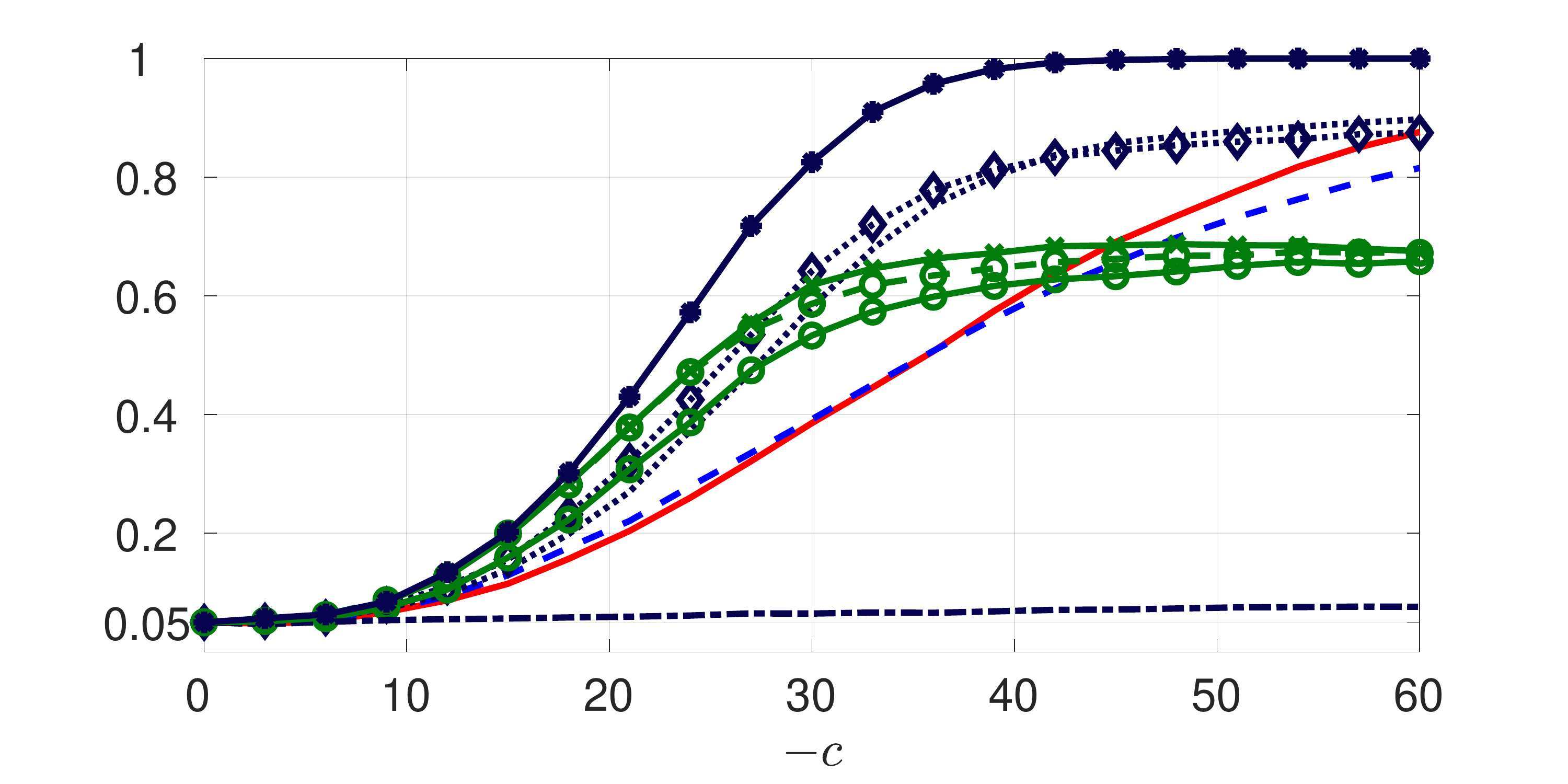}
\end{subfigure}\begin{subfigure}{0.25\textheight}
	\centering
	\caption*{ARMA, $\phi=0.6,\theta=0.3$}
	\vspace{-1ex}
	\includegraphics[trim={2cm 0.2cm 2cm 0.5cm},width=0.98\textwidth,clip]{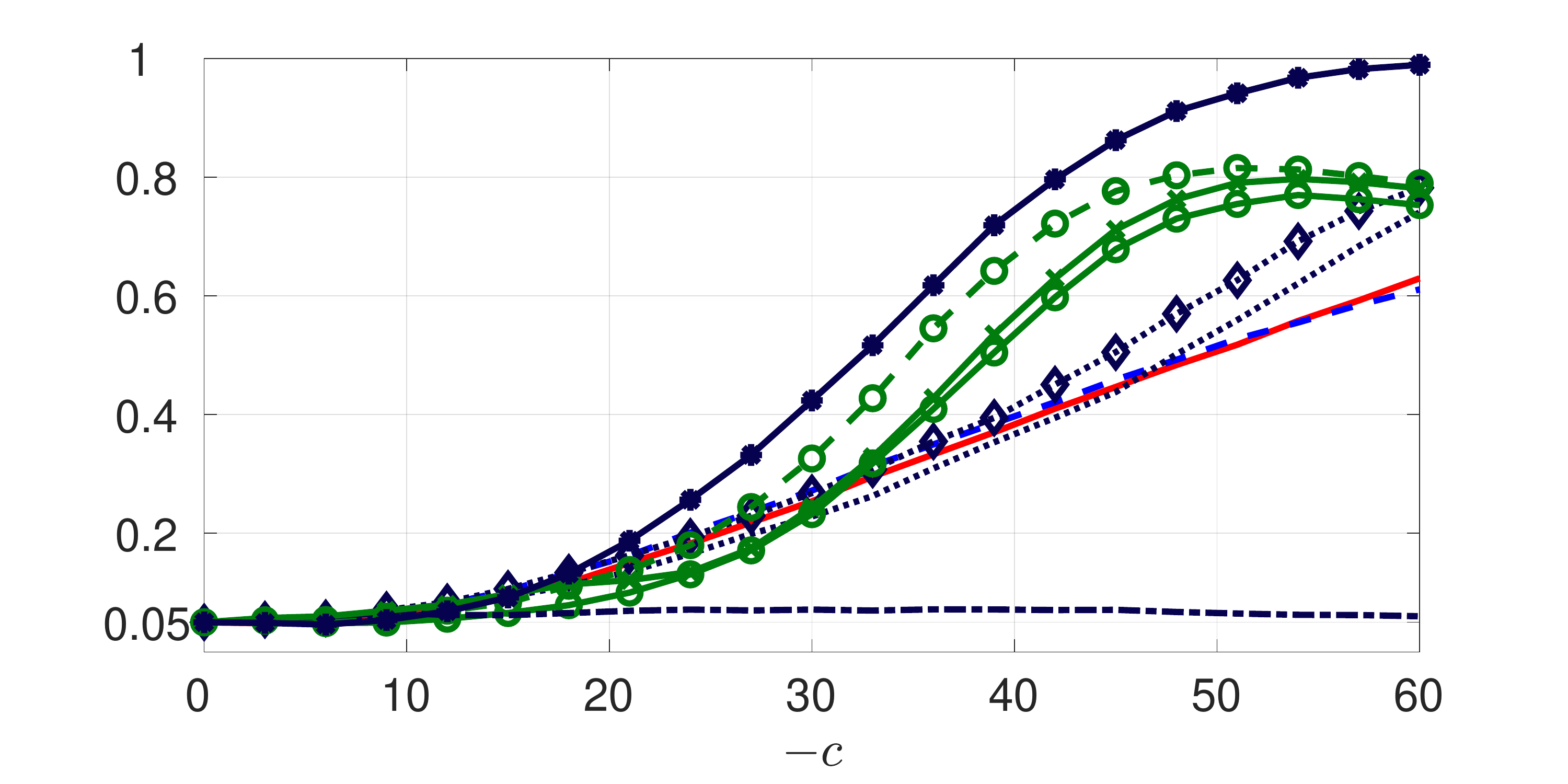}
\end{subfigure}\begin{subfigure}{0.25\textheight}
	\centering
	\caption*{GARCH, $a_1=0.05,a_2=0.93$}
	\vspace{-1ex}
	\includegraphics[trim={2cm 0.2cm 2cm 0.5cm},width=0.98\textwidth,clip]{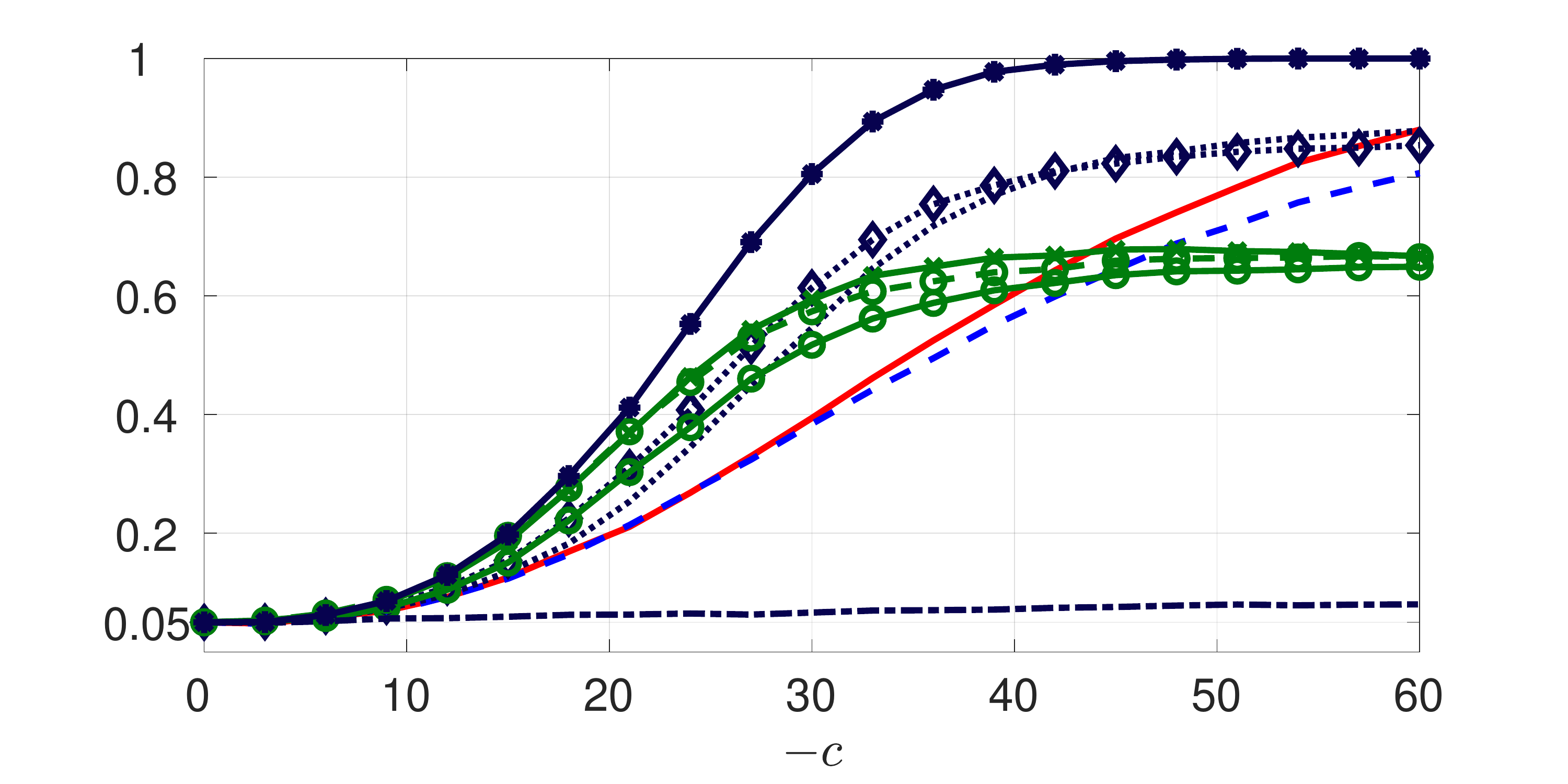}
\end{subfigure}\begin{subfigure}{0.25\textheight}
	\centering
	\caption*{GARCH, $a_1=0.01,a_2=0.98$}
	\vspace{-1ex}
	\includegraphics[trim={2cm 0.2cm 2cm 0.5cm},width=0.98\textwidth,clip]{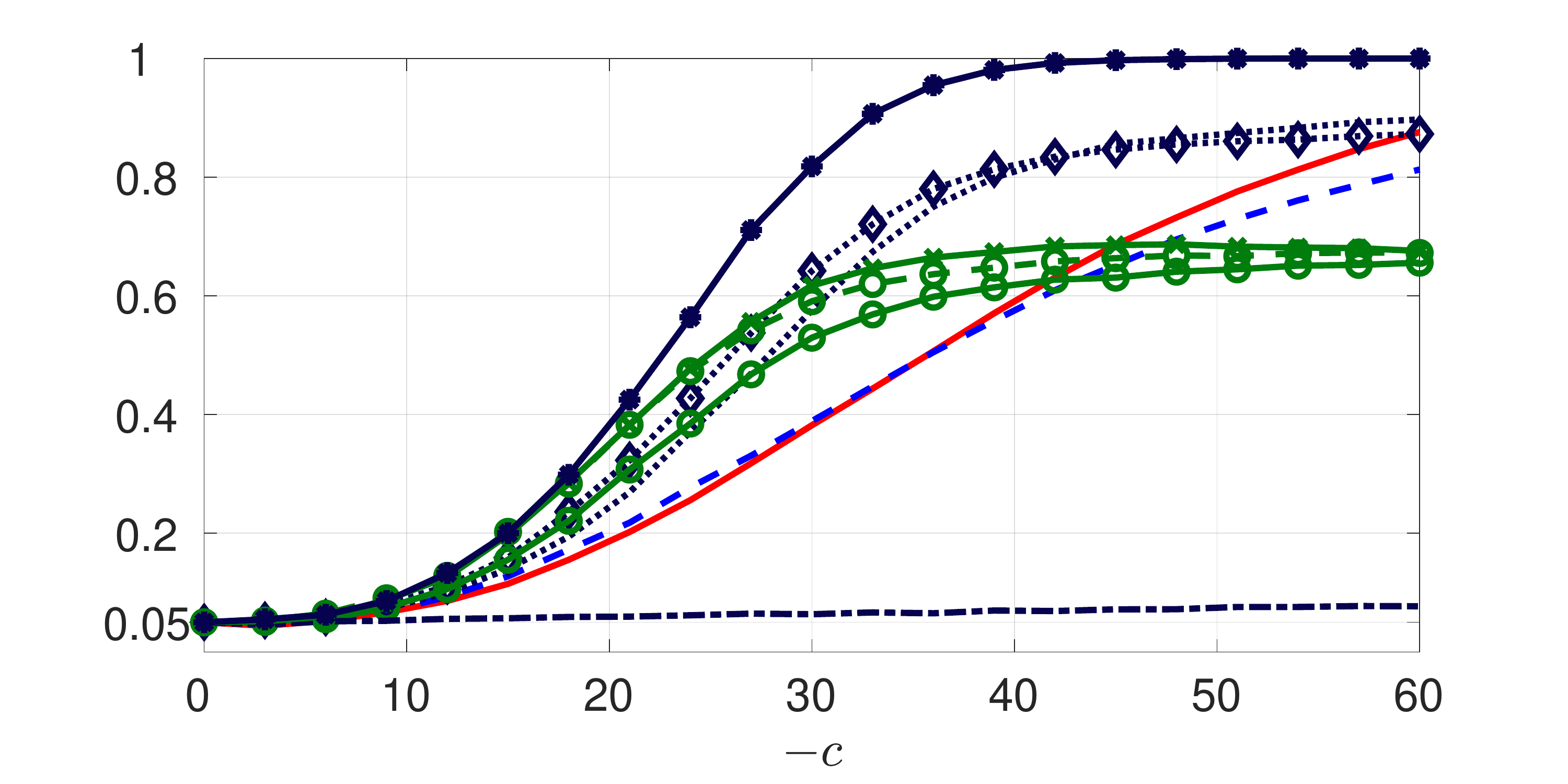}
\end{subfigure}

\end{center}
\vspace{-3ex}
\caption{Size-corrected power of the tests at the nominal $5\%$ level for $\text{$\text{H}_0$:}\ \rho = 1$ under the alternative $\rho = 1+c/T$ in case D2, for $T=100$ and $R^2=0.4$\\{\footnotesize Note: The superscripts ``(GLS)'' and ``$*$'' indicate GLS detrending instead of OLS detrending and the use of MAIC instead of AIC, respectively.}}
\label{fig:size_adjusted_power_m_1_T_100_deter_2_R2_04_AIC}
%\end{figure}
\end{sidewaysfigure}

Size-corrected power of the tests decreases when $R^2$ becomes larger. However, for all three deterministic specifications different short-run dynamics in $\xi_t$ can have more adverse effects on the performance of the tests than increasing $R^2$ from zero to $0.8$. Figure~\ref{fig:size_adjusted_power_m_1_T_100_deter_2_R2_04_AIC} presents the size-corrected power curves of the tests for the different short-run dynamics in case D2 with $T=100$ and $R^2=0.4$.\footnote{Results in case D0 and D1 are qualitatively similar. The Supplementary Material (Figures~\ref{SM-fig:size_adjusted_power_m_1_T_100_deter_0_R2_0_AIC} --~\ref{SM-fig:size_adjusted_power_m_1_T_250_deter_2_R2_08_AIC} and~\ref{SM-fig:size_adjusted_power_m_1_T_100_deter_0_R2_0_BIC} --~\ref{SM-fig:size_adjusted_power_m_1_T_250_deter_2_R2_08_BIC}) provides figures similar to Figure~\ref{fig:size_adjusted_power_m_1_T_100_deter_2_R2_04_AIC} for all three deterministic specifications and all combinations of $R^2$ and $T$.} For most short-run dynamics, the $\widehat Z_\alpha$ test performs best, while the MSB test performs worse. The ranking of the remaining tests is highly case dependent. 

In general, the variance ratio test performs relatively well and much better than suggested by the local asymptotic power results. In particular, it outperforms the $\text{MSB}^{(\GLS)*}$ test and both the OLS and GLS versions of the $\text{ADF}^*$ test -- thus its biggest competitors under the null hypothesis -- in the MA cases and in the ARMA(0.3,0.6). In the MA case with $\theta=0.9$, the variance ratio test is even the most powerful test, but this result should be interpreted with caution, as all tests are heavily size-distorted in this case. In other cases, the variance ratio test is less powerful than its competitors for small deviations from the null hypothesis but outperforms the $\text{MSB}^{(\GLS)*}$ test and both the OLS and GLS versions of the $\text{ADF}^*$ test for larger deviations from the null hypothesis.\footnote{For some short-run dynamics the power curves of all tests considered can decline slightly below $\alpha$ for small deviations from the null, especially for $R^2=0.8$. This is in line with the local asymptotic power curves in Figure~\ref{fig:local_power_m_1}.} 

Importantly, for some short-run dynamics, the power curves of the $\text{MSB}^{(\GLS)*}$ test and both the OLS and GLS versions of the $\text{ADF}^*$ test reveal power reversal problems. This phenomenon is well known in the unit root literature for the MSB and ADF tests in combination with the AIC criterion in the MA case with a large $\theta$. However, the MAIC criterion has been introduced as a possible way to prevent this degeneracy \citeaffixed{PeQu07}{cf. the discussion in}. Comparing the power curves of the $\text{ADF}^*$ and $\text{ADF}^{(\GLS)*}$ tests with those of the ADF and $\text{ADF}^{(\GLS)}$ tests reveals that using the MAIC criterion rather than the AIC criterion indeed prevents power reversal problems associated with the ADF test in the MA case with $\theta=0.9$.\footnote{In the MA case with $\theta=0.9$ the $\text{MSB}^{(\GLS)*}$ test still suffers from power reversal problems.} In other cases, however, the tests based on MAIC still suffer from power reversal problems (\eg, in the AR case with $\phi=0.6$ and in the ARMA case with $\phi=0.6$ and $\theta=0.3$) and perform worse than those based on AIC (\eg, in the IID and GARCH cases). 

With respect to preliminary data detrending, the results suggest that the GLS versions of the ADF and $\text{ADF}^*$ tests perform similarly to the OLS versions of the test, whereas the GLS version of the variance ratio test is often (considerably) less powerful than the OLS version.\footnote{The local asymptotic power results suggest that the GLS version of the variance ratio test is generally more powerful than its OLS version for $R^2=0.8$. However, the finite sample results do not reveal a similar advantage of the GLS version over the OLS version in terms of size-corrected power, compare Figure~\ref{fig:size_adjusted_power_m_1_T_100_deter_2_R2_08_AIC} in Appendix~\ref{app:addresults} for the size-corrected power curves in case D2 for $T=100$ and note that results in case D1 are similar, compare Figure~\ref{SM-fig:size_adjusted_power_m_1_T_100_deter_1_R2_08_AIC} in the Supplementary Material. For $T=250$, GLS detrending even seems to be disadvantageous in cases D1 and D2, compare Figures~\ref{SM-fig:size_adjusted_power_m_1_T_250_deter_1_R2_08_AIC} and~\ref{SM-fig:size_adjusted_power_m_1_T_250_deter_2_R2_08_AIC} in the Supplementary Material.} On the other hand, the $\text{MSB}^{(\GLS)*}$ test is in most cases clearly more powerful than the $\text{MSB}$ test. 

Finally, increasing the sample size is beneficial for the power of the tests, but results for $T=250$ are qualitatively similar, see, \eg, Figure~\ref{fig:size_adjusted_power_m_1_T_250_deter_2_R2_04_AIC} in Appendix~\ref{app:addresults} in case D2 with $R^2=0.4$. It is noteworthy that short-run dynamics in $\xi_t$ even have adverse effects -- including power reversal problems for the ADF tests and the $\text{MSB}^{(\GLS)*}$ test -- on the power of the tests for larger sample sizes, especially in case D2 with $R^2=0.8$. However, as $T$ increases, short-run dynamics in $\xi_t$ become less pervasive such that the size-corrected power results approach the local asymptotic power results analyzed in Section~\ref{sec:Th-LAPower}, see, \eg, the results for $T=1{,}000$ in case D2 with $R^2=0.4$ in Figure~\ref{fig:size_adjusted_power_m_1_T_1000_deter_2_R2_04_AIC} in Appendix~\ref{app:addresults}.\footnote{Figures~\ref{SM-fig:size_adjusted_power_m_1_T_1000_deter_0_R2_0_AIC} --~\ref{SM-fig:size_adjusted_power_m_1_T_1000_deter_2_R2_08_AIC} in the Supplementary Material show the results for $T=1{,}000$ and all values of $R^2$ in cases D0, D1, and D2. Moreover, Figures~\ref{SM-fig:size_adjusted_power_m_1_T_1000_deter_0_R2_0_BIC} --~\ref{SM-fig:size_adjusted_power_m_1_T_1000_deter_2_R2_08_BIC} show the results for the ADF and MSB tests based on (M)BIC.} In general, we notice that for small to medium sample sizes, the finite sample performance of the tests are dominated by the short-run dynamics in $\xi_t$ such that size-corrected power results deviate considerably from local asymptotic power results. \citeasnoun{Pe07} and \citeasnoun{PeRo16} do not detect these discrepancies between the finite sample performance of the tests and the local asymptotic power results because their DGPs only allow for very mild effects of short-run dynamics (compare the discussion in Remark~\ref{rem:DGPs}). As the sample size increases, the short-run dynamics in $\xi_t$ become less important and the performance of the tests is then dominated by $R^2$. In this case, size-corrected power results are more similar to local asymptotic power results. The sample size required for a sufficient degree of similarity between finite sample results and local asymptotic power results generally increases with the order of the deterministic component and $R^2$, \ie, the required sample size is much smaller in case D0 with $R^2=0$ than in case D2 with $R^2=0.8$.

To provide an overall assessment, it is important to note that residuals-based no-cointegration tests are used to avoid analyzing the relationship between stochastically trending (economic) variables in a spurious regression framework. For practitioners, small upward size distortions may thus have a higher weight in the overall assessment of the general performance of the tests than high power under the alternative. We thus conclude that both the variance ratio test and the $\text{ADF}^*$ test perform best among the tests considered in this simulation study and thus may be deemed useful for testing for cointegration in applications. However, practitioners should be aware of the individual specific shortcomings of the tests.\footnote{To make the analysis more robust against individual specific shortcomings of the tests, practitioners could use a Fisher-type combination test or a ``union-of-rejections'' decision rule \citeaffixed{BaHa13}{cf.}.}

%\begin{landscape}
%\end{landscape}

\subsection{A Large Initial Value $u_0$}\label{sec:FiniteSample_u0}
Several contributions in related literature highlight strong effects of a large initial value $u_0$ on the power of unit root \citeaffixed[and references therein]{HLT09}{see, \eg,} and residuals-based no-cointegration \citeaffixed{PeRo16}{see, \eg,} tests. To comment briefly on the effect of a large (in a well-defined sense) initial value $u_0$ on the power of the variance ratio test, we use the same set-up as before but generate initial values $u_0$ of order $T^{1/2}$.\footnote{It is clear from the proof of Proposition~\ref{prop:VR_H0} that $u_0$ does not affect the distribution of the VR statistic under the null hypothesis in cases D1 and D2, neither in the limit nor in finite samples. In case D0, however, an initial value $u_0$ of order $T^{1/2}$ does affect the distribution of the VR statistic under the null hypothesis both in finite samples and in the limit. Since in applications an intercept is typically included in~\eqref{eq:y}, we do not comment upon this issue any further.} To this end, we follow \citeasnoun{HLT09} and \citeasnoun{PeRo16} and generate $u_0$ as $u_0=\lambda_u/(1-\rho_T^2)^{1/2}$, where $\rho_T = 1+c/T$, $c<0$, and $\lambda_u$ is a fixed constant.\footnote{Effects in case $u_0$ is drawn from a normal distribution with mean zero and variance $\lambda_u^2/(1-\rho_T^2)$ are less pronounced.} We compare the performance of the variance ratio test with the performance of the tests already analyzed in the previous subsections. To simplify the comparison with the size-corrected power curves discussed in Section~\ref{sec:FiniteSample_Power}, we again consider size-corrected power results of the tests, using the same empirical critical values as in the previous subsection. 

\begin{sidewaysfigure}[!ht]
%\begin{figure}[!ht]
\begin{center}
\begin{subfigure}{0.25\textheight}
	\centering
	\caption*{IID}
	\vspace{-1ex}
	\includegraphics[trim={2cm 0.2cm 2cm 0.5cm},width=0.98\textwidth,clip]{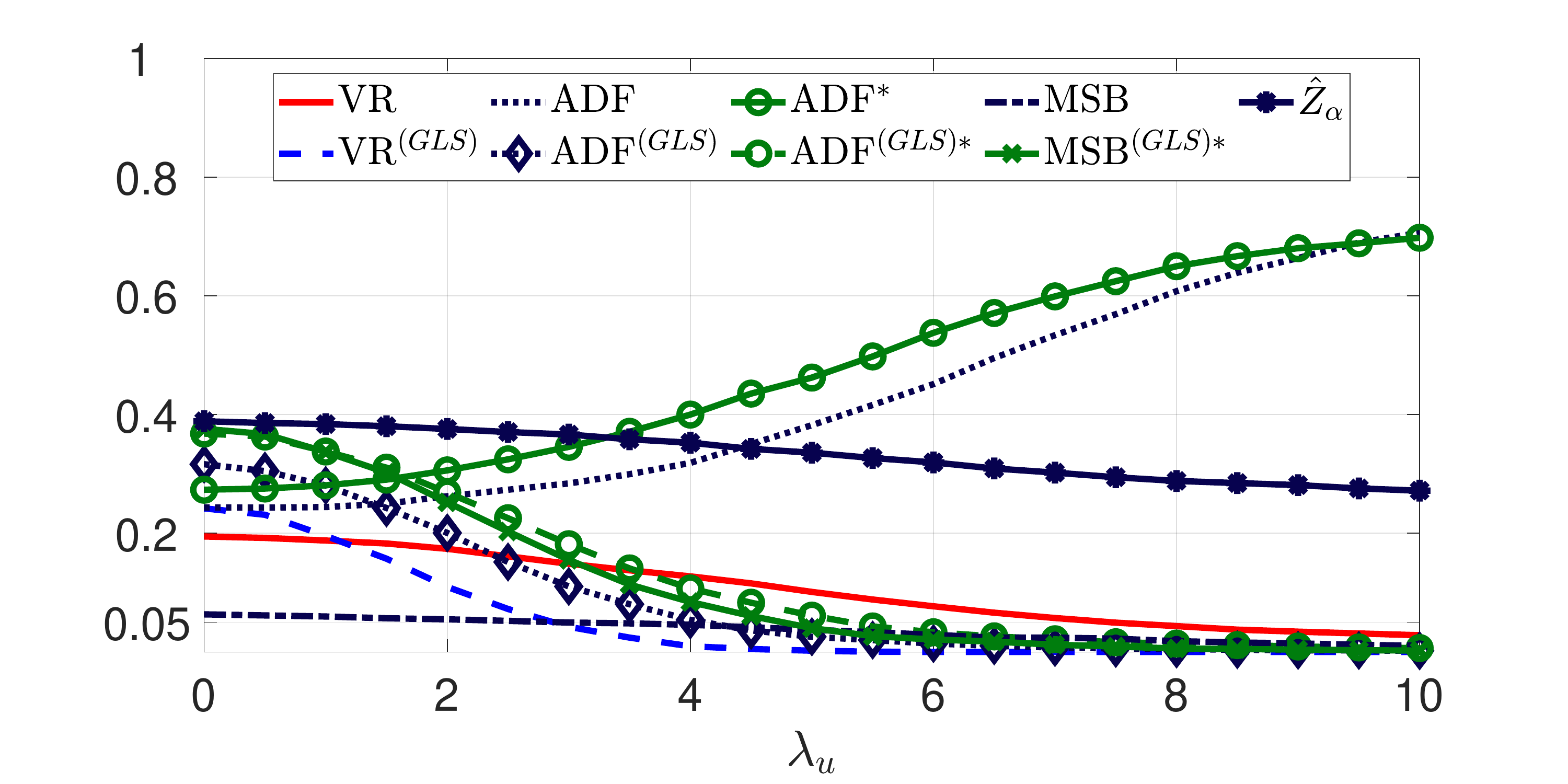}
\end{subfigure}\begin{subfigure}{0.25\textheight}
	\centering
	\caption*{AR, $\phi=0.3$}
	\vspace{-1ex}
	\includegraphics[trim={2cm 0.2cm 2cm 0.5cm},width=0.98\textwidth,clip]{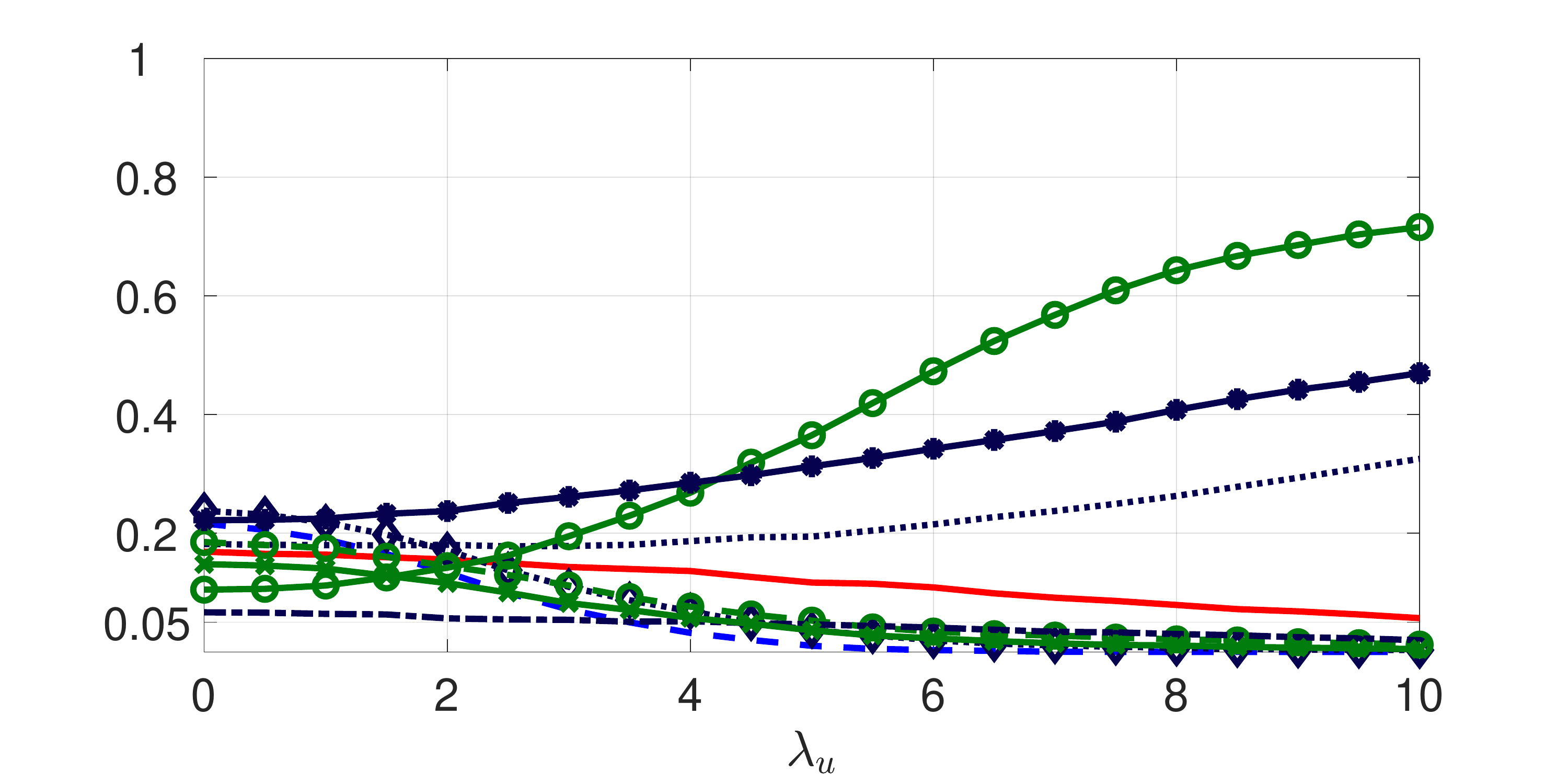}
\end{subfigure}\begin{subfigure}{0.25\textheight}
	\centering
	\caption*{AR, $\phi=0.6$}
	\vspace{-1ex}
	\includegraphics[trim={2cm 0.2cm 2cm 0.5cm},width=0.98\textwidth,clip]{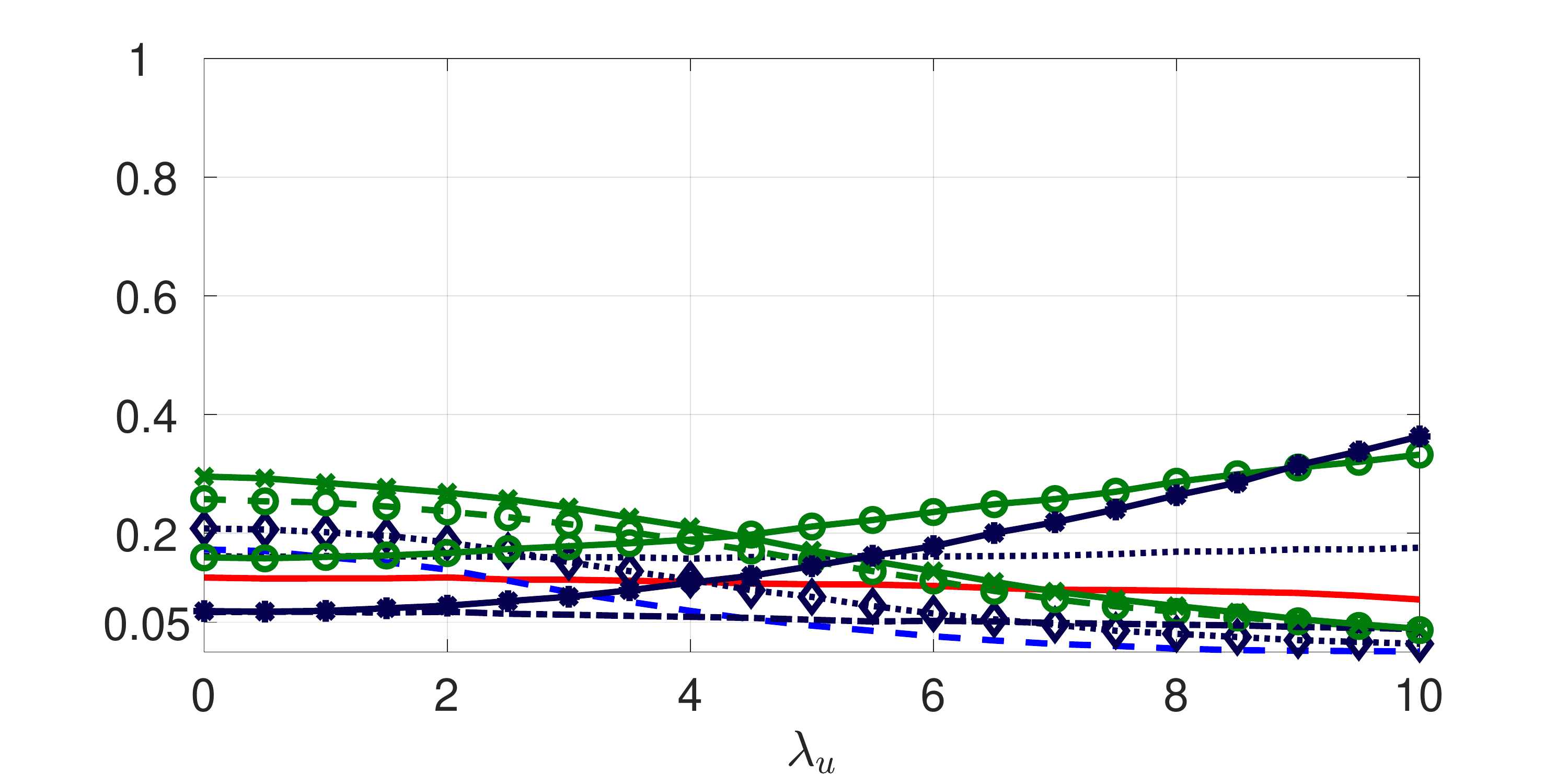}
\end{subfigure}\begin{subfigure}{0.25\textheight}
	\centering
	\caption*{AR, $\phi=0.9$}
	\vspace{-1ex}
	\includegraphics[trim={2cm 0.2cm 2cm 0.5cm},width=0.98\textwidth,clip]{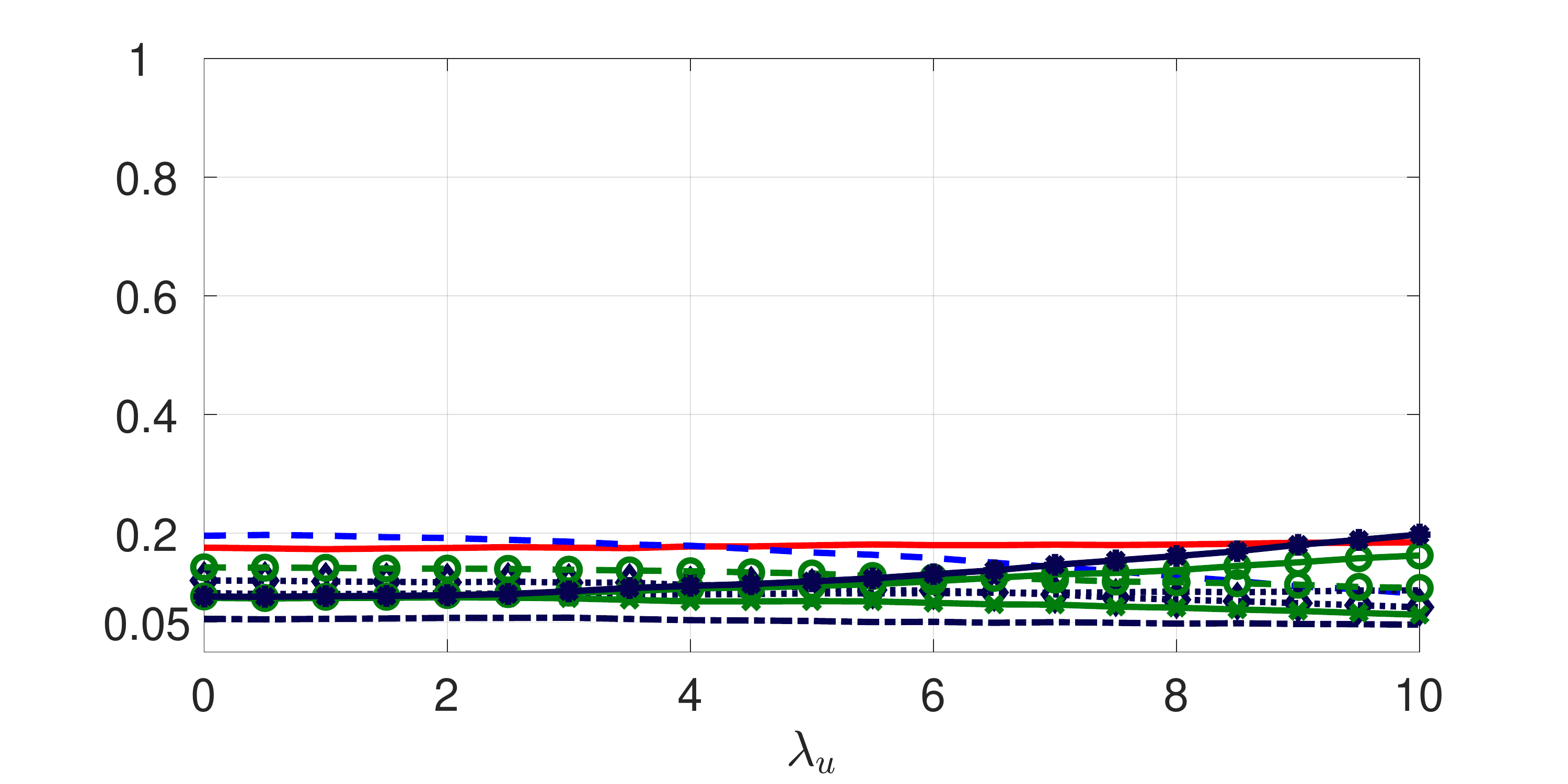}
\end{subfigure}

\vspace{1ex}

\begin{subfigure}{0.25\textheight}
	\centering
	\caption*{MA, $\theta=0.3$}
	\vspace{-1ex}
	\includegraphics[trim={2cm 0.2cm 2cm 0.5cm},width=0.98\textwidth,clip]{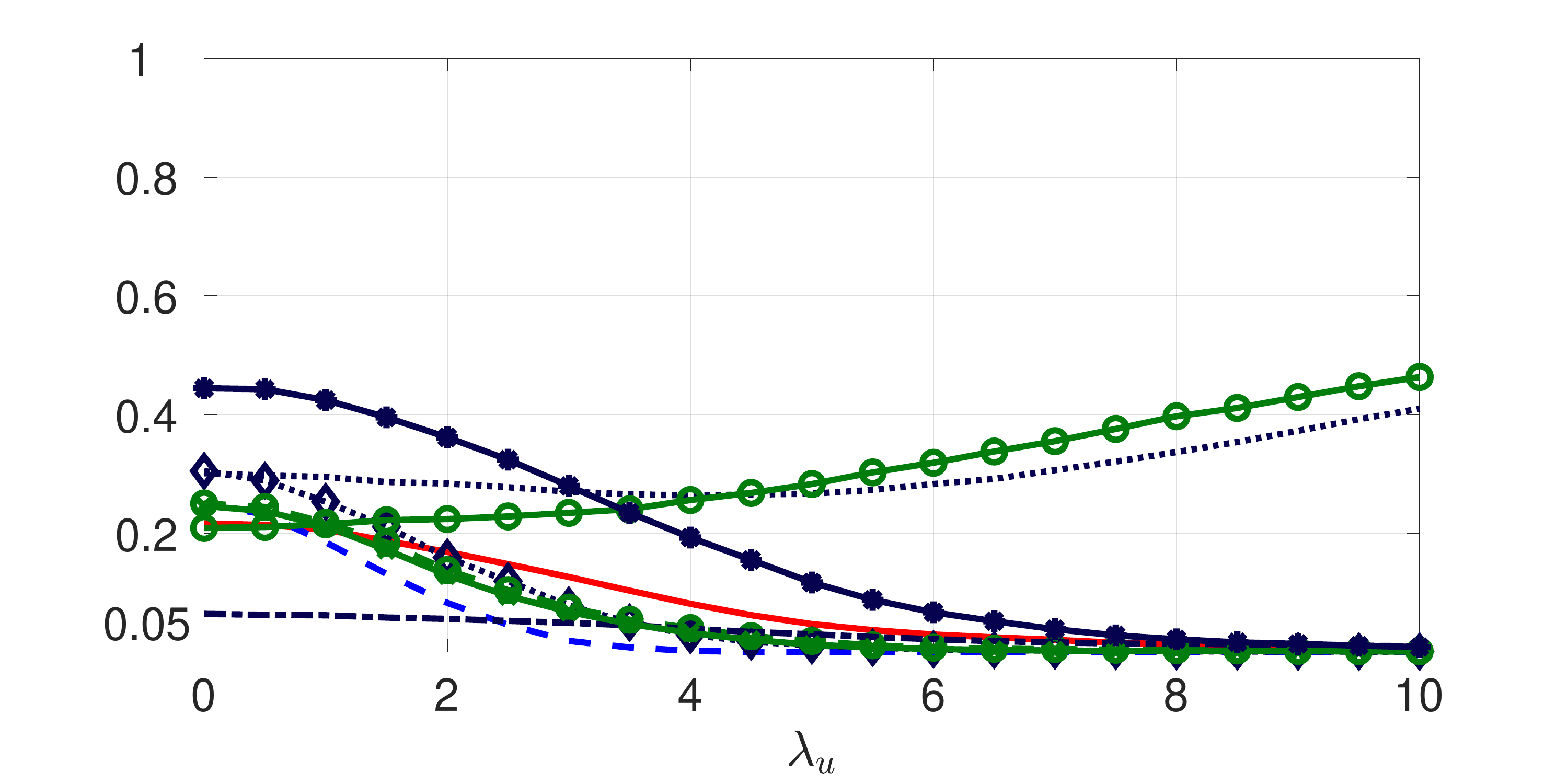}
\end{subfigure}\begin{subfigure}{0.25\textheight}
	\centering
	\caption*{MA, $\theta=0.6$}
	\vspace{-1ex}
	\includegraphics[trim={2cm 0.2cm 2cm 0.5cm},width=0.98\textwidth,clip]{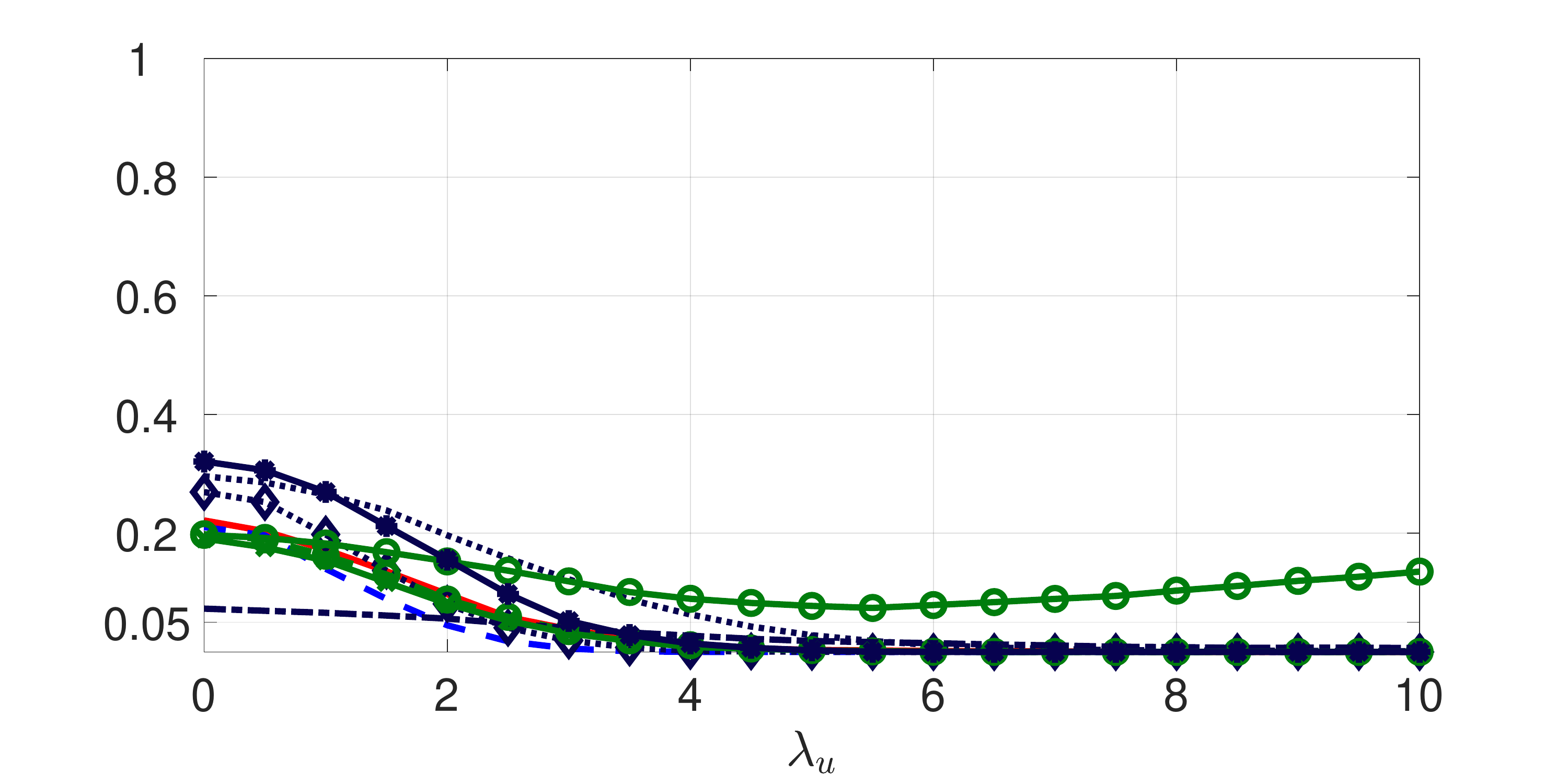}
\end{subfigure}\begin{subfigure}{0.25\textheight}
	\centering
	\caption*{MA, $\theta=0.9$}
	\vspace{-1ex}
	\includegraphics[trim={2cm 0.2cm 2cm 0.5cm},width=0.98\textwidth,clip]{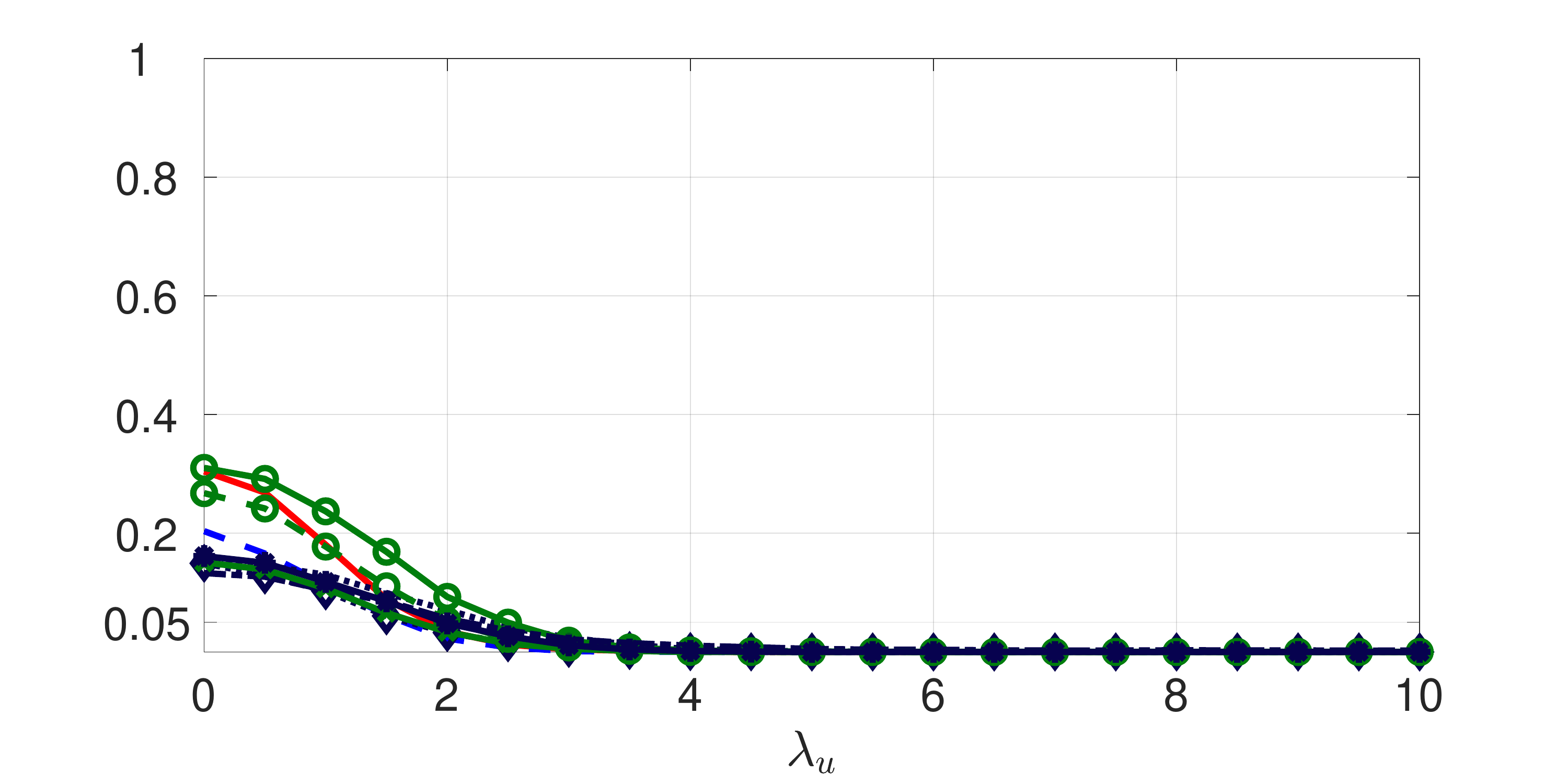}
\end{subfigure}\begin{subfigure}{0.25\textheight}
	\centering
	\caption*{ARMA, $\phi=0.3,\theta=0.6$}
	\vspace{-1ex}
	\includegraphics[trim={2cm 0.2cm 2cm 0.5cm},width=0.98\textwidth,clip]{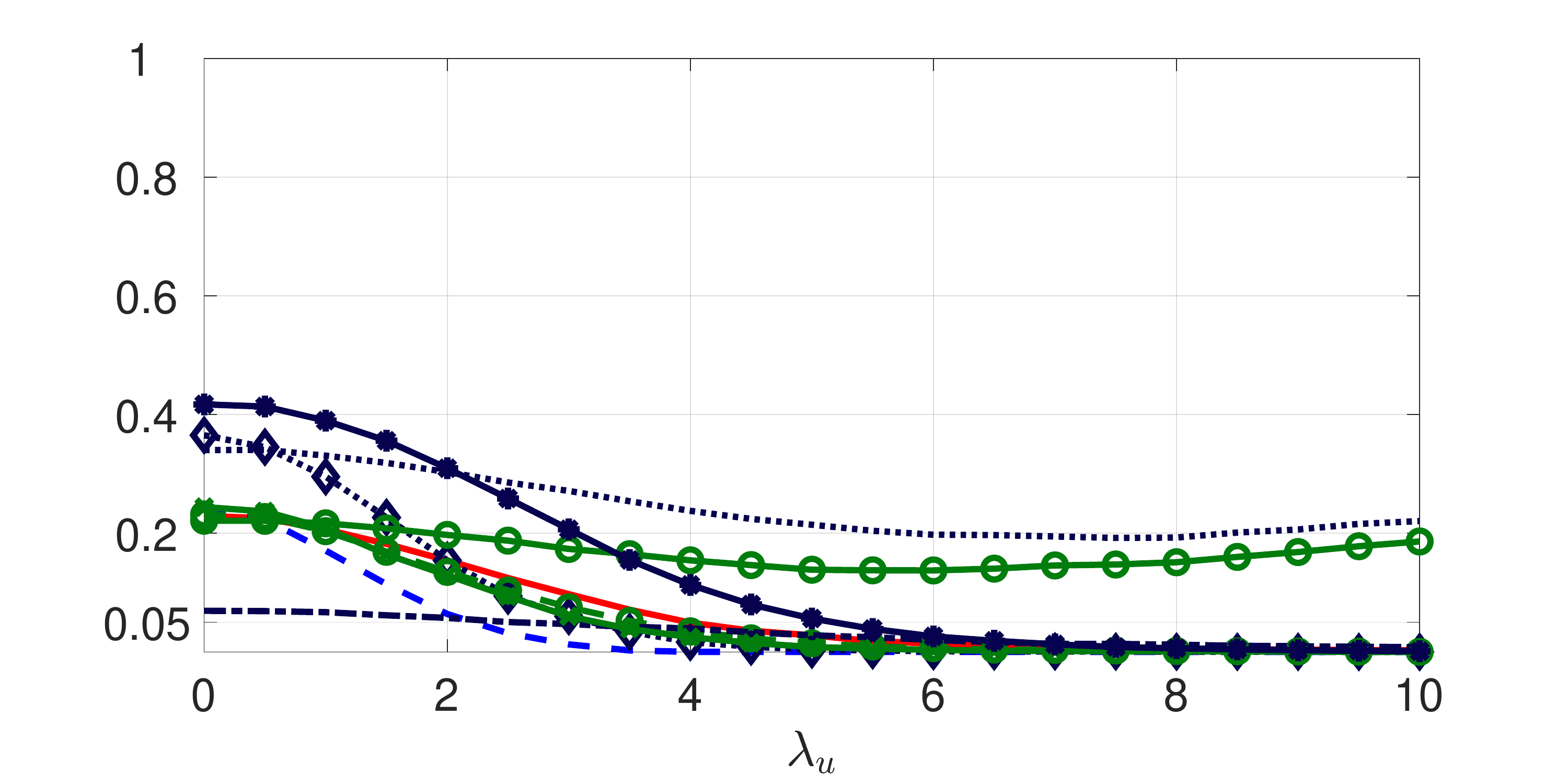}
\end{subfigure}

\vspace{1ex}

\begin{subfigure}{0.25\textheight}
	\centering
	\caption*{ARMA, $\phi=0.3,\theta=0.3$}
	\vspace{-1ex}
	\includegraphics[trim={2cm 0.2cm 2cm 0.5cm},width=0.98\textwidth,clip]{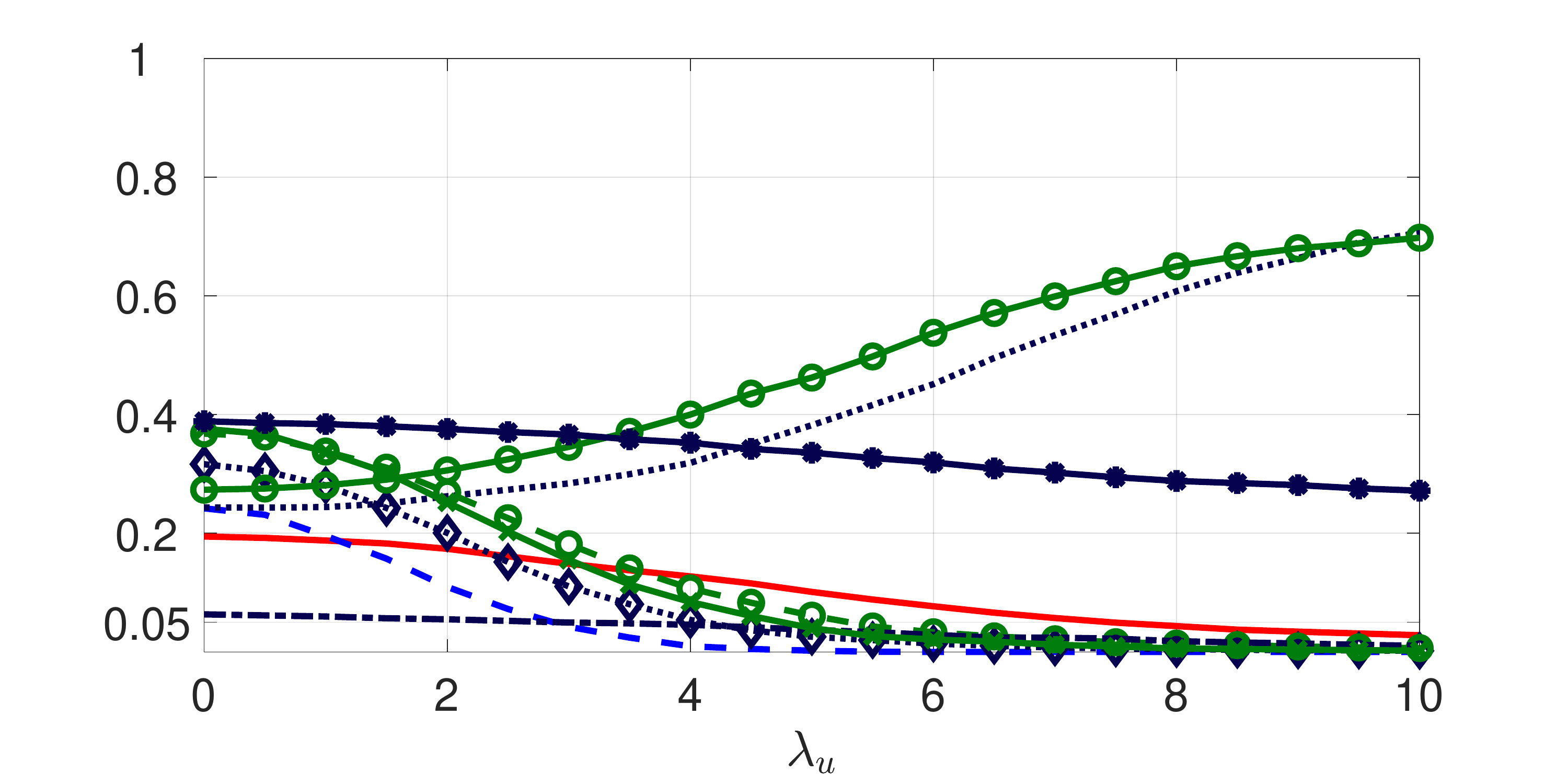}
\end{subfigure}\begin{subfigure}{0.25\textheight}
	\centering
	\caption*{ARMA, $\phi=0.6,\theta=0.3$}
	\vspace{-1ex}
	\includegraphics[trim={2cm 0.2cm 2cm 0.5cm},width=0.98\textwidth,clip]{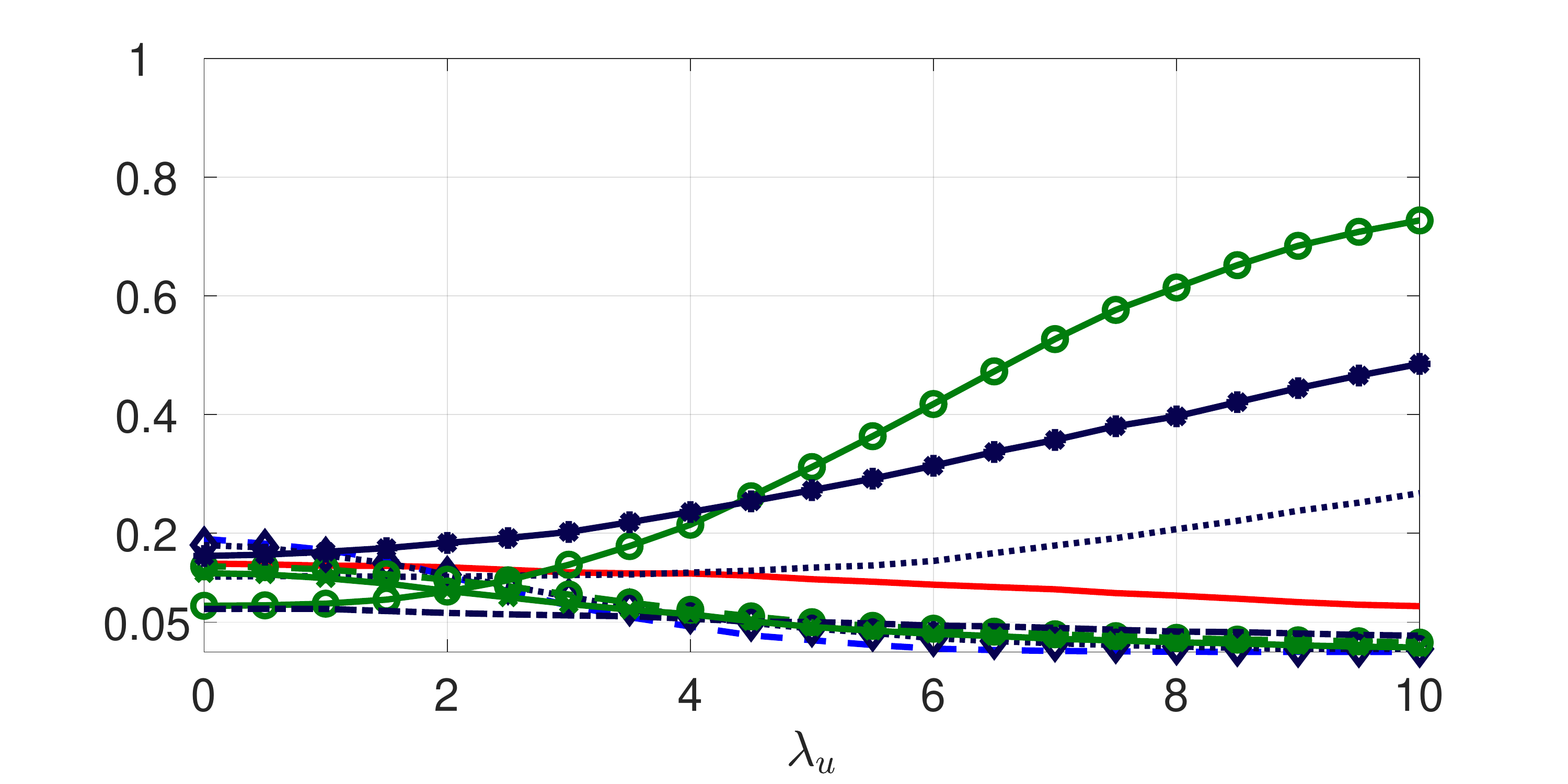}
\end{subfigure}\begin{subfigure}{0.25\textheight}
	\centering
	\caption*{GARCH, $a_1=0.05,a_2=0.93$}
	\vspace{-1ex}
	\includegraphics[trim={2cm 0.2cm 2cm 0.5cm},width=0.98\textwidth,clip]{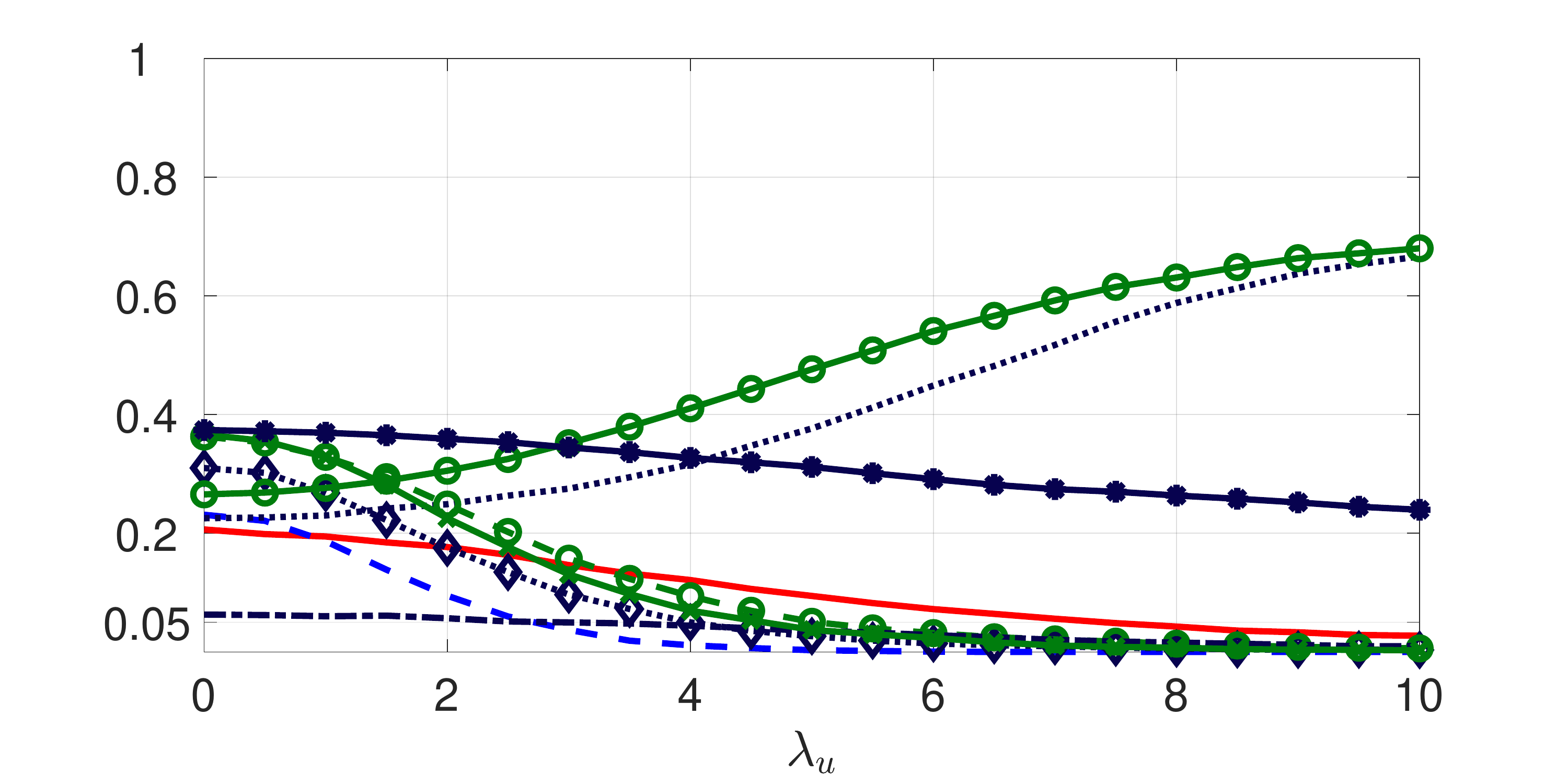}
\end{subfigure}\begin{subfigure}{0.25\textheight}
	\centering
	\caption*{GARCH, $a_1=0.01,a_2=0.98$}
	\vspace{-1ex}
	\includegraphics[trim={2cm 0.2cm 2cm 0.5cm},width=0.98\textwidth,clip]{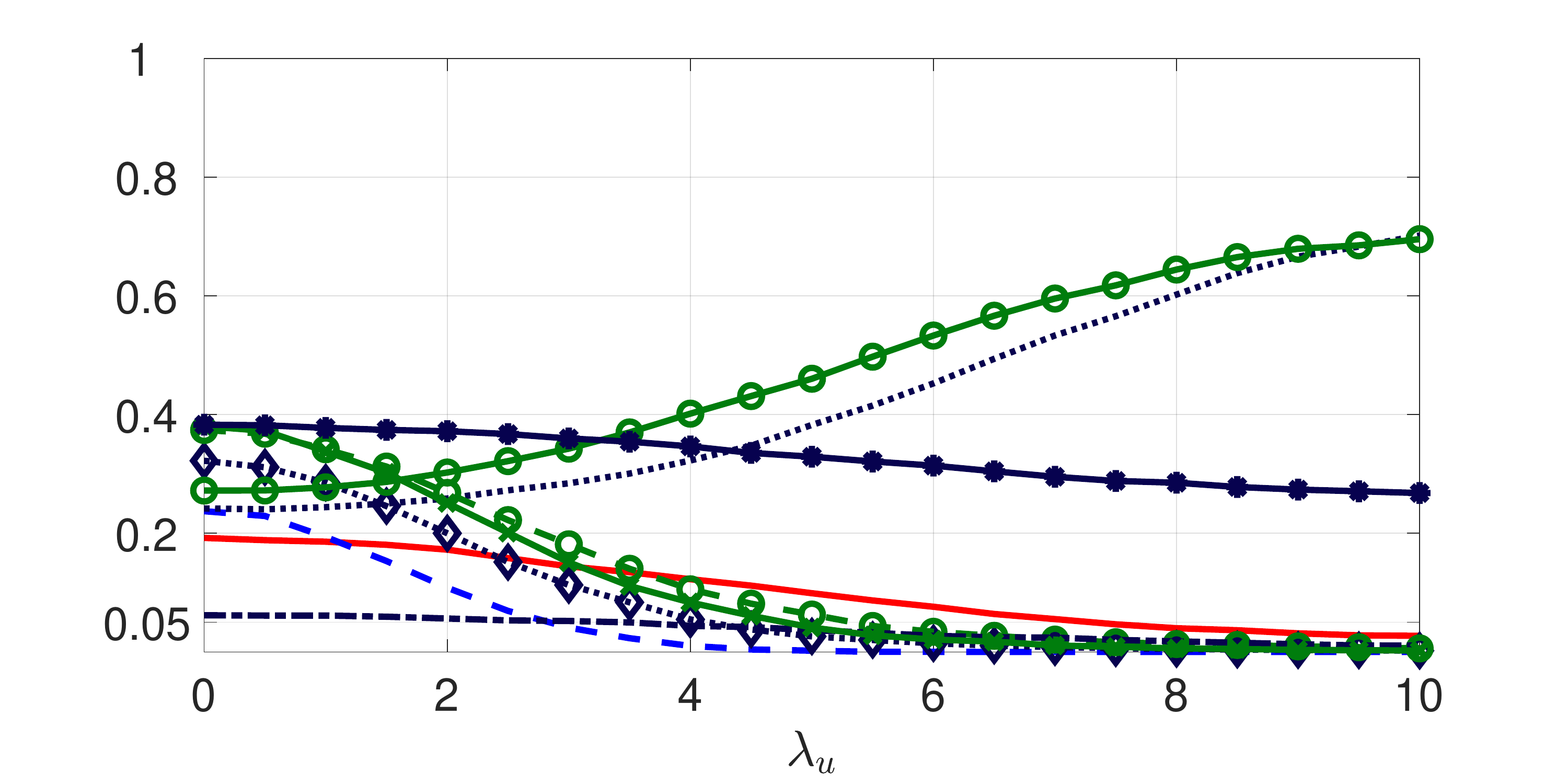}
\end{subfigure}

\end{center}
\vspace{-3ex}
\caption{Size-corrected power of the tests at the nominal $5\%$ level for $\text{$\text{H}_0$:}\ \rho = 1$ in the presence of a large non-random initial value $u_0$ under the alternative $\rho = 1-20/T$ in case D2, for $T=100$ and $R^2=0.4$\\{\footnotesize See note to Figure~\ref{fig:size_adjusted_power_m_1_T_100_deter_2_R2_04_AIC}.}}
\label{fig:size_adjusted_power_m_1_T_100_deter_2_R2_04_AIC_u0_fixed_c_20}
%\end{figure}
\end{sidewaysfigure}

Figure~\ref{fig:size_adjusted_power_m_1_T_100_deter_2_R2_04_AIC_u0_fixed_c_20} presents the results for $c=-20$, $T=100$ and $R^2=0.4$ in case D2. Again, the performance of the tests highly depends on the short-run dynamics in $\xi_t$. In the MA case with $\theta=0.9$ and in the ARMA(0.3,0.6) case, size-corrected power of all tests decreases as $\lambda_u$ moves away from zero, whereas in the AR case with $\phi=0.9$, size-corrected power of the tests is rather unaffected by changes in $\lambda_u$. In the majority of cases, however, size-corrected power of the ADF and of the $\text{ADF}^*$ tests increases as $\lambda_u$ moves away from zero, with clear performance advantages of $\text{ADF}^*$ over ADF, whereas power of the other tests decreases for larger initial values $u_0$.\footnote{Replacing (M)AIC with (M)BIC in the construction of the ADF and MSB tests yields similar results.} These adverse effects are more pronounced for the GLS-versions of the tests than for their OLS counterparts, which fits well to the findings in \citeasnoun{PeRo16}. Noticeably, the VR test is, except for the MA and ARMA(0.3,0.6) cases, rather unaffected by initial values $u_0$ corresponding to small to medium values of $\lambda_u$. Results for $T=250$ and/or other choices of $c$ and $R^2$ are qualitatively similar and we observe comparable effects also in case D1.

%\begin{landscape}
%\end{landscape}

\newpage
\clearpage

\section{Empirical Illustration}\label{sec:Illu}
\begin{figure}[!t]%01.11.2022
	\centering
	\includegraphics[trim={2cm 0.2cm 2.5cm 0.5cm},width=0.8\textwidth,clip]{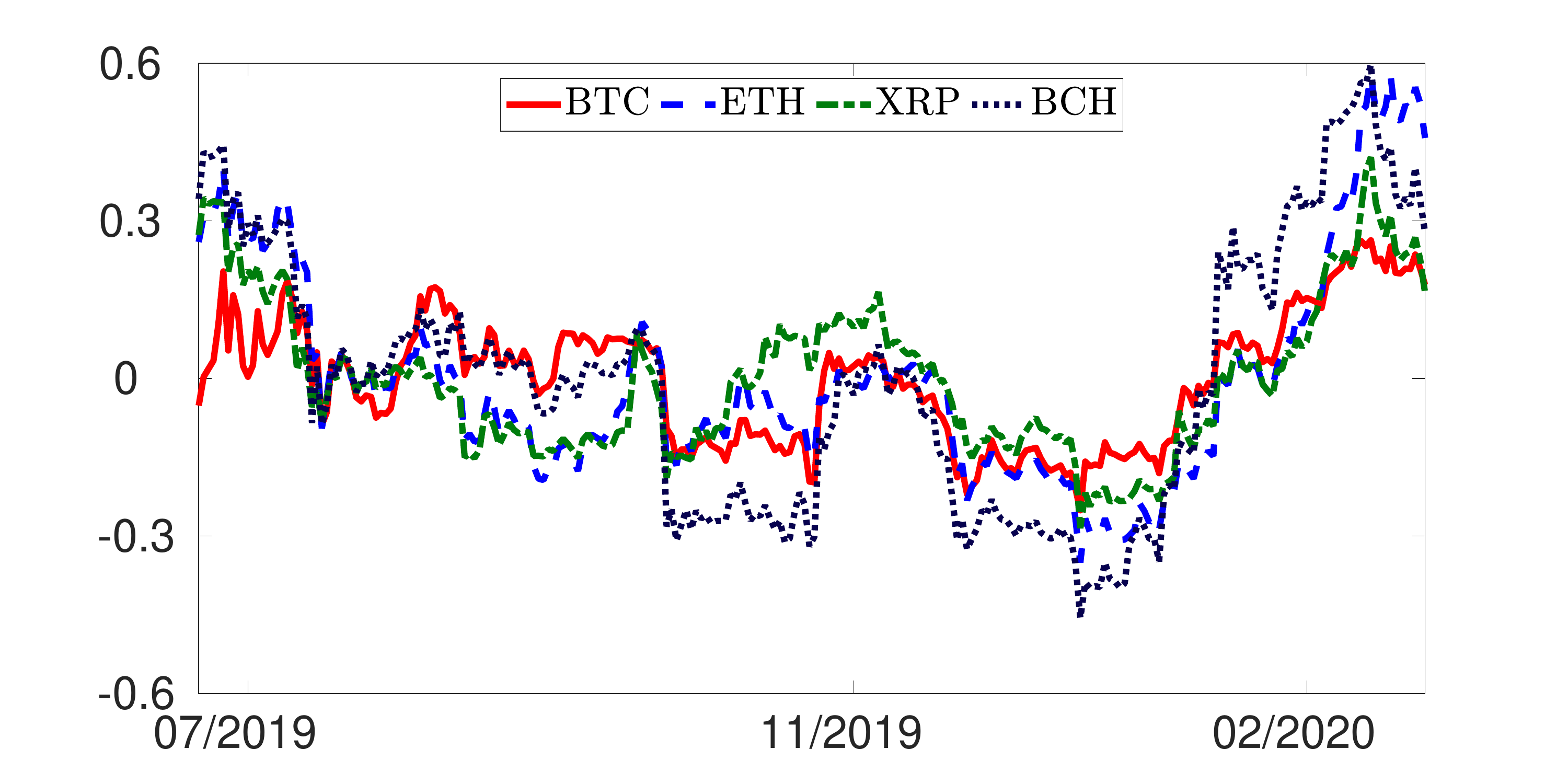}
	\caption{OLS detrended log prices of cryptocurrencies from June 21, 2019 to February 25, 2020.}
	\label{fig:crypto_D2_T250}
\end{figure}

This illustration considers daily closing prices (in U.S. dollar) of cryptocurrencies. In particular, we test for cointegration between the four cryptocurrencies with the highest market capitalization (as of February 25, 2020, excluding stable coins), namely Bitcoin (BTC), Ethereum (ETH), XRP and Bitcoin Cash (BCH). We focus on the latest $T=100$, $T=200$ and $T=250$ time points of the data set analyzed in detail in \citeasnoun{KeZh21}, which ends in February 25, 2020.\footnote{The data set is available on \href{https://github.com/QuantLet/CryptoDynamics/blob/master/CryptoDynamics_Series/logprice.csv}{https://github.com/QuantLet/CryptoDynamics/blob/master/\\CryptoDynamics\_Series/logprice.csv} (accessed: September 4, 2022). Choosing the three (nested) periods is merely to analyze the relationship for different sample sizes similar to those used in Section~\ref{sec:FiniteSample}. This should not be interpreted as a monitoring strategy.} Figure~\ref{fig:crypto_D2_T250} shows the OLS detrended logarithms of the $T=250$ daily prices of the four cryptocurrencies between June 21, 2019 and February 25, 2020. \citeasnoun{KeZh21} find evidence that the four series are integrated of order one. In addition, maybe apart from the beginning of the period, Figure~\ref{fig:crypto_D2_T250} indicates a strong co-movement of the four series.

We choose Bitcoin (\ie, the cryptocurrency with the highest market capitalization) as the left-hand side variable and calculate the nine test statistics considered in the previous section allowing for a deterministic time trend (D2). Table~\ref{tab:results} provides the results for the three periods.
\begin{table}[!ht]%01.11.2022
\centering
\adjustbox{max width=\textwidth}{\begin{threeparttable}
	\caption{Realizations of test statistics}
	\label{tab:results}
	\begin{tabular}{cccccccccc}
		\toprule[1pt]\midrule[0.3pt]
		& $\text{VR}$ & $\text{VR}^{(\GLS)}$ & ADF & $\text{ADF}^{(\GLS)}$ & $\text{ADF}^*$ & $\text{ADF}^{(\GLS)*}$
		 & MSB & $\text{MSB}^{(\GLS)*}$ & $\widehat{Z}_{\alpha}$\\
		 \cmidrule(lr){2-10}
		$T=100$ & \textbf{0.0010} & \textbf{0.0020} & \textbf{-4.5413} & -4.0909 & \textbf{-4.5413} & -4.0909 & 0.1430 & 0.1465 & -31.6678 \\
		$T=200$ & \textbf{0.0012} & 0.0087 & \textbf{-5.0965} & \textbf{-4.3508} & \textbf{-4.5956} & -3.9285 & \textbf{0.1050} & 0.1362 & \textbf{-43.6046} \\
		$T=250$ & 0.0045 & 0.0420 & \textbf{-5.2109} & -3.5599 & \textbf{-5.3029} & -3.5599 & 0.1214 & 0.1486 & \textbf{-42.7777} \\
		\midrule[0.3pt]\bottomrule[1pt]
	\end{tabular}
	\begin{tablenotes}
		\item Notes: \textbf{Bold} numbers indicate significance at the $5\%$ level. The superscripts ``(GLS)'' and ``$*$'' indicate GLS detrending instead of OLS detrending and the use of MAIC instead of AIC, respectively. Except for the $\text{ADF}^{(\GLS)}$ test and the MSB test in the $T=200$ period, significance of results does not change when (M)BIC replaces (M)AIC. The $\widehat{Z}_{\alpha}$ test is based on the QS kernel. Using the Bartlett kernel instead leads to similar results. 
	\end{tablenotes}
\end{threeparttable}}
\end{table}
For each period, test decisions at the $5\%$ significance level are heterogeneous across tests. The ADF and $\text{ADF}^*$ tests reject the null hypothesis of no cointegration in all three periods, whereas the $\text{ADF}^{(\GLS)*}$ test and the $\text{MSB}^{(\GLS)*}$ test never reject. For the $T=100$ and $T=200$ periods, the test decisions of the VR test are in line with the those of the ADF and $\text{ADF}^*$ tests, but for the $T=250$ period the VR test leads to an opposing result. Comparing the values of the test statistics between the $T=200$ and the $T=250$ periods reveals that the ADF and $\text{ADF}^*$ test statistics decrease (\ie, become more significant) as the sample size increases, whereas the remaining seven test statistics -- including the VR statistic -- increase (\ie, become less significant). Thus, it seems that a more detailed analysis is needed to decide whether the series are cointegrated in the $T=250$ period or not. On the other hand, in light of the simulation results in Section~\ref{sec:FiniteSample}, the results in the $T=100$ and $T=200$ periods seem to provide reliable evidence for the presence of cointegration.\footnote{Using different methods and a larger number of cryptocurrencies \citeasnoun{KeZh21} and \citeasnoun{ByGo22} also find evidence for cointegration in the cryptocurrency market.}

For investors, the presence of a cointegrating relationship between the four cryptocurrencies with the highest market capitalization clearly complicates diversification of their cryptocurrency portfolios. On the other hand, it allows them to use a cointegration-based trading strategy to increase profits \citeaffixed{LeNg19,KeZh21}{cf, \eg,}.

\section{Summary and Conclusion}\label{sec:Conc}
This paper derives asymptotic theory for \citename{Br02}'s \citeyear{Br02} nonparametric variance ratio unit root test when applied to regression residuals and analyzes its asymptotic and finite sample properties in this case. The results reveal that the variance ratio test has smaller local asymptotic power than its competitors. However, in finite samples, short-run dynamics in the errors can have severe effects on the performance of the tests both under the null hypothesis and under the alternative. In terms of empirical size, the variance ratio test and the ADF test based on a modified AIC (BIC) perform best and, in particular, outperform the ADF test based on the (unmodified) AIC (BIC). As both tests also perform relatively well in terms of size-corrected power, they may be deemed useful for testing for cointegration in applications. In particular, both test behave nicely in the important case where the regression errors have a moving average component with a small negative coefficient. However, practitioners should be aware that under some short-run dynamics the variance ratio test can be less powerful than its competitors under small deviations from the null hypothesis (but then quickly catches up), whereas the ADF test is prone to power reversal problems for larger deviations from the null hypothesis -- whether or not the AIC (BIC) is modified. Finally, an empirical illustration testing for cointegration between daily prices of cryptocurrencies shows the usefulness of the variance ratio test in practice.

Future research could strive to make the ADF test (and the MSB test) more robust against power reversal problems. Moreover, it might be possible to increase local asymptotic power of the variance ratio test by extending \citename{Ni09}'s \citeyear{Ni09} family of unit root tests -- which contains \citename{Br02}'s \citeyear{Br02} test as a special case -- to cointegration testing. However, this comes at the cost of introducing an index parameter to be chosen by the practitioner. A detailed simulation study then needs to assess whether increasing local asymptotic power of the variance ratio test is also beneficial for its finite sample performance both under the null hypothesis and under the alternative. 

%\section*{Acknowledgements}
%I thank Christoph Hanck, Carsten Jentsch and Martin Wagner for helpful comments.

\section*{Declaration of Interest}
The author has no conflicts of interest to declare.

\bibliographystyle{ifac}

\begin{appendices}
\setcounter{equation}{0}
\renewcommand\theequation{\Alph{section}.\arabic{equation}}	
\setcounter{table}{0}
\renewcommand\thetable{\Alph{section}.\arabic{table}}
\setcounter{figure}{0}
\renewcommand\thefigure{\Alph{section}.\arabic{figure}}

\section{Values of $\bar{c}$ and Asymptotic Critical Values}\label{app:Critvals}

\begin{table}[!ht]%13.07.2022
\centering
\adjustbox{max width=\textwidth}{\begin{threeparttable}
\caption{Values of $\bar{c}$ for the variance ratio GLS test}
\label{tab:cbar}
\begin{tabular}{cccccc}
	\toprule[1pt]\midrule[0.3pt]
	&\multicolumn{5}{c}{$m$}\\
	\cmidrule(lr){2-6}
	Deterministic specification & 1 & 2 & 3 & 4 & 5 \\\midrule
	D1 & -40.25 & -46.25 & -53.75 & -55.75 & -60.00 \\
	D2 & -48.25 & -55.25 & -56.50 & -65.00 & -68.75 \\
	\midrule[0.3pt]\bottomrule[1pt]
\end{tabular}
\begin{tablenotes}
\item \footnotesize{Notes: The values of $\bar c$ correspond to the local alternatives against which the variance ratio test based on GLS detrended data has asymptotic power equal to one-half at the nominal 5\% level when $R^2=0.4$. The results are based on $10{,}000$ Monte Carlo replications and standard Brownian motions are approximated by normalized partial sums of $10{,}000$ i.i.d.\ standard normal random variables.}
\end{tablenotes}
\end{threeparttable}}
\end{table}

\begin{table}[!ht]%25.04.2022
\centering
\adjustbox{max width=\textwidth}{\begin{threeparttable}
\caption{$\alpha$-quantiles of the limiting null distribution of the variance ratio test statistic}
\label{tab:critvalsVR}
\begin{tabular}{ccccccc}
	\toprule[1pt]\midrule[0.3pt]
	$m$& $1.0\%$ & $2.5\%$ & $5.0\%$ & $7.5\%$ & $10.0\%$ & $15.0\%$ \\
	\midrule
	\multicolumn{7}{l}{D0 (without detrending) and D1 with GLS detrending}\\
	\midrule
	1 & 0.00487 & 0.00672 & 0.00908 & 0.01139 & 0.01364 & 0.01818 \\
	2 & 0.00367 & 0.00484 & 0.00619 & 0.00735 & 0.00863 & 0.01077 \\
	3 & 0.00258 & 0.00328 & 0.00422 & 0.00509 & 0.00597 & 0.00745 \\
	4 & 0.00207 & 0.00261 & 0.00327 & 0.00387 & 0.00446 & 0.00547 \\
	5 & 0.00158 & 0.00201 & 0.00256 & 0.00299 & 0.00342 & 0.00422 \\
	\midrule
	\multicolumn{7}{l}{D1 with OLS detrending}\\
	\midrule
	1 & 0.00344 & 0.00458 & 0.00579 & 0.00680 & 0.00772 & 0.00936 \\
	2 & 0.00242 & 0.00313 & 0.00379 & 0.00437 & 0.00491 & 0.00587 \\
	3 & 0.00175 & 0.00224 & 0.00278 & 0.00314 & 0.00349 & 0.00418 \\
	4 & 0.00141 & 0.00174 & 0.00211 & 0.00241 & 0.00267 & 0.00310 \\
	5 & 0.00112 & 0.00137 & 0.00164 & 0.00185 & 0.00204 & 0.00242 \\
	\midrule
	\multicolumn{7}{l}{D2 with OLS detrending}\\
	\midrule
	1 & 0.00166 & 0.00213 & 0.00259 & 0.00296 & 0.00328 & 0.00384 \\
	2 & 0.00130 & 0.00168 & 0.00201 & 0.00228 & 0.00253 & 0.00291 \\
	3 & 0.00106 & 0.00131 & 0.00159 & 0.00179 & 0.00197 & 0.00228 \\
	4 & 0.00092 & 0.00111 & 0.00130 & 0.00146 & 0.00159 & 0.00184 \\
	5 & 0.00077 & 0.00092 & 0.00110 & 0.00122 & 0.00132 & 0.00152 \\
	\midrule
	\multicolumn{7}{l}{D2 with GLS detrending}\\
	\midrule
	1 & 0.00363 & 0.00512 & 0.00668 & 0.00807 & 0.00926 & 0.01164 \\
	2 & 0.00274 & 0.00354 & 0.00468 & 0.00563 & 0.00649 & 0.00807 \\
	3 & 0.00220 & 0.00278 & 0.00354 & 0.00415 & 0.00468 & 0.00582 \\
	4 & 0.00165 & 0.00209 & 0.00267 & 0.00318 & 0.00363 & 0.00442 \\
	5 & 0.00133 & 0.00168 & 0.00214 & 0.00255 & 0.00287 & 0.00348 \\
	\midrule[0.3pt]\bottomrule[1pt]
\end{tabular}
\begin{tablenotes}
	\item \footnotesize{Notes: The variance ratio test is a left-tailed test rejecting the null hypothesis of no cointegration for realizations of the test statistic \textit{smaller} than the $\alpha$-quantile. $m$ denotes the number of stochastic regressors in~\eqref{eq:y}. Under GLS detrending, critical values in case D1 do not depend on $\bar c$, whereas critical values in case D2 depend on $\bar c$ and those reported here correspond to the values of $\bar{c}$ given in Table~\ref{tab:cbar}. Critical values are based on $10{,}000$ Monte Carlo replications and standard Brownian motions are approximated by normalized partial sums of $10{,}000$ i.i.d.\ standard normal random variables.}
\end{tablenotes}
\end{threeparttable}}
\end{table}

\newpage
\clearpage

\section{Additional Results}\label{app:addresults}

\begin{table}[!ht] %14.07.2022
\centering
\adjustbox{max width=\textwidth}{\begin{threeparttable}
\caption{Empirical sizes of the tests in case D0 for $T=100$ and $T=250$.}
\label{tab:RejectionRates-m1-D0}
\begin{tabular}{clcccccccccccc}
	\toprule[1pt]\midrule[0.3pt]
	\multicolumn{2}{c}{}&\multicolumn{1}{c}{}&\multicolumn{3}{c}{AR}&\multicolumn{3}{c}{MA}&\multicolumn{3}{c}{ARMA}&\multicolumn{2}{c}{GARCH}\\
	\cmidrule(lr){4-6}
	\cmidrule(lr){7-9}
	\cmidrule(lr){10-12}
	\cmidrule(lr){13-14}
	$R^2$&Test & IID & 0.3 & 0.6 & 0.9 & 0.3 & 0.6 & 0.9 & (0.3,0.6) & (0.3,0.3) & (0.6,0.3) & (0.05,0.94) & (0.01,0.98) \\\midrule
	\multicolumn{14}{l}{$T=100$}\\
	\midrule
	0&$\text{VR}$ & 0.02 & 0.01 & 0.01 & 0.01 & 0.02 & 0.03 & 0.17 & 0.02 & 0.02 & 0.01 & 0.02 & 0.02 \\
	&ADF & 0.08 & 0.07 & 0.07 & 0.06 & 0.10 & 0.14 & 0.43 & 0.14 & 0.08 & 0.07 & 0.08 & 0.08 \\
	&$\text{ADF}^*$ & 0.05 & 0.04 & 0.04 & 0.04 & 0.06 & 0.06 & 0.14 & 0.07 & 0.05 & 0.04 & 0.05 & 0.05 \\
	&MSB & 0.04 & 0.04 & 0.04 & 0.05 & 0.05 & 0.07 & 0.33 & 0.07 & 0.04 & 0.04 & 0.04 & 0.04 \\
	&$\widehat{Z}_{\alpha}$ & 0.03 & 0.02 & 0.01 & 0.01 & 0.07 & 0.30 & 0.89 & 0.13 & 0.03 & 0.01 & 0.03 & 0.03 \\\midrule
	0.4&$\text{VR}$ & 0.01 & 0.01 & 0.01 & 0.01 & 0.01 & 0.03 & 0.17 & 0.02 & 0.01 & 0.01 & 0.01 & 0.01 \\
	&ADF & 0.07 & 0.08 & 0.07 & 0.07 & 0.10 & 0.13 & 0.45 & 0.13 & 0.07 & 0.07 & 0.07 & 0.07 \\
	&$\text{ADF}^*$ & 0.05 & 0.04 & 0.05 & 0.05 & 0.06 & 0.06 & 0.15 & 0.07 & 0.05 & 0.05 & 0.05 & 0.05 \\
	&MSB & 0.03 & 0.04 & 0.04 & 0.04 & 0.04 & 0.07 & 0.34 & 0.07 & 0.03 & 0.04 & 0.03 & 0.03 \\
	&$\widehat{Z}_{\alpha}$ & 0.02 & 0.02 & 0.02 & 0.02 & 0.07 & 0.31 & 0.89 & 0.12 & 0.03 & 0.02 & 0.02 & 0.02 \\\midrule
	0.8&$\text{VR}$ & 0.01 & 0.01 & 0.01 & 0.01 & 0.01 & 0.03 & 0.18 & 0.01 & 0.01 & 0.01 & 0.01 & 0.01 \\
	&ADF & 0.08 & 0.08 & 0.10 & 0.09 & 0.09 & 0.14 & 0.47 & 0.13 & 0.07 & 0.08 & 0.08 & 0.07 \\
	&$\text{ADF}^*$ & 0.05 & 0.06 & 0.07 & 0.06 & 0.06 & 0.06 & 0.17 & 0.07 & 0.05 & 0.05 & 0.05 & 0.05 \\
	&MSB & 0.02 & 0.03 & 0.03 & 0.04 & 0.03 & 0.07 & 0.37 & 0.06 & 0.02 & 0.03 & 0.02 & 0.02 \\
	&$\widehat{Z}_{\alpha}$ & 0.02 & 0.02 & 0.04 & 0.03 & 0.07 & 0.34 & 0.89 & 0.13 & 0.02 & 0.02 & 0.02 & 0.02 \\
	\midrule
	\multicolumn{14}{l}{$T=250$}\\
	\midrule
	0&$\text{VR}$ & 0.02 & 0.02 & 0.02 & 0.02 & 0.02 & 0.03 & 0.14 & 0.03 & 0.02 & 0.02 & 0.02 & 0.02 \\
	&ADF & 0.06 & 0.06 & 0.05 & 0.04 & 0.07 & 0.09 & 0.32 & 0.09 & 0.06 & 0.05 & 0.06 & 0.06 \\
	&$\text{ADF}^*$ & 0.04 & 0.04 & 0.03 & 0.03 & 0.05 & 0.05 & 0.11 & 0.05 & 0.04 & 0.03 & 0.05 & 0.04 \\
	&MSB & 0.03 & 0.03 & 0.03 & 0.03 & 0.04 & 0.05 & 0.17 & 0.06 & 0.03 & 0.03 & 0.04 & 0.03 \\
	&$\widehat{Z}_{\alpha}$ & 0.03 & 0.02 & 0.02 & 0.01 & 0.07 & 0.31 & 0.94 & 0.14 & 0.03 & 0.02 & 0.03 & 0.03 \\\midrule
	0.4&$\text{VR}$ & 0.02 & 0.02 & 0.02 & 0.02 & 0.02 & 0.03 & 0.16 & 0.02 & 0.02 & 0.02 & 0.02 & 0.02 \\
	&ADF & 0.06 & 0.06 & 0.06 & 0.06 & 0.07 & 0.10 & 0.36 & 0.09 & 0.06 & 0.05 & 0.06 & 0.06 \\
	&$\text{ADF}^*$ & 0.04 & 0.04 & 0.04 & 0.04 & 0.05 & 0.06 & 0.14 & 0.05 & 0.04 & 0.04 & 0.05 & 0.05 \\
	&MSB & 0.02 & 0.03 & 0.03 & 0.03 & 0.03 & 0.04 & 0.21 & 0.05 & 0.03 & 0.03 & 0.03 & 0.03 \\
	&$\widehat{Z}_{\alpha}$ & 0.03 & 0.02 & 0.02 & 0.03 & 0.07 & 0.35 & 0.95 & 0.15 & 0.03 & 0.02 & 0.03 & 0.03 \\\midrule
	0.8&$\text{VR}$ & 0.01 & 0.01 & 0.01 & 0.02 & 0.02 & 0.03 & 0.18 & 0.02 & 0.01 & 0.01 & 0.01 & 0.01 \\
	&ADF & 0.06 & 0.07 & 0.10 & 0.12 & 0.08 & 0.11 & 0.42 & 0.10 & 0.06 & 0.08 & 0.06 & 0.06 \\
	&$\text{ADF}^*$ & 0.05 & 0.05 & 0.06 & 0.09 & 0.05 & 0.06 & 0.20 & 0.06 & 0.05 & 0.06 & 0.05 & 0.05 \\
	&MSB & 0.02 & 0.02 & 0.05 & 0.07 & 0.03 & 0.03 & 0.28 & 0.04 & 0.02 & 0.03 & 0.02 & 0.02 \\
	&$\widehat{Z}_{\alpha}$ & 0.02 & 0.03 & 0.08 & 0.09 & 0.08 & 0.43 & 0.96 & 0.18 & 0.02 & 0.04 & 0.03 & 0.02 \\
	\midrule[0.3pt]\bottomrule[1pt]
\end{tabular}
\begin{tablenotes}
	\item Note: The superscript ``$*$'' indicates the use of MAIC instead of AIC.
\end{tablenotes}
\end{threeparttable}}
\end{table}

\begin{table}[!ht]%14.07.2022
\adjustbox{max width=\textwidth}{\begin{threeparttable}
\caption{Empirical sizes of the tests in cases D1 and D2 for $T=250$.}
\label{tab:RejectionRates-m1-T250-D1D2}
\begin{tabular}{clcccccccccccc}
	\toprule[1pt]\midrule[0.3pt]
	\multicolumn{2}{c}{}&\multicolumn{1}{c}{}&\multicolumn{3}{c}{AR}&\multicolumn{3}{c}{MA}&\multicolumn{3}{c}{ARMA}&\multicolumn{2}{c}{GARCH}\\
	\cmidrule(lr){4-6}
	\cmidrule(lr){7-9}
	\cmidrule(lr){10-12}
	\cmidrule(lr){13-14}
	$R^2$&Test & IID & 0.3 & 0.6 & 0.9 & 0.3 & 0.6 & 0.9 & (0.3,0.6) & (0.3,0.3) & (0.6,0.3) & (0.05,0.94) & (0.01,0.98) \\\midrule
	\multicolumn{14}{l}{Deterministic specification D1}\\
	\midrule
	0&$\text{VR}$ & 0.05 & 0.05 & 0.04 & 0.02 & 0.06 & 0.10 & 0.56 & 0.08 & 0.05 & 0.04 & 0.06 & 0.05 \\
	&$\text{VR}^{(\GLS)}$ & 0.10 & 0.10 & 0.09 & 0.06 & 0.11 & 0.15 & 0.55 & 0.12 & 0.10 & 0.09 & 0.10 & 0.10 \\
	&ADF & 0.06 & 0.06 & 0.05 & 0.04 & 0.08 & 0.12 & 0.61 & 0.12 & 0.06 & 0.05 & 0.07 & 0.06 \\
	&$\text{ADF}^{(\GLS)}$ & 0.06 & 0.06 & 0.05 & 0.04 & 0.08 & 0.12 & 0.43 & 0.13 & 0.06 & 0.05 & 0.07 & 0.06 \\
	&$\text{ADF}^*$ & 0.04 & 0.03 & 0.02 & 0.01 & 0.04 & 0.04 & 0.16 & 0.05 & 0.04 & 0.02 & 0.04 & 0.04 \\
	&$\text{ADF}^{(\GLS)*}$ & 0.04 & 0.04 & 0.03 & 0.01 & 0.05 & 0.05 & 0.16 & 0.06 & 0.04 & 0.02 & 0.04 & 0.04 \\
	&MSB & 0.07 & 0.08 & 0.08 & 0.08 & 0.09 & 0.11 & 0.48 & 0.14 & 0.07 & 0.08 & 0.08 & 0.07 \\
	&$\text{MSB}^{(\GLS)*}$ & 0.03 & 0.03 & 0.03 & 0.02 & 0.04 & 0.03 & 0.04 & 0.05 & 0.03 & 0.02 & 0.04 & 0.03 \\
	&$\widehat{Z}_{\alpha}$ & 0.06 & 0.03 & 0.02 & 0.00 & 0.15 & 0.64 & 1.00 & 0.31 & 0.06 & 0.02 & 0.06 & 0.06 \\\midrule
	0.4&$\text{VR}$ & 0.05 & 0.05 & 0.05 & 0.03 & 0.06 & 0.12 & 0.66 & 0.08 & 0.05 & 0.05 & 0.06 & 0.05 \\
	&$\text{VR}^{(\GLS)}$ & 0.10 & 0.10 & 0.10 & 0.07 & 0.12 & 0.17 & 0.62 & 0.13 & 0.10 & 0.10 & 0.11 & 0.10 \\
	&ADF & 0.06 & 0.06 & 0.06 & 0.05 & 0.08 & 0.14 & 0.71 & 0.14 & 0.06 & 0.05 & 0.07 & 0.06 \\
	&$\text{ADF}^{(\GLS)}$ & 0.07 & 0.07 & 0.07 & 0.06 & 0.09 & 0.14 & 0.49 & 0.14 & 0.07 & 0.06 & 0.08 & 0.07 \\
	&$\text{ADF}^*$ & 0.04 & 0.03 & 0.03 & 0.03 & 0.04 & 0.05 & 0.22 & 0.05 & 0.04 & 0.03 & 0.04 & 0.04 \\
	&$\text{ADF}^{(\GLS)*}$ & 0.05 & 0.04 & 0.04 & 0.03 & 0.05 & 0.06 & 0.22 & 0.06 & 0.05 & 0.03 & 0.04 & 0.05 \\
	&MSB & 0.07 & 0.08 & 0.09 & 0.08 & 0.10 & 0.12 & 0.60 & 0.15 & 0.07 & 0.08 & 0.09 & 0.07 \\
	&$\text{MSB}^{(\GLS)*}$ & 0.04 & 0.04 & 0.04 & 0.03 & 0.05 & 0.03 & 0.07 & 0.05 & 0.04 & 0.03 & 0.03 & 0.03 \\
	&$\widehat{Z}_{\alpha}$ & 0.06 & 0.04 & 0.04 & 0.03 & 0.18 & 0.74 & 1.00 & 0.36 & 0.06 & 0.03 & 0.06 & 0.06 \\\midrule
	0.8&$\text{VR}$ & 0.05 & 0.05 & 0.06 & 0.06 & 0.07 & 0.19 & 0.83 & 0.10 & 0.05 & 0.06 & 0.06 & 0.05 \\
	&$\text{VR}^{(\GLS)}$ & 0.10 & 0.10 & 0.11 & 0.12 & 0.12 & 0.25 & 0.74 & 0.15 & 0.10 & 0.10 & 0.10 & 0.10 \\
	&ADF & 0.06 & 0.07 & 0.14 & 0.13 & 0.09 & 0.21 & 0.89 & 0.17 & 0.06 & 0.09 & 0.07 & 0.06 \\
	&$\text{ADF}^{(\GLS)}$ & 0.07 & 0.08 & 0.15 & 0.18 & 0.10 & 0.20 & 0.61 & 0.17 & 0.07 & 0.10 & 0.08 & 0.07 \\
	&$\text{ADF}^*$ & 0.04 & 0.04 & 0.06 & 0.08 & 0.04 & 0.05 & 0.41 & 0.06 & 0.04 & 0.04 & 0.04 & 0.03 \\
	&$\text{ADF}^{(\GLS)*}$ & 0.04 & 0.05 & 0.08 & 0.12 & 0.05 & 0.07 & 0.39 & 0.07 & 0.04 & 0.06 & 0.04 & 0.04 \\
	&MSB & 0.07 & 0.09 & 0.16 & 0.17 & 0.10 & 0.15 & 0.83 & 0.18 & 0.07 & 0.11 & 0.09 & 0.07 \\
	&$\text{MSB}^{(\GLS)*}$ & 0.03 & 0.04 & 0.07 & 0.11 & 0.04 & 0.02 & 0.18 & 0.05 & 0.03 & 0.05 & 0.04 & 0.03 \\
	&$\widehat{Z}_{\alpha}$ & 0.06 & 0.08 & 0.19 & 0.16 & 0.29 & 0.94 & 1.00 & 0.57 & 0.06 & 0.10 & 0.06 & 0.06 \\
	\midrule
	\multicolumn{14}{l}{Deterministic specification D2}\\
	\midrule
	0&$\text{VR}$ & 0.05 & 0.05 & 0.03 & 0.01 & 0.07 & 0.15 & 0.83 & 0.10 & 0.05 & 0.04 & 0.06 & 0.05 \\
	&$\text{VR}^{(\GLS)}$ & 0.13 & 0.12 & 0.11 & 0.05 & 0.15 & 0.23 & 0.65 & 0.18 & 0.13 & 0.11 & 0.14 & 0.13 \\
	&ADF & 0.07 & 0.07 & 0.06 & 0.04 & 0.10 & 0.18 & 0.79 & 0.19 & 0.07 & 0.06 & 0.08 & 0.07 \\
	&$\text{ADF}^{(\GLS)}$ & 0.08 & 0.08 & 0.07 & 0.05 & 0.11 & 0.18 & 0.62 & 0.19 & 0.08 & 0.07 & 0.09 & 0.08 \\
	&$\text{ADF}^*$ & 0.04 & 0.03 & 0.03 & 0.01 & 0.04 & 0.05 & 0.24 & 0.06 & 0.04 & 0.01 & 0.04 & 0.04 \\
	&$\text{ADF}^{(\GLS)*}$ & 0.05 & 0.04 & 0.03 & 0.01 & 0.06 & 0.07 & 0.25 & 0.07 & 0.05 & 0.02 & 0.04 & 0.04 \\
	&MSB & 0.09 & 0.11 & 0.12 & 0.14 & 0.12 & 0.15 & 0.68 & 0.19 & 0.09 & 0.12 & 0.10 & 0.09 \\
	&$\text{MSB}^{(\GLS)*}$ & 0.03 & 0.04 & 0.03 & 0.03 & 0.05 & 0.03 & 0.11 & 0.06 & 0.03 & 0.01 & 0.03 & 0.03 \\
	&$\widehat{Z}_{\alpha}$ & 0.05 & 0.03 & 0.01 & 0.00 & 0.21 & 0.83 & 1.00 & 0.43 & 0.05 & 0.01 & 0.06 & 0.06 \\\midrule
	0.4&$\text{VR}$ & 0.05 & 0.05 & 0.04 & 0.01 & 0.07 & 0.19 & 0.91 & 0.10 & 0.05 & 0.04 & 0.05 & 0.05 \\
	&$\text{VR}^{(\GLS)}$ & 0.13 & 0.12 & 0.11 & 0.07 & 0.15 & 0.25 & 0.69 & 0.19 & 0.13 & 0.11 & 0.13 & 0.13 \\
	&ADF & 0.08 & 0.08 & 0.07 & 0.05 & 0.11 & 0.20 & 0.87 & 0.19 & 0.08 & 0.06 & 0.09 & 0.08 \\
	&$\text{ADF}^{(\GLS)}$ & 0.08 & 0.08 & 0.08 & 0.06 & 0.12 & 0.21 & 0.68 & 0.20 & 0.08 & 0.07 & 0.09 & 0.09 \\
	&$\text{ADF}^*$ & 0.04 & 0.03 & 0.03 & 0.02 & 0.04 & 0.05 & 0.29 & 0.06 & 0.04 & 0.02 & 0.04 & 0.04 \\
	&$\text{ADF}^{(\GLS)*}$ & 0.05 & 0.04 & 0.03 & 0.02 & 0.06 & 0.06 & 0.29 & 0.08 & 0.05 & 0.02 & 0.05 & 0.05 \\
	&MSB & 0.08 & 0.11 & 0.11 & 0.11 & 0.12 & 0.16 & 0.79 & 0.20 & 0.08 & 0.11 & 0.11 & 0.09 \\
	&$\text{MSB}^{(\GLS)*}$ & 0.03 & 0.03 & 0.03 & 0.02 & 0.04 & 0.03 & 0.15 & 0.06 & 0.03 & 0.02 & 0.03 & 0.03 \\
	&$\widehat{Z}_{\alpha}$ & 0.06 & 0.03 & 0.03 & 0.01 & 0.25 & 0.91 & 1.00 & 0.51 & 0.06 & 0.02 & 0.06 & 0.06 \\\midrule
	0.8&$\text{VR}$ & 0.05 & 0.06 & 0.07 & 0.03 & 0.09 & 0.32 & 0.99 & 0.15 & 0.05 & 0.06 & 0.05 & 0.06 \\
	&$\text{VR}^{(\GLS)}$ & 0.13 & 0.13 & 0.15 & 0.13 & 0.16 & 0.35 & 0.73 & 0.22 & 0.13 & 0.13 & 0.13 & 0.13 \\
	&ADF & 0.07 & 0.09 & 0.18 & 0.09 & 0.12 & 0.32 & 0.96 & 0.25 & 0.07 & 0.11 & 0.08 & 0.07 \\
	&$\text{ADF}^{(\GLS)}$ & 0.08 & 0.10 & 0.19 & 0.12 & 0.13 & 0.29 & 0.76 & 0.25 & 0.08 & 0.13 & 0.09 & 0.08 \\
	&$\text{ADF}^*$ & 0.04 & 0.04 & 0.07 & 0.04 & 0.04 & 0.06 & 0.44 & 0.06 & 0.04 & 0.05 & 0.04 & 0.04 \\
	&$\text{ADF}^{(\GLS)*}$ & 0.05 & 0.05 & 0.08 & 0.07 & 0.06 & 0.07 & 0.42 & 0.08 & 0.05 & 0.06 & 0.05 & 0.04 \\
	&MSB & 0.09 & 0.11 & 0.19 & 0.12 & 0.13 & 0.22 & 0.92 & 0.25 & 0.09 & 0.13 & 0.10 & 0.09 \\
	&$\text{MSB}^{(\GLS)*}$ & 0.03 & 0.04 & 0.07 & 0.05 & 0.04 & 0.02 & 0.24 & 0.06 & 0.03 & 0.05 & 0.03 & 0.03 \\
	&$\widehat{Z}_{\alpha}$ & 0.06 & 0.08 & 0.21 & 0.07 & 0.42 & 1.00 & 1.00 & 0.76 & 0.06 & 0.10 & 0.06 & 0.05 \\
	\midrule[0.3pt]\bottomrule[1pt]
\end{tabular}
\begin{tablenotes}
	\item Note: See note to Table~\ref{tab:RejectionRates-m1-T100-D1D2}.
\end{tablenotes}
\end{threeparttable}}
\end{table}

%\begin{landscape}
\begin{sidewaysfigure}[!ht]
%\begin{figure}[!ht]
\begin{center}
	\begin{subfigure}{0.25\textheight}
		\centering
		\caption*{IID}
		\vspace{-1ex}
		\includegraphics[trim={2cm 0.2cm 2cm 0.5cm},width=0.98\textwidth,clip]{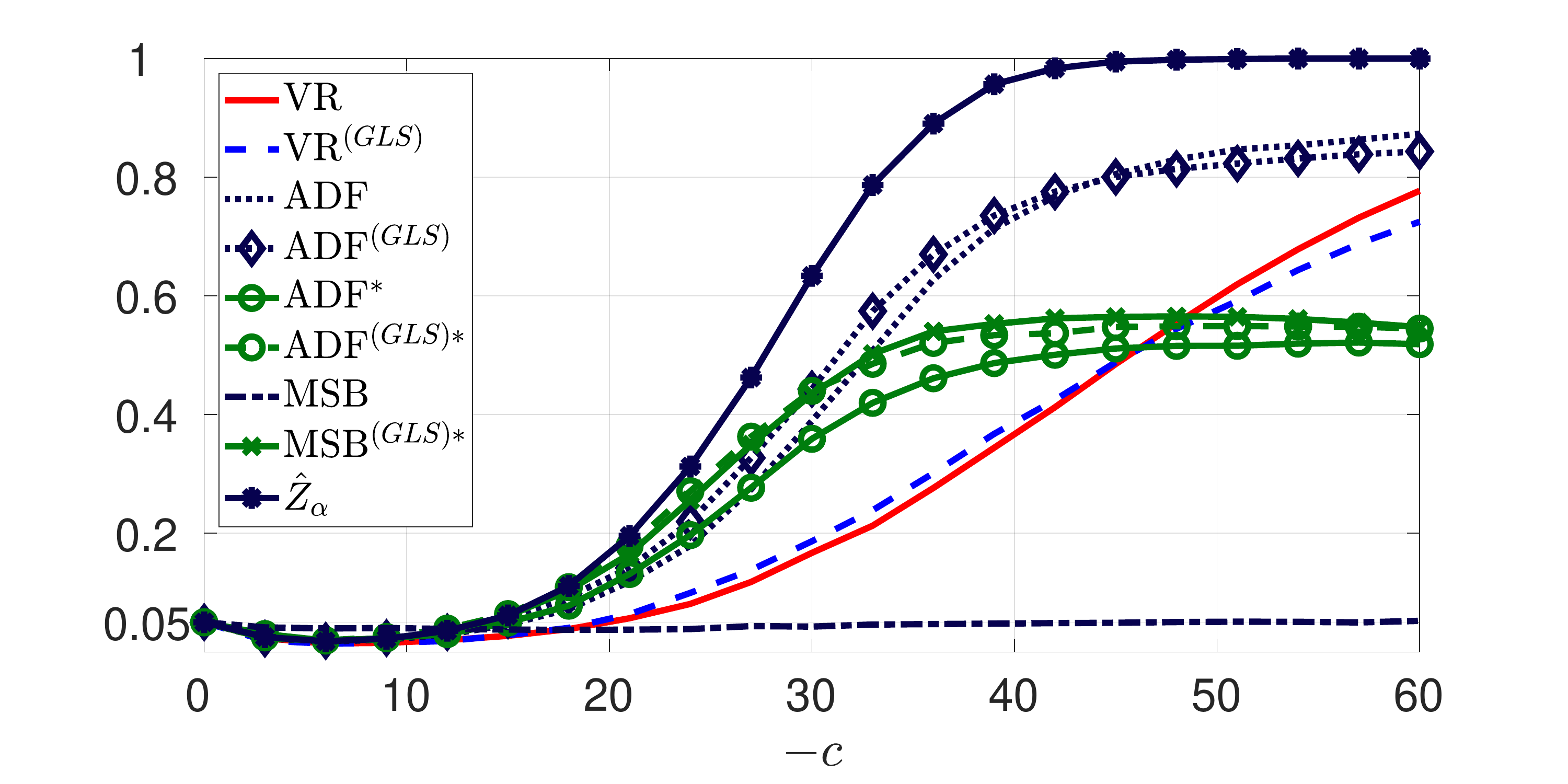}
	\end{subfigure}\begin{subfigure}{0.25\textheight}
		\centering
		\caption*{AR, $\phi=0.3$}
		\vspace{-1ex}
		\includegraphics[trim={2cm 0.2cm 2cm 0.5cm},width=0.98\textwidth,clip]{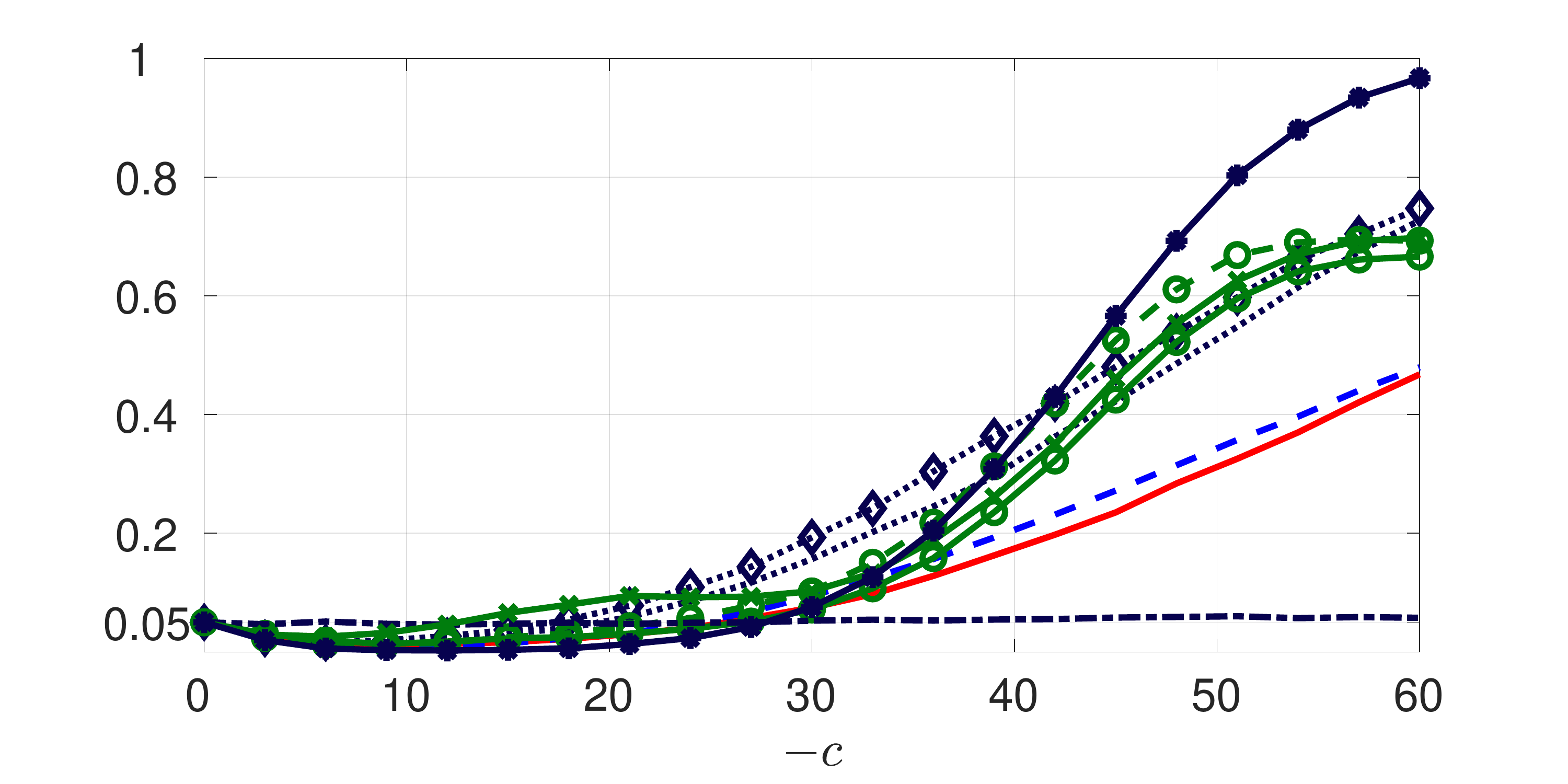}
	\end{subfigure}\begin{subfigure}{0.25\textheight}
		\centering
		\caption*{AR, $\phi=0.6$}
		\vspace{-1ex}
		\includegraphics[trim={2cm 0.2cm 2cm 0.5cm},width=0.98\textwidth,clip]{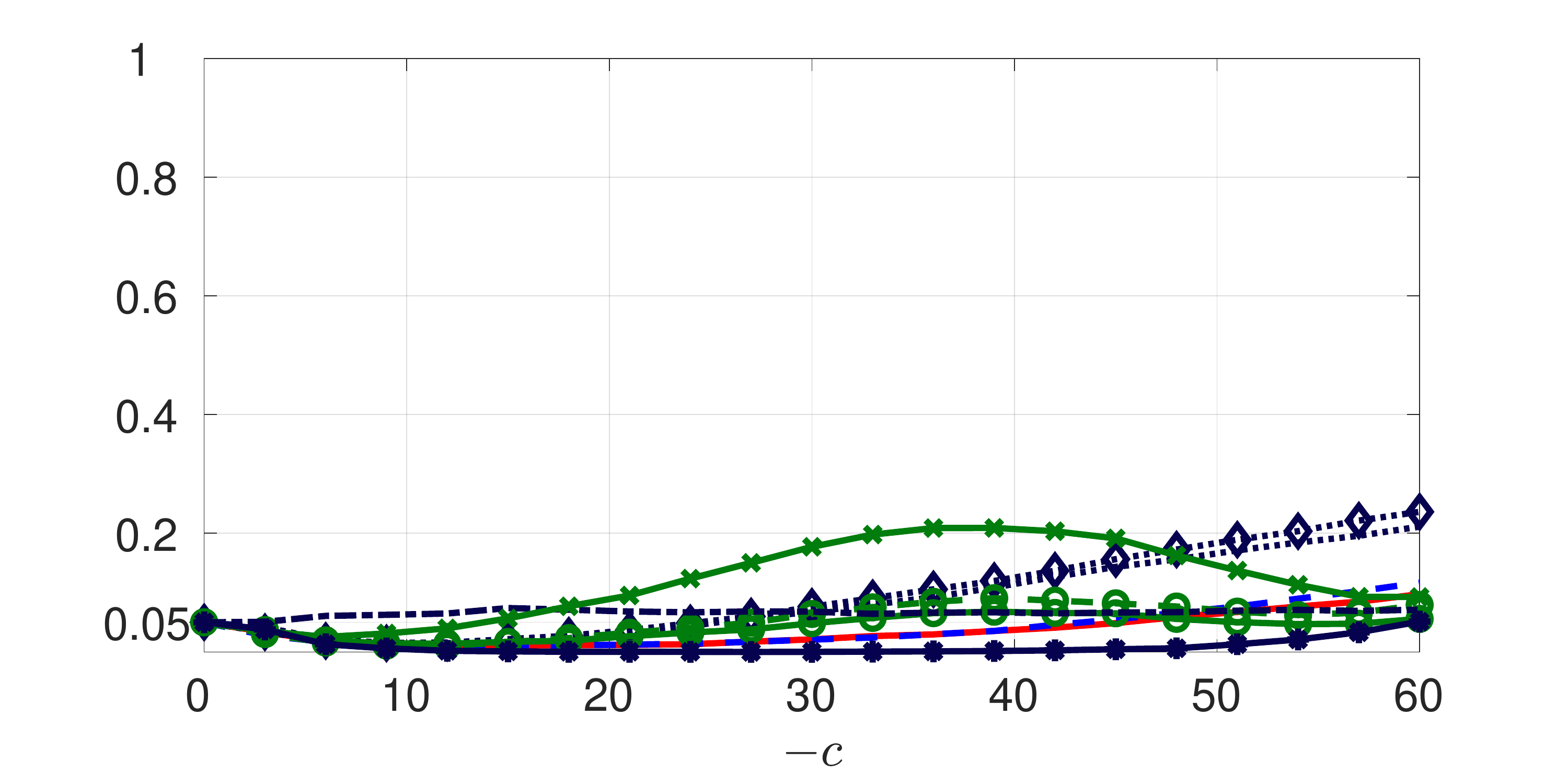}
	\end{subfigure}\begin{subfigure}{0.25\textheight}
		\centering
		\caption*{AR, $\phi=0.9$}
		\vspace{-1ex}
		\includegraphics[trim={2cm 0.2cm 2cm 0.5cm},width=0.98\textwidth,clip]{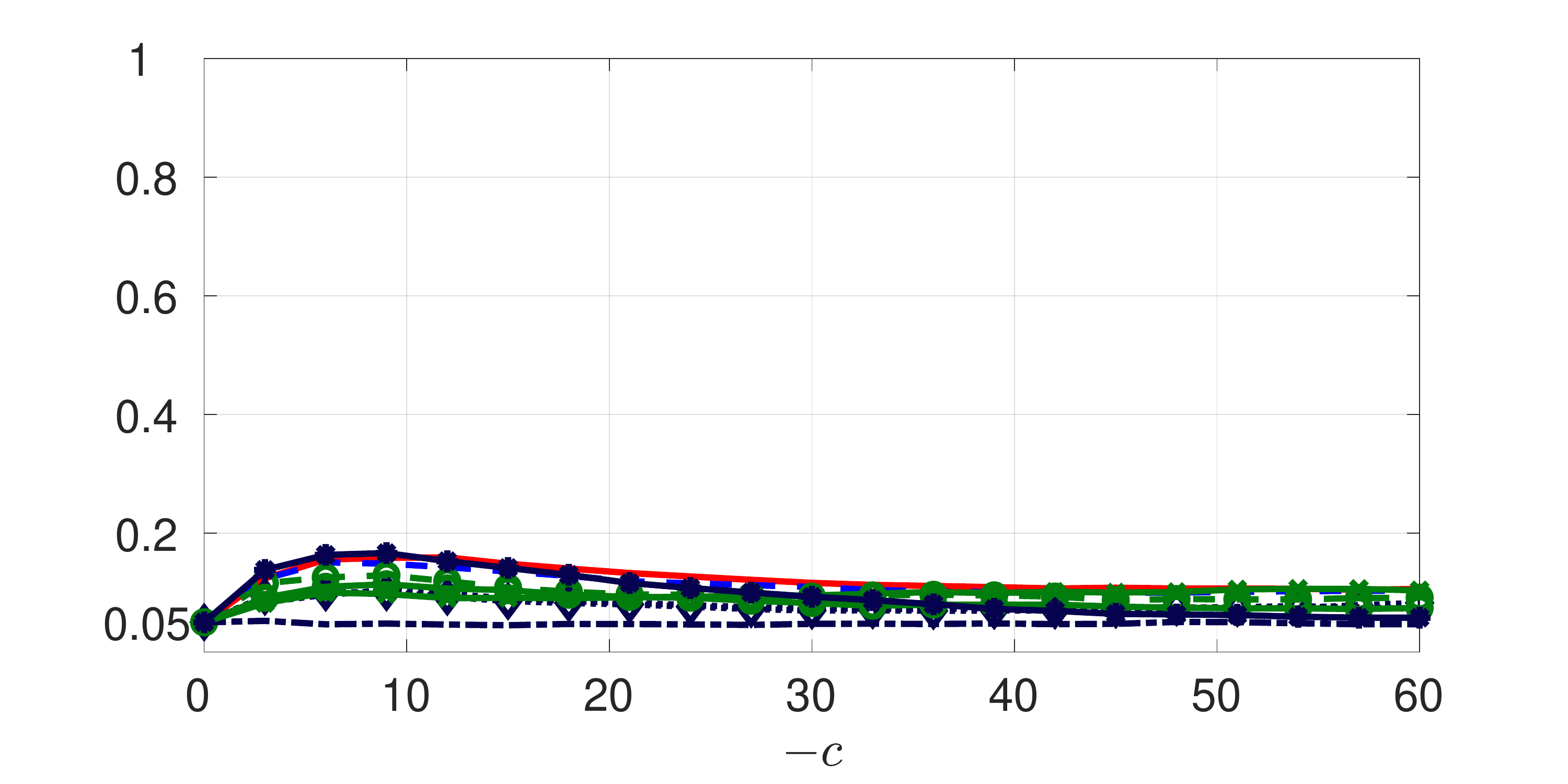}
	\end{subfigure}
	
	\vspace{1ex}
	
	\begin{subfigure}{0.25\textheight}
		\centering
		\caption*{MA, $\theta=0.3$}
		\vspace{-1ex}
		\includegraphics[trim={2cm 0.2cm 2cm 0.5cm},width=0.98\textwidth,clip]{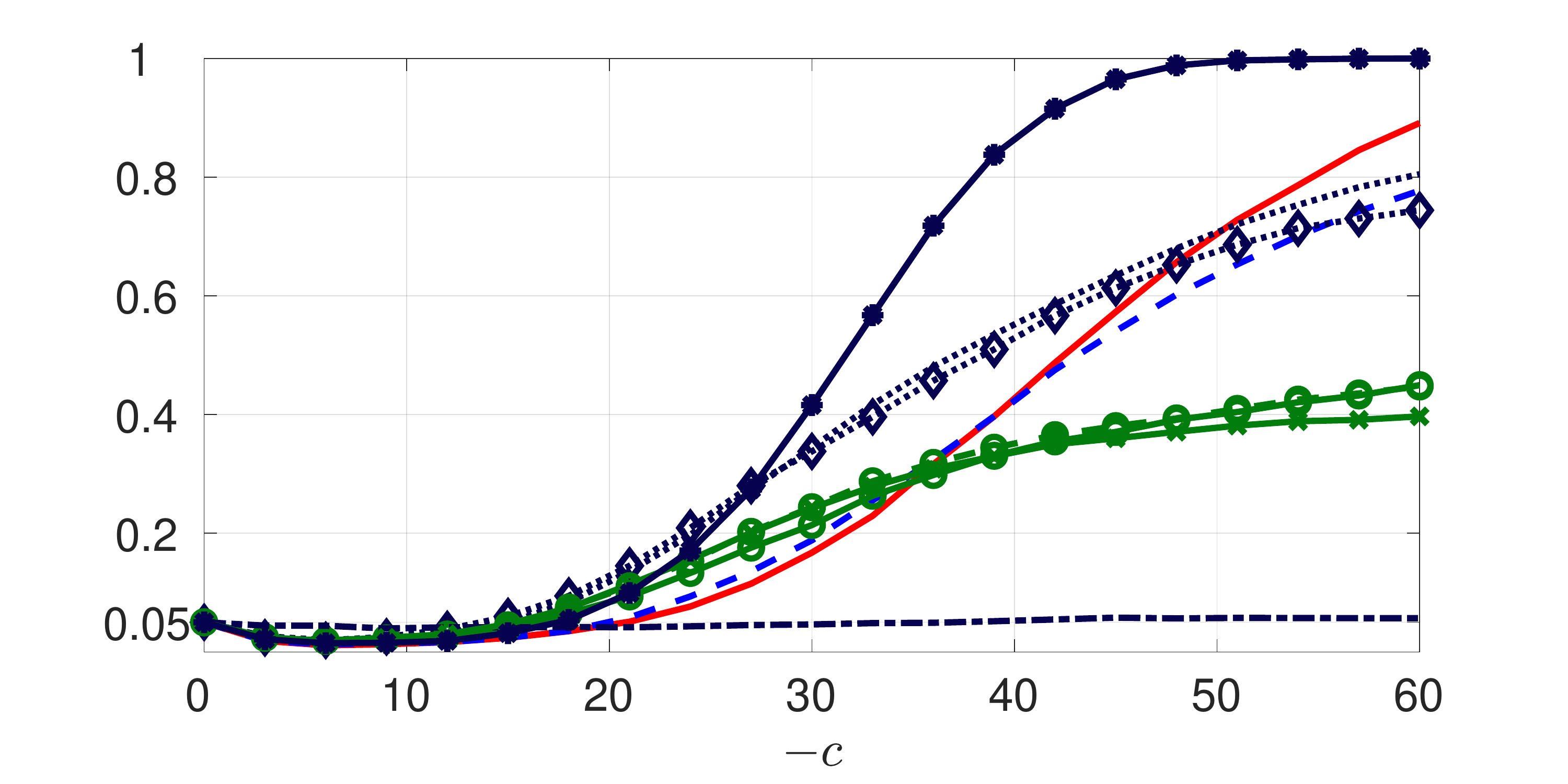}
	\end{subfigure}\begin{subfigure}{0.25\textheight}
		\centering
		\caption*{MA, $\theta=0.6$}
		\vspace{-1ex}
		\includegraphics[trim={2cm 0.2cm 2cm 0.5cm},width=0.98\textwidth,clip]{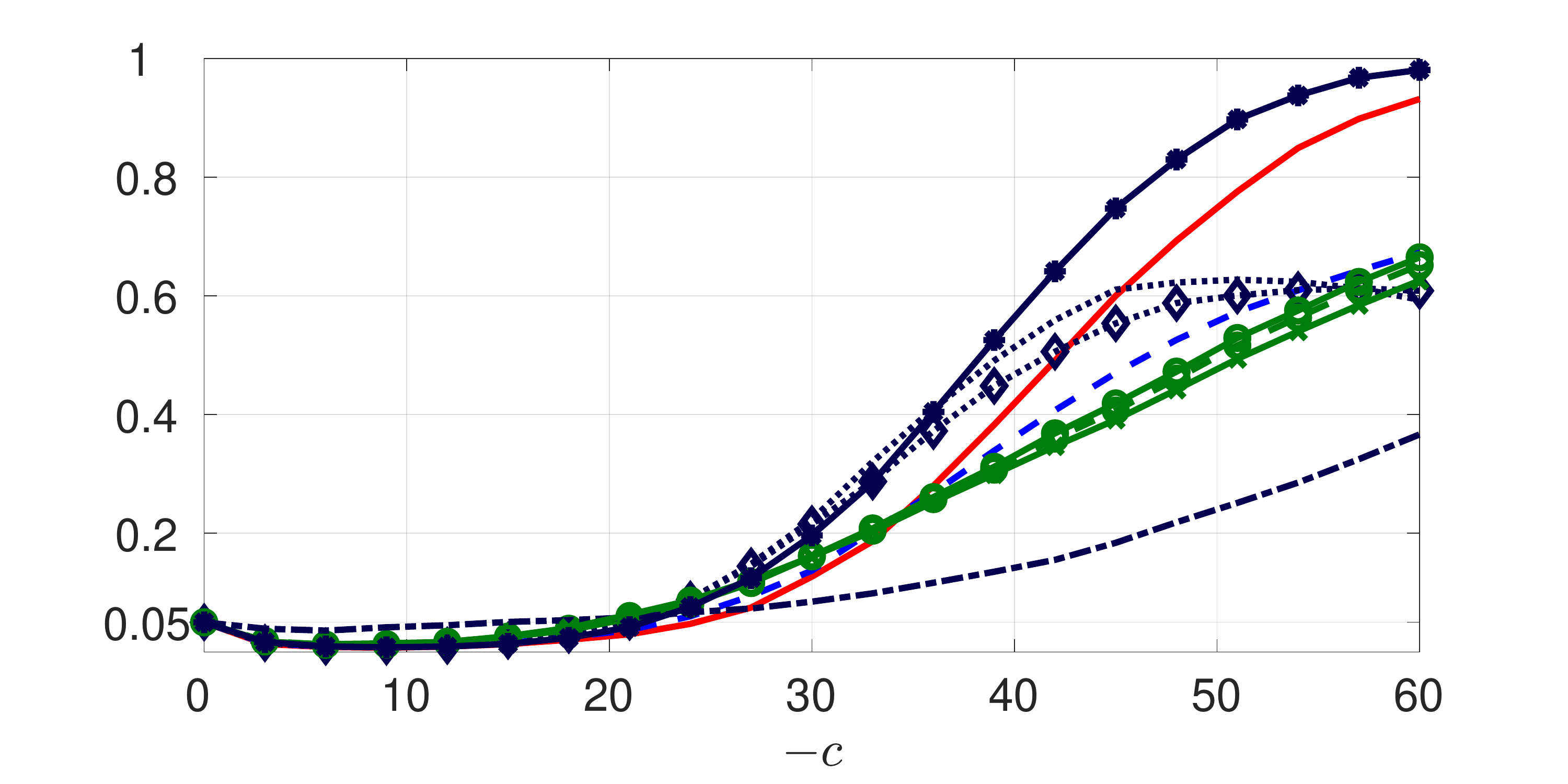}
	\end{subfigure}\begin{subfigure}{0.25\textheight}
		\centering
		\caption*{MA, $\theta=0.9$}
		\vspace{-1ex}
		\includegraphics[trim={2cm 0.2cm 2cm 0.5cm},width=0.98\textwidth,clip]{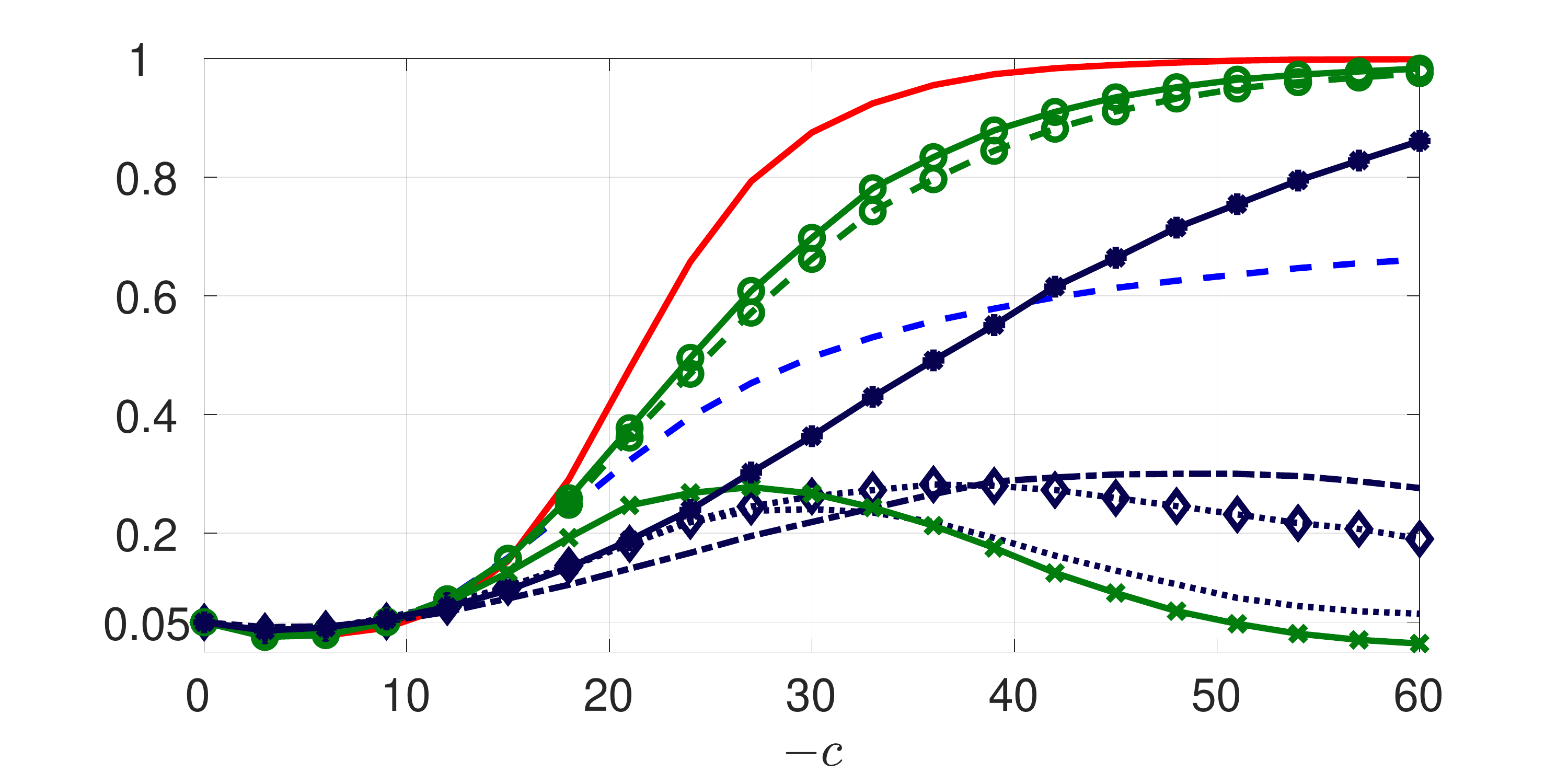}
	\end{subfigure}\begin{subfigure}{0.25\textheight}
		\centering
		\caption*{ARMA, $\phi=0.3,\theta=0.6$}
		\vspace{-1ex}
		\includegraphics[trim={2cm 0.2cm 2cm 0.5cm},width=0.98\textwidth,clip]{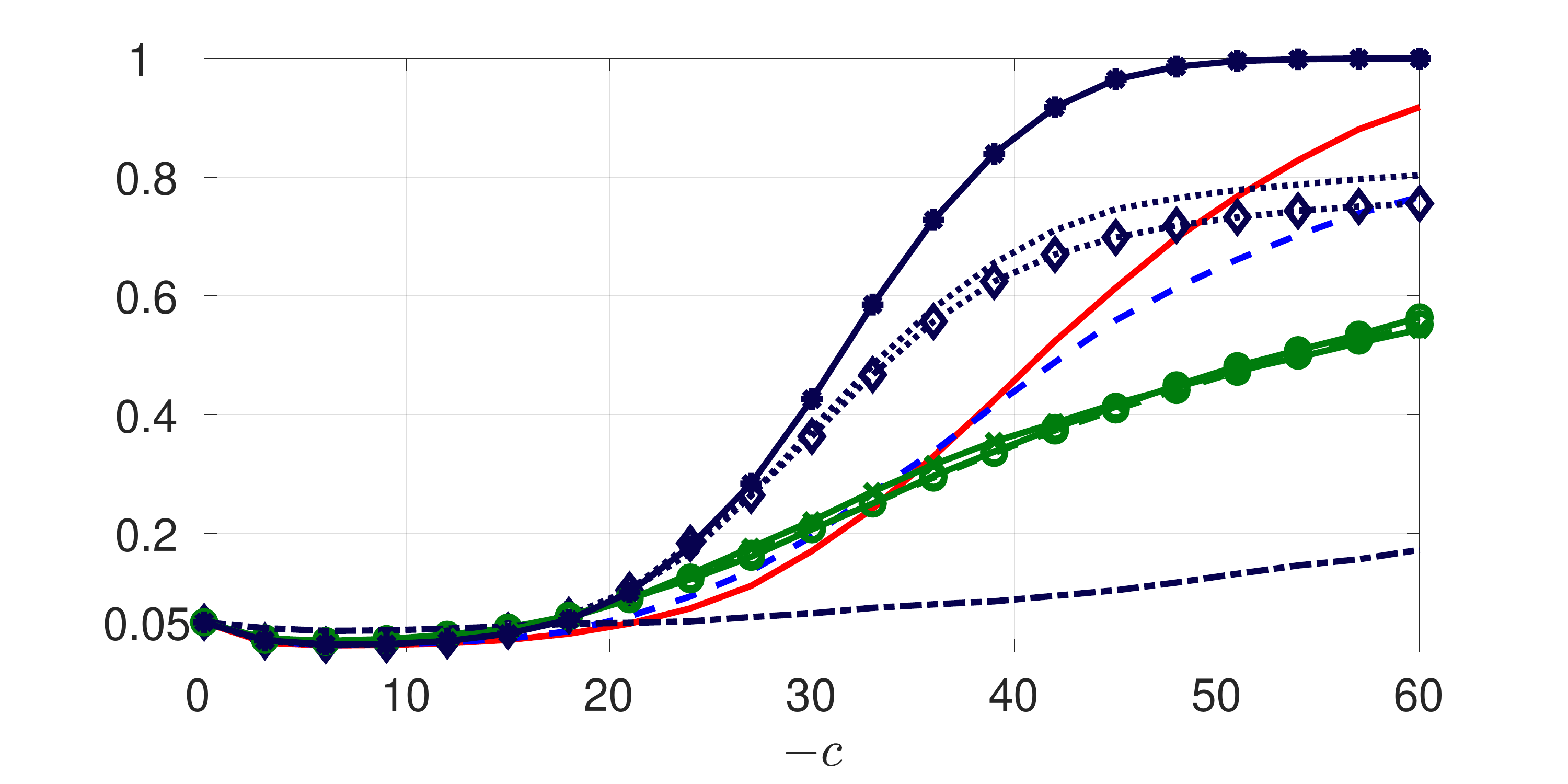}
	\end{subfigure}
	
	\vspace{1ex}
	
	\begin{subfigure}{0.25\textheight}
		\centering
		\caption*{ARMA, $\phi=0.3,\theta=0.3$}
		\vspace{-1ex}
		\includegraphics[trim={2cm 0.2cm 2cm 0.5cm},width=0.98\textwidth,clip]{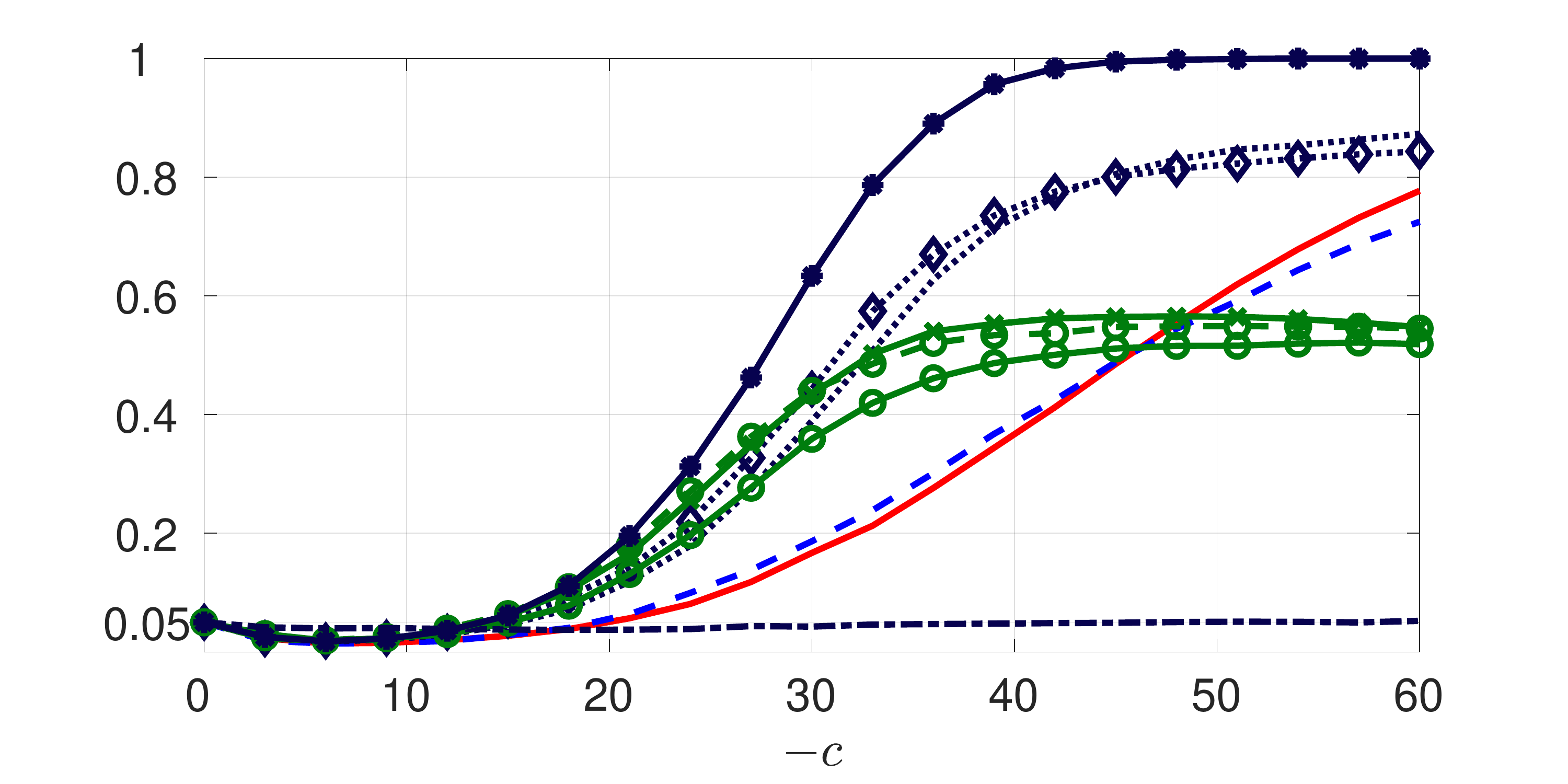}
	\end{subfigure}\begin{subfigure}{0.25\textheight}
		\centering
		\caption*{ARMA, $\phi=0.6,\theta=0.3$}
		\vspace{-1ex}
		\includegraphics[trim={2cm 0.2cm 2cm 0.5cm},width=0.98\textwidth,clip]{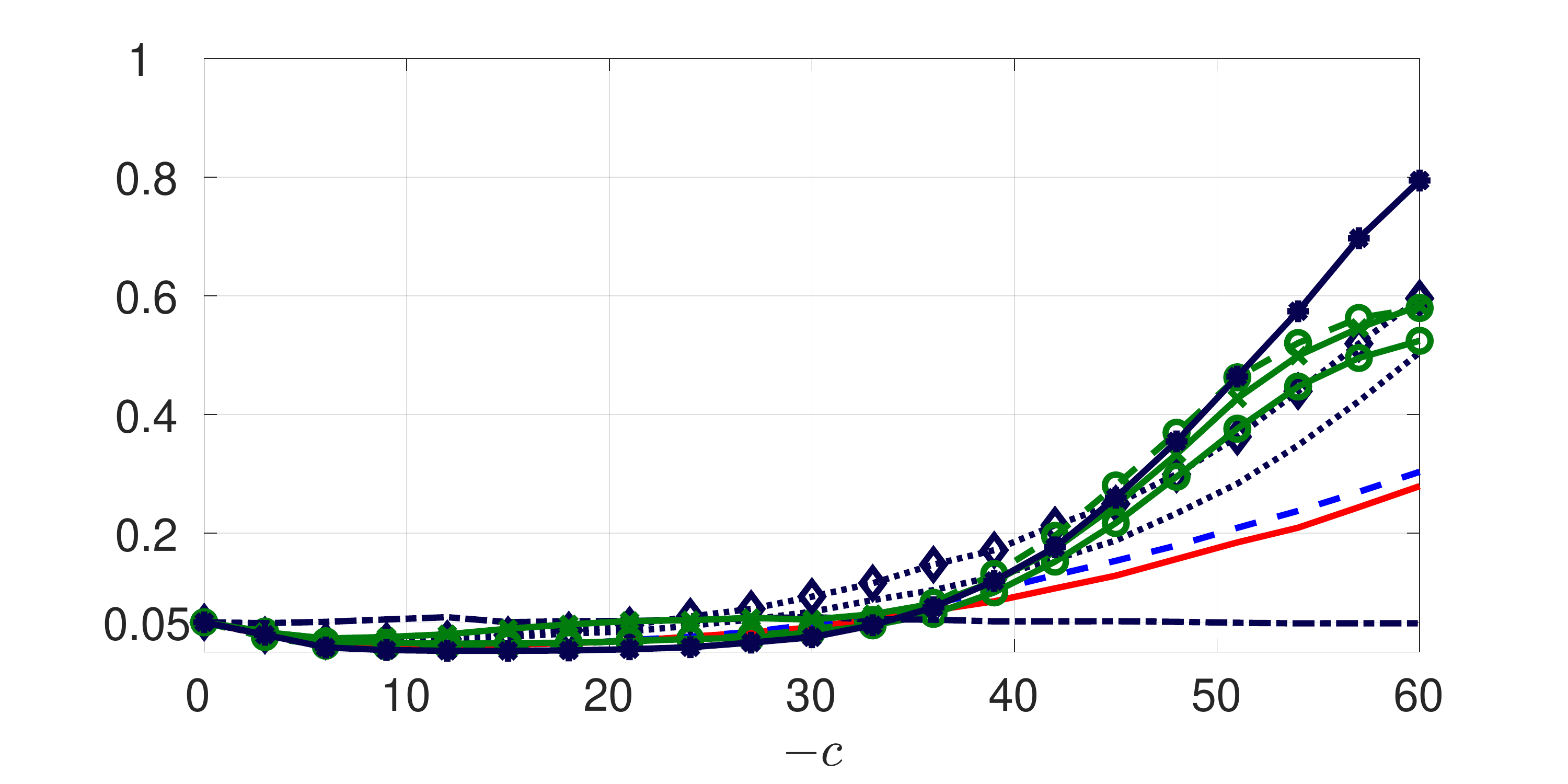}
	\end{subfigure}\begin{subfigure}{0.25\textheight}
		\centering
		\caption*{GARCH, $a_1=0.05,a_2=0.93$}
		\vspace{-1ex}
		\includegraphics[trim={2cm 0.2cm 2cm 0.5cm},width=0.98\textwidth,clip]{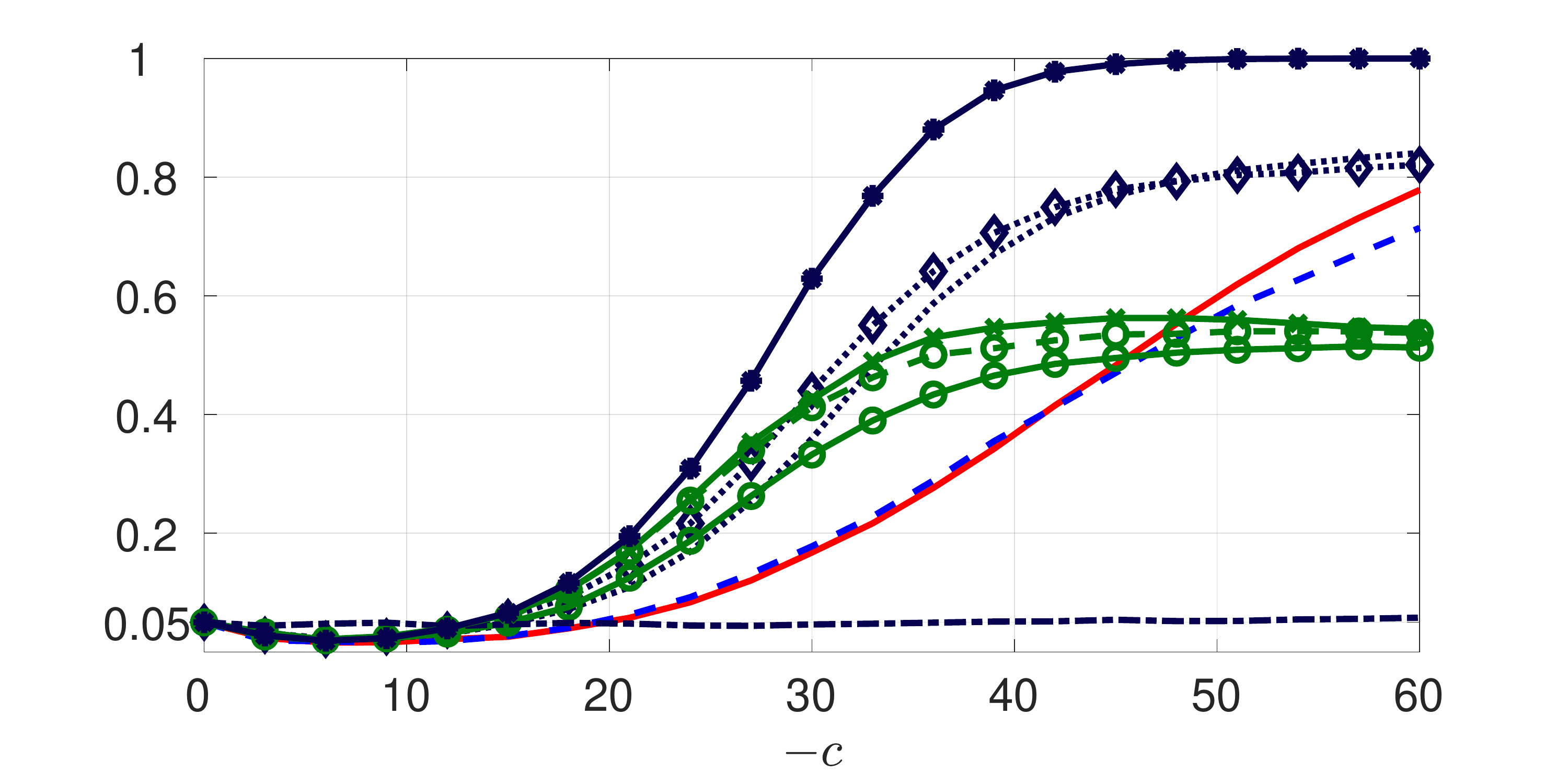}
	\end{subfigure}\begin{subfigure}{0.25\textheight}
		\centering
		\caption*{GARCH, $a_1=0.01,a_2=0.98$}
		\vspace{-1ex}
		\includegraphics[trim={2cm 0.2cm 2cm 0.5cm},width=0.98\textwidth,clip]{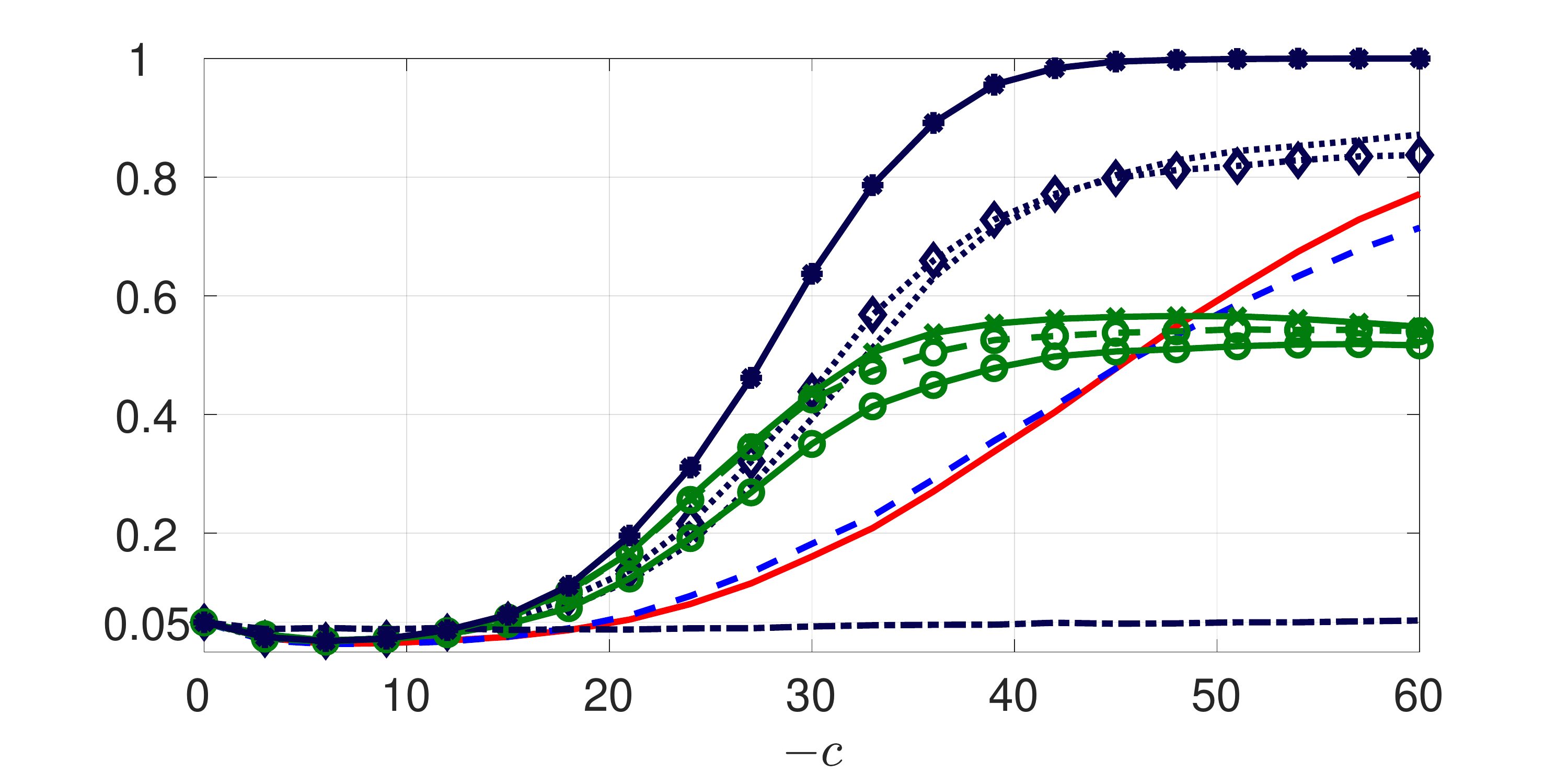}
	\end{subfigure}
	
\end{center}
\vspace{-3ex}
\caption{Size-corrected power of the tests at the nominal $5\%$ level for $\text{$\text{H}_0$:}\ \rho = 1$ under the alternative $\rho = 1+c/T$ in case D2, for $T=100$ and $R^2=0.8$\\{\footnotesize Note: See note to Figure~\ref{fig:size_adjusted_power_m_1_T_100_deter_2_R2_04_AIC}.}}
\label{fig:size_adjusted_power_m_1_T_100_deter_2_R2_08_AIC}
%\end{figure}
\end{sidewaysfigure}
%\end{landscape}

%\begin{landscape}
\begin{sidewaysfigure}[!ht]
%\begin{figure}[!ht]
\begin{center}
	\begin{subfigure}{0.25\textheight}
		\centering
		\caption*{IID}
		\vspace{-1ex}
		\includegraphics[trim={2cm 0.2cm 2cm 0.5cm},width=0.98\textwidth,clip]{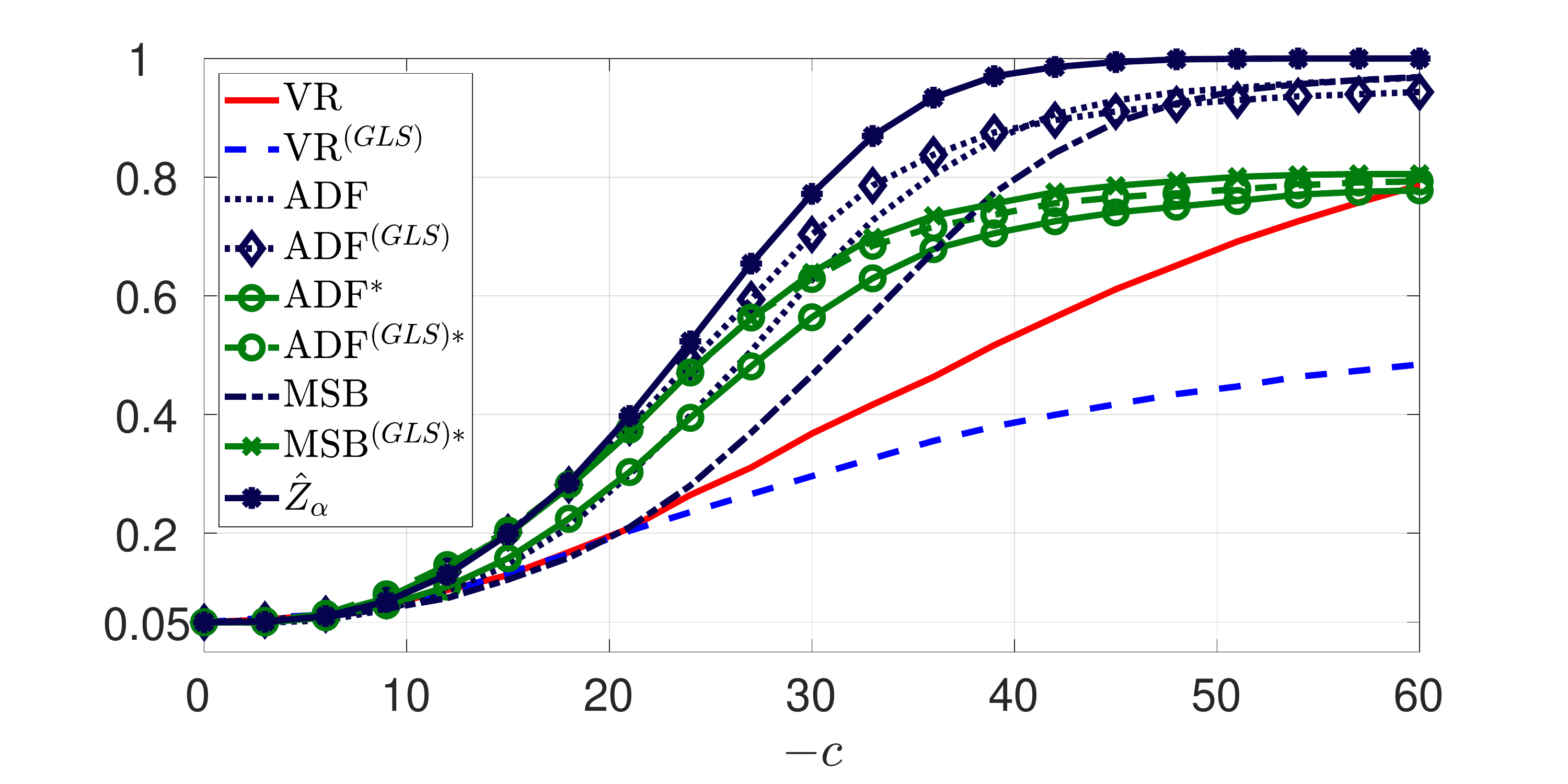}
	\end{subfigure}\begin{subfigure}{0.25\textheight}
		\centering
		\caption*{AR, $\phi=0.3$}
		\vspace{-1ex}
		\includegraphics[trim={2cm 0.2cm 2cm 0.5cm},width=0.98\textwidth,clip]{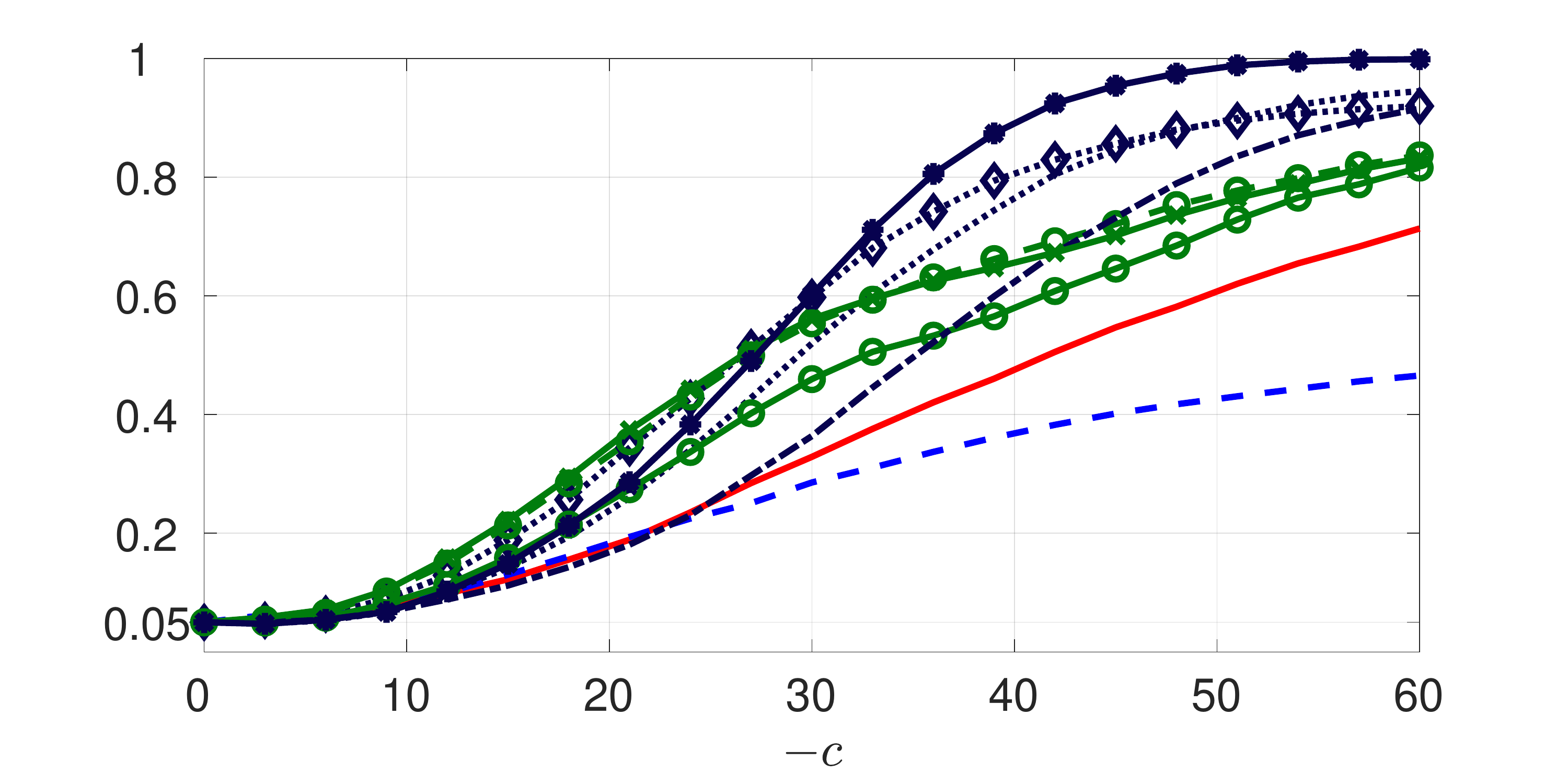}
	\end{subfigure}\begin{subfigure}{0.25\textheight}
		\centering
		\caption*{AR, $\phi=0.6$}
		\vspace{-1ex}
		\includegraphics[trim={2cm 0.2cm 2cm 0.5cm},width=0.98\textwidth,clip]{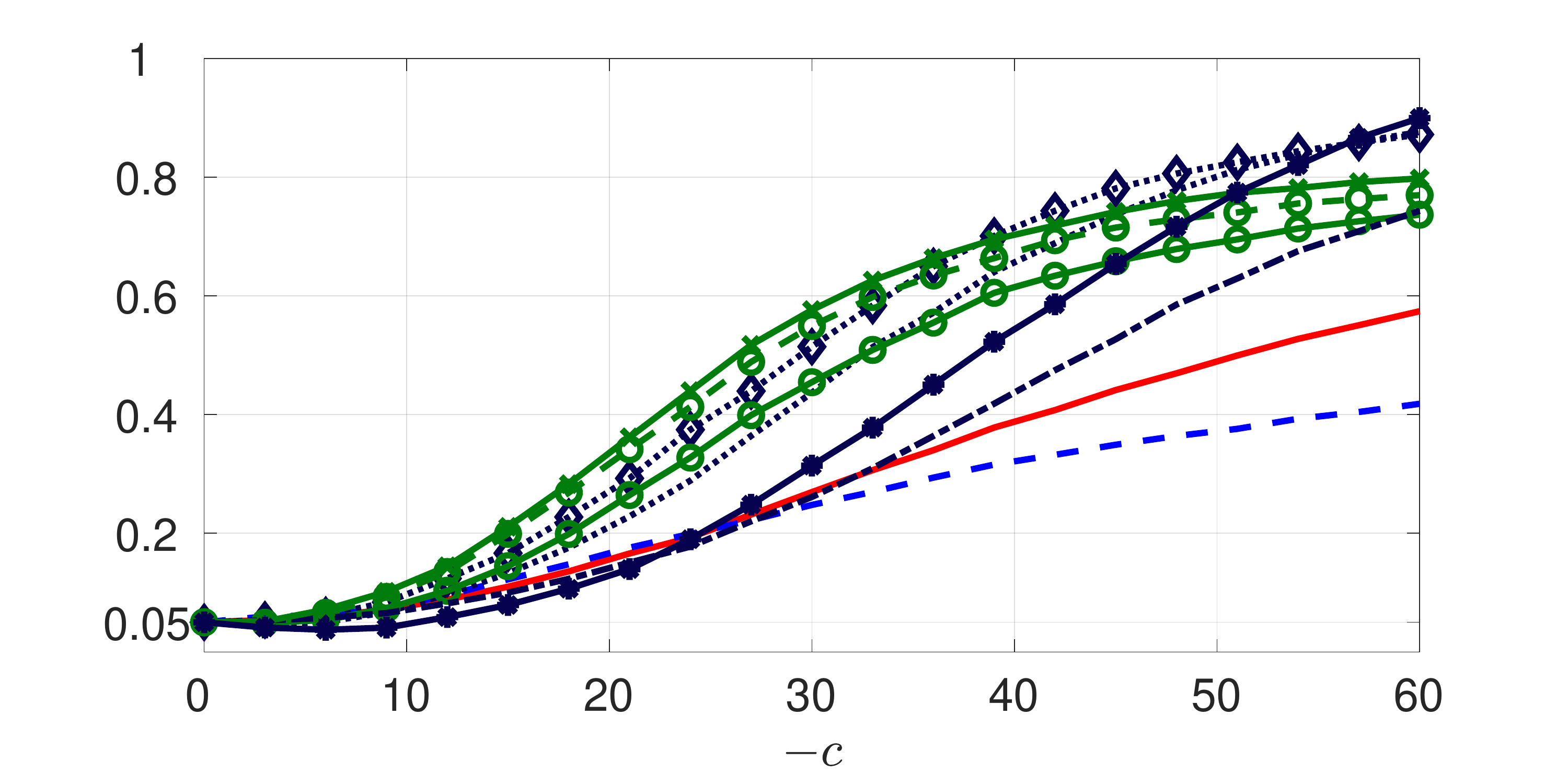}
	\end{subfigure}\begin{subfigure}{0.25\textheight}
		\centering
		\caption*{AR, $\phi=0.9$}
		\vspace{-1ex}
		\includegraphics[trim={2cm 0.2cm 2cm 0.5cm},width=0.98\textwidth,clip]{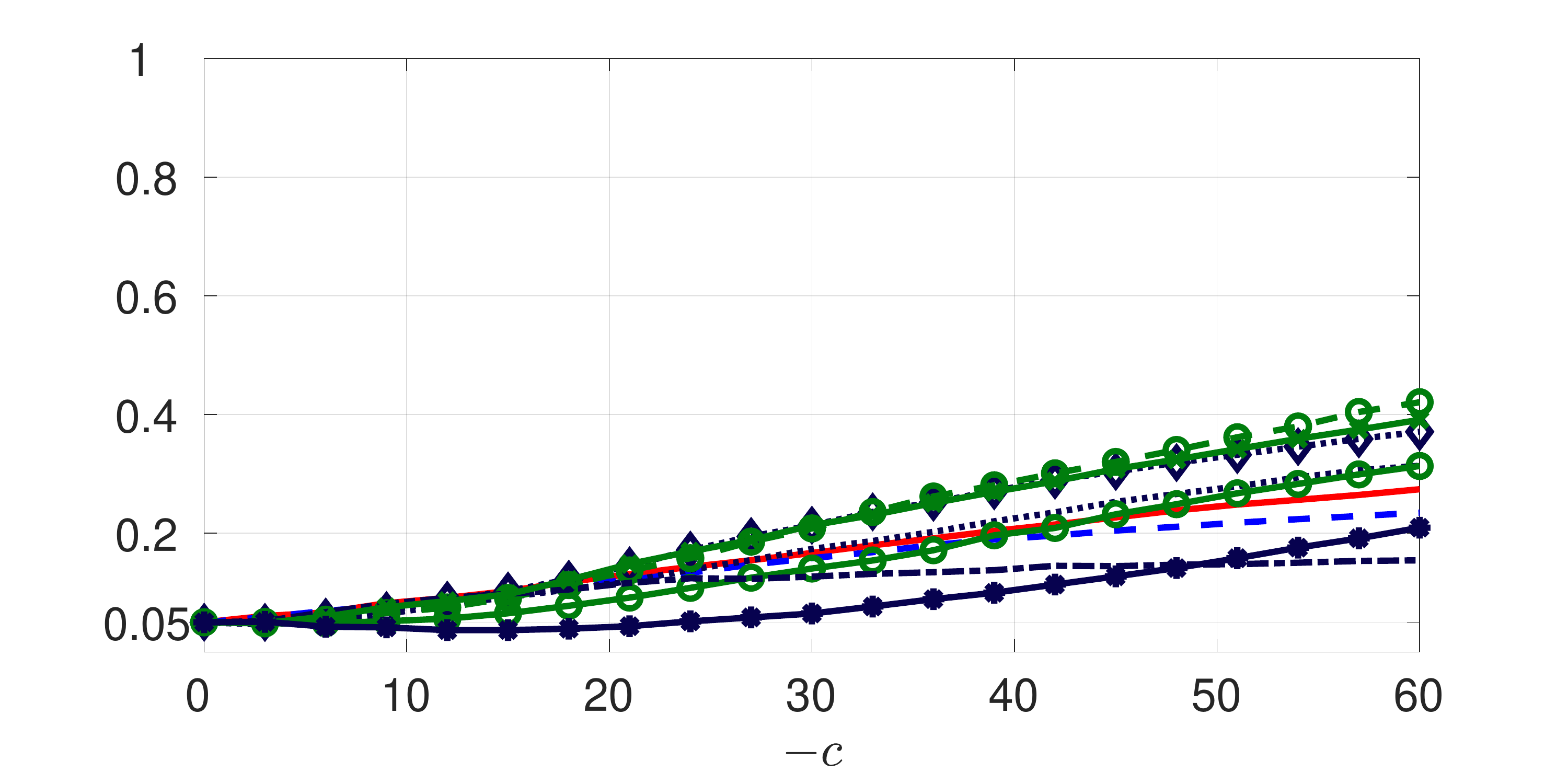}
	\end{subfigure}
	
	\vspace{1ex}
	
	\begin{subfigure}{0.25\textheight}
		\centering
		\caption*{MA, $\theta=0.3$}
		\vspace{-1ex}
		\includegraphics[trim={2cm 0.2cm 2cm 0.5cm},width=0.98\textwidth,clip]{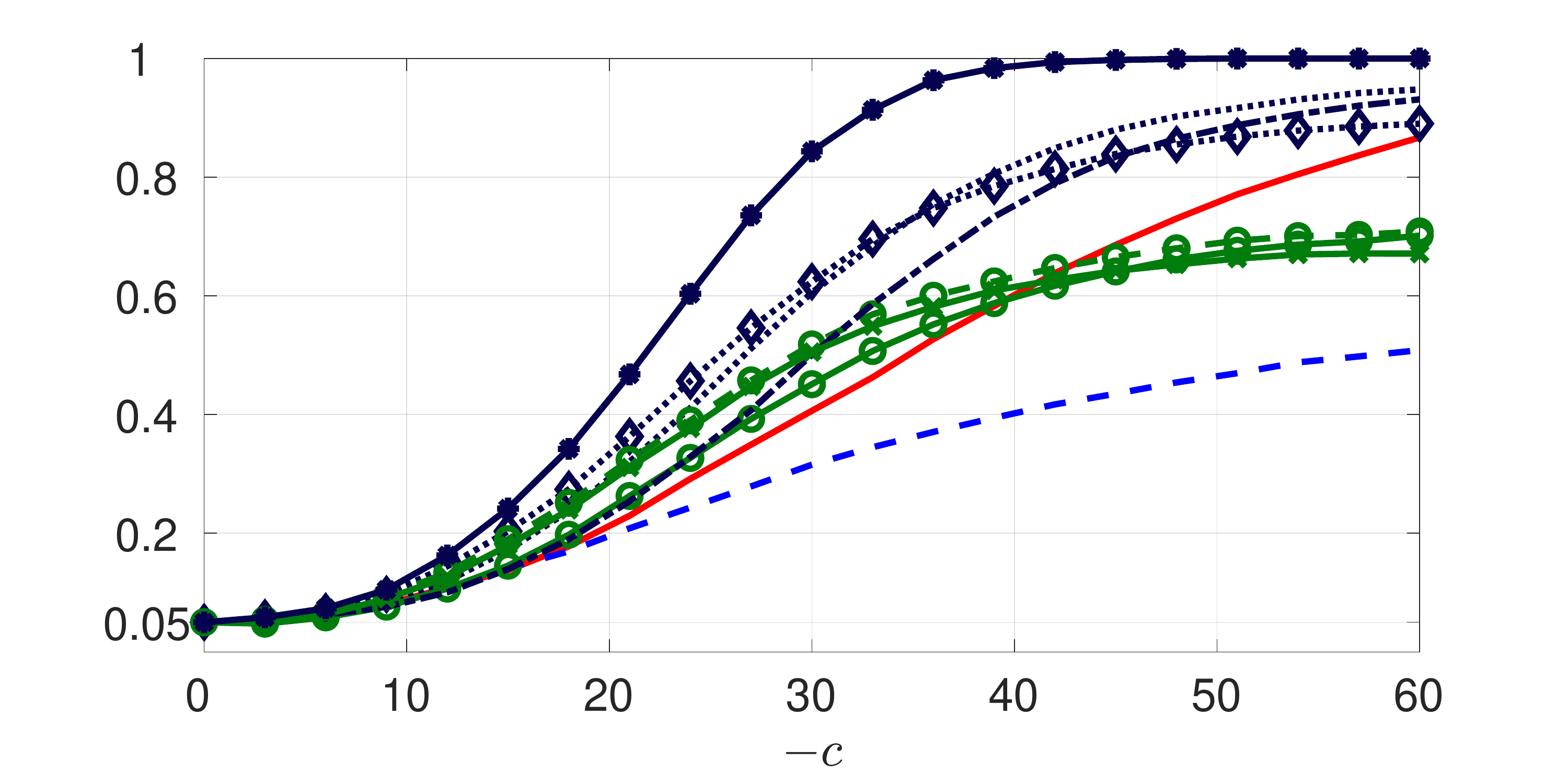}
	\end{subfigure}\begin{subfigure}{0.25\textheight}
		\centering
		\caption*{MA, $\theta=0.6$}
		\vspace{-1ex}
		\includegraphics[trim={2cm 0.2cm 2cm 0.5cm},width=0.98\textwidth,clip]{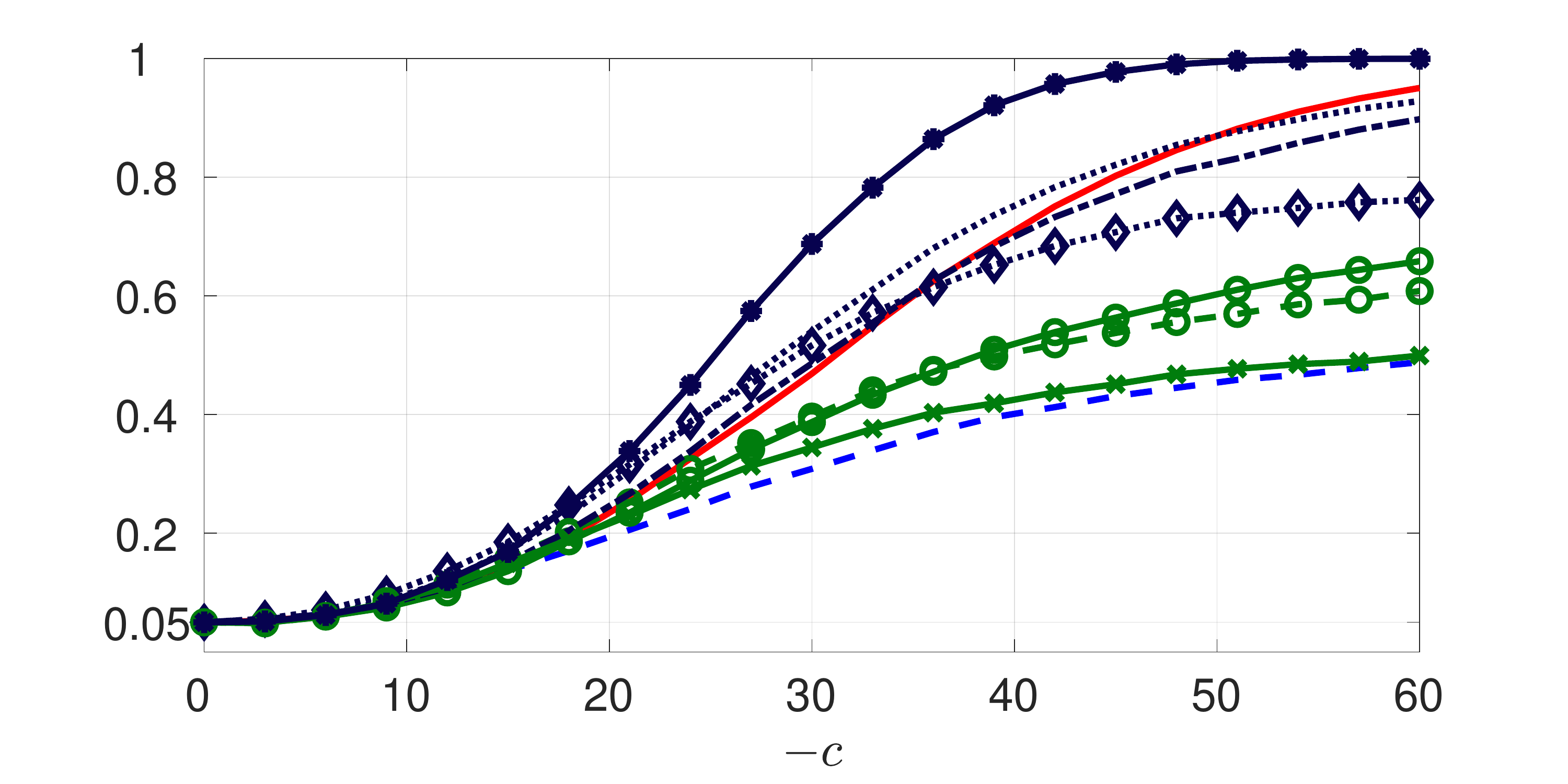}
	\end{subfigure}\begin{subfigure}{0.25\textheight}
		\centering
		\caption*{MA, $\theta=0.9$}
		\vspace{-1ex}
		\includegraphics[trim={2cm 0.2cm 2cm 0.5cm},width=0.98\textwidth,clip]{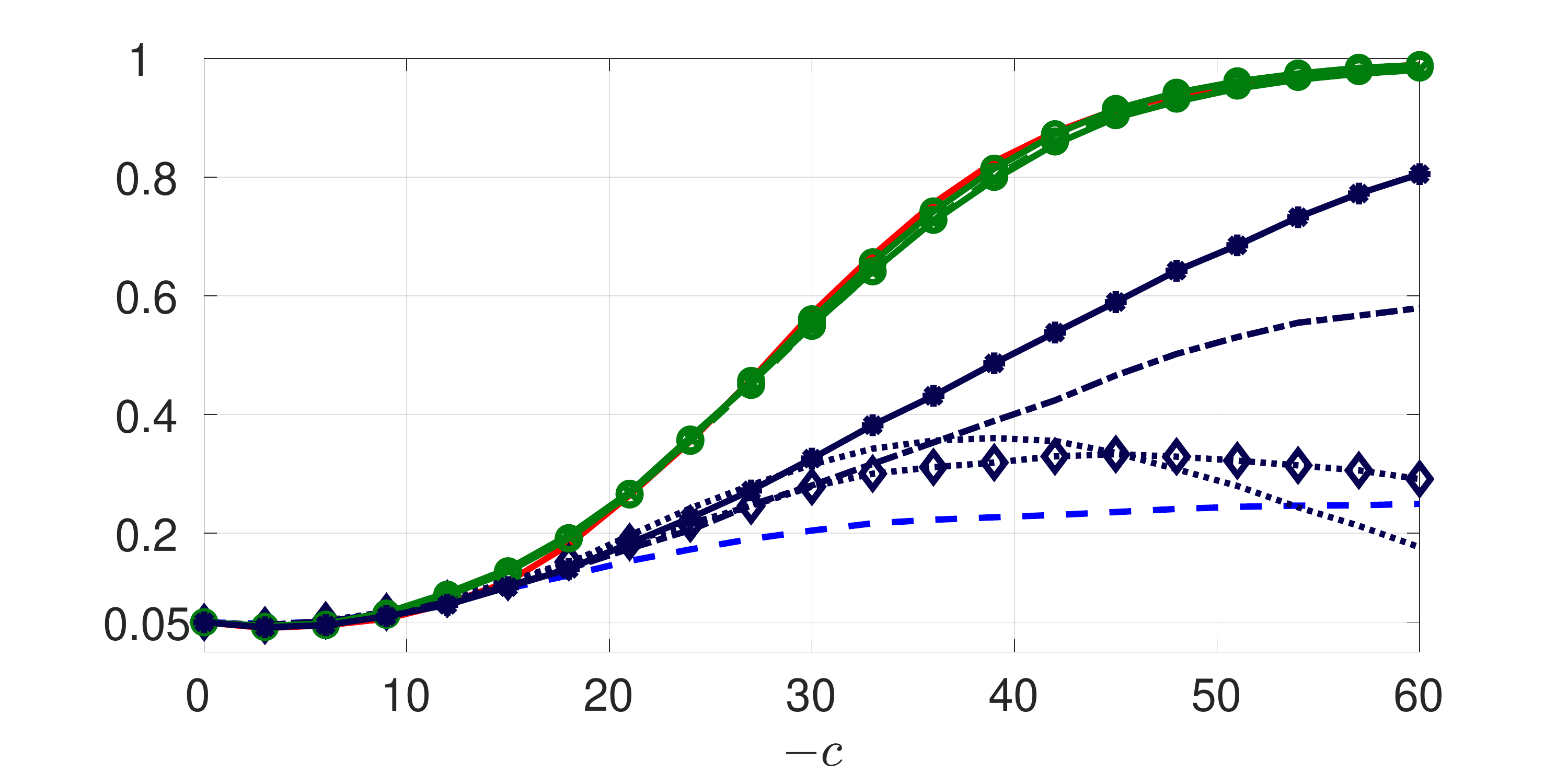}
	\end{subfigure}\begin{subfigure}{0.25\textheight}
		\centering
		\caption*{ARMA, $\phi=0.3,\theta=0.6$}
		\vspace{-1ex}
		\includegraphics[trim={2cm 0.2cm 2cm 0.5cm},width=0.98\textwidth,clip]{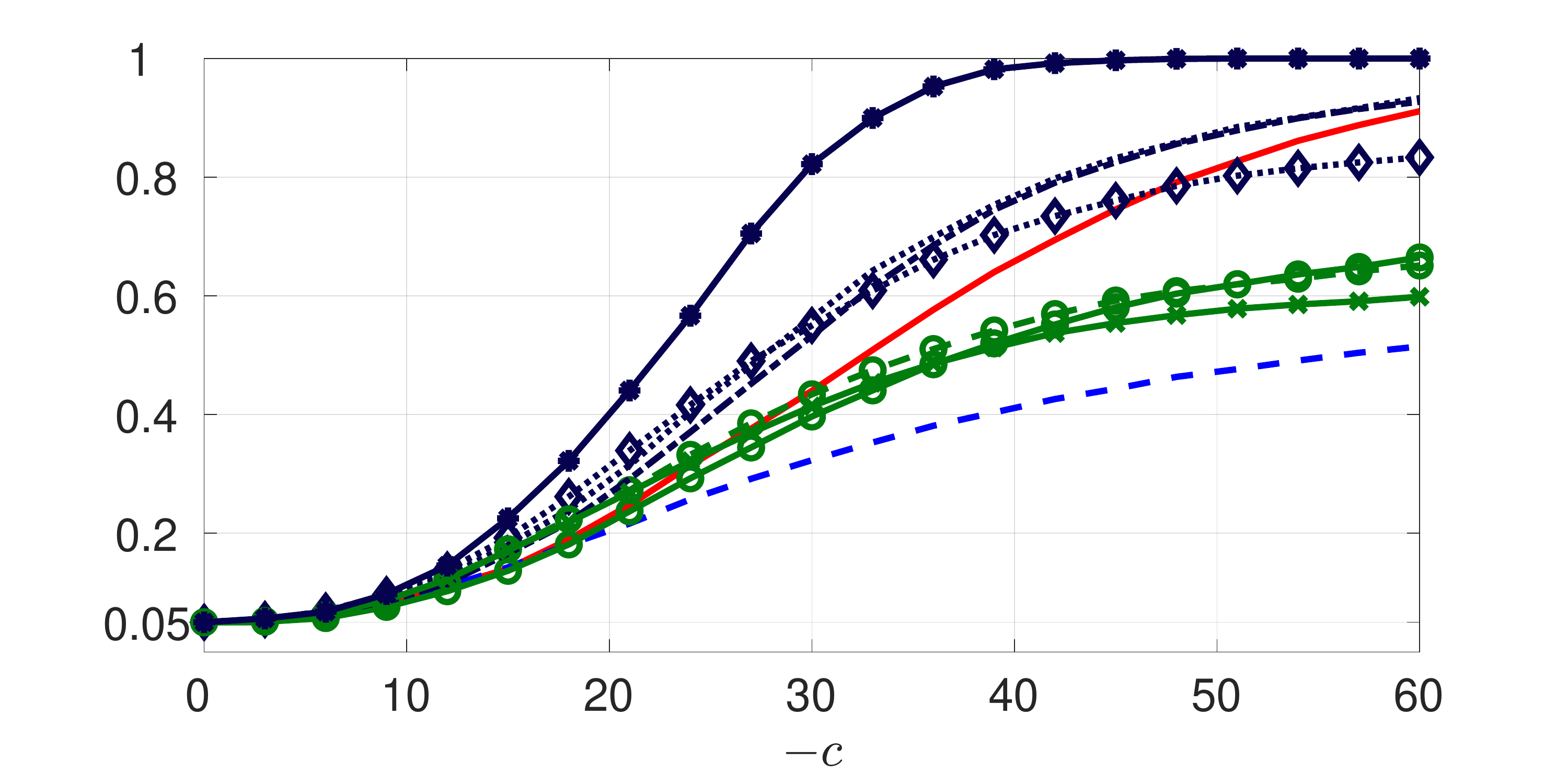}
	\end{subfigure}
	
	\vspace{1ex}
	
	\begin{subfigure}{0.25\textheight}
		\centering
		\caption*{ARMA, $\phi=0.3,\theta=0.3$}
		\vspace{-1ex}
		\includegraphics[trim={2cm 0.2cm 2cm 0.5cm},width=0.98\textwidth,clip]{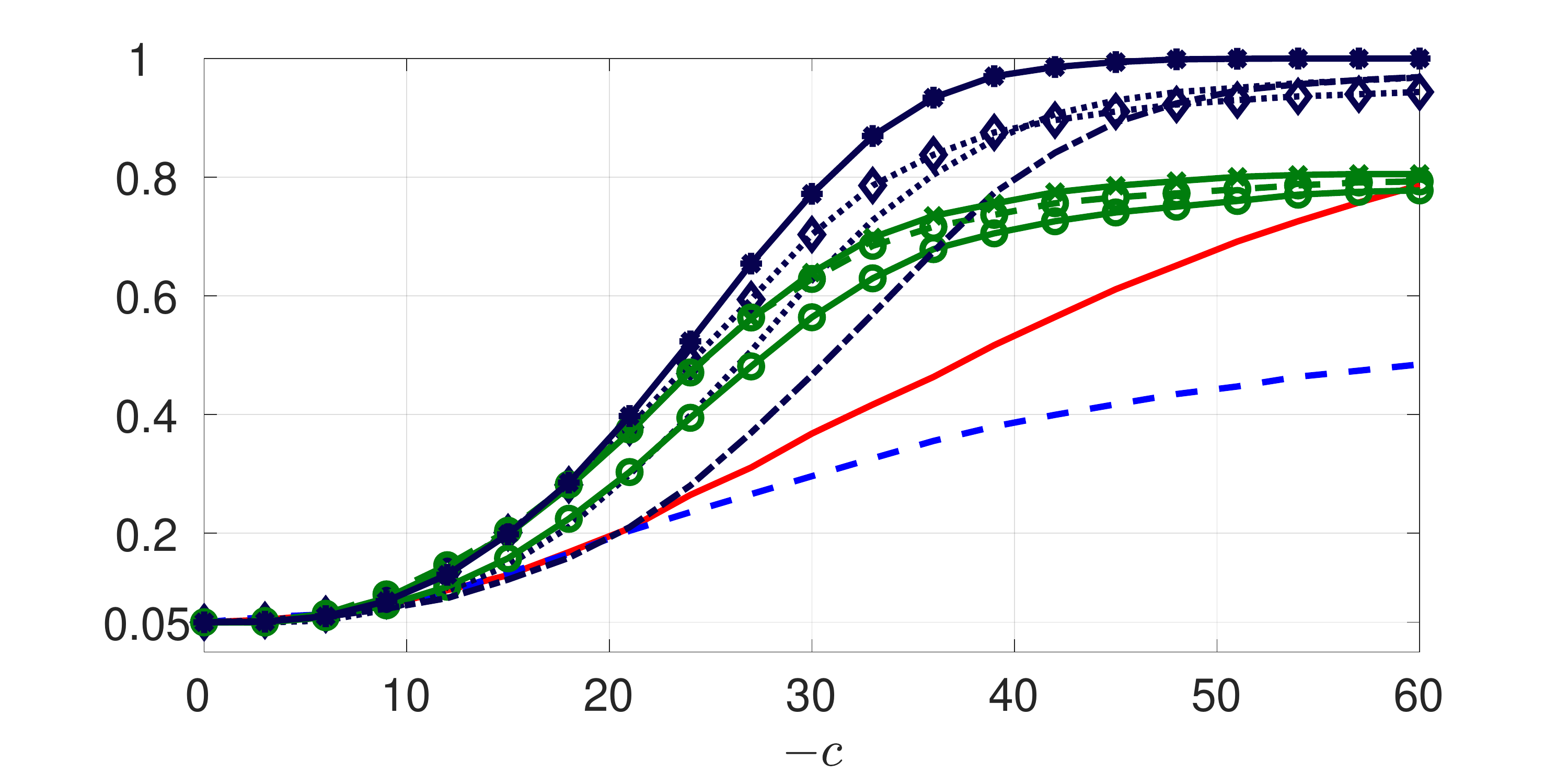}
	\end{subfigure}\begin{subfigure}{0.25\textheight}
		\centering
		\caption*{ARMA, $\phi=0.6,\theta=0.3$}
		\vspace{-1ex}
		\includegraphics[trim={2cm 0.2cm 2cm 0.5cm},width=0.98\textwidth,clip]{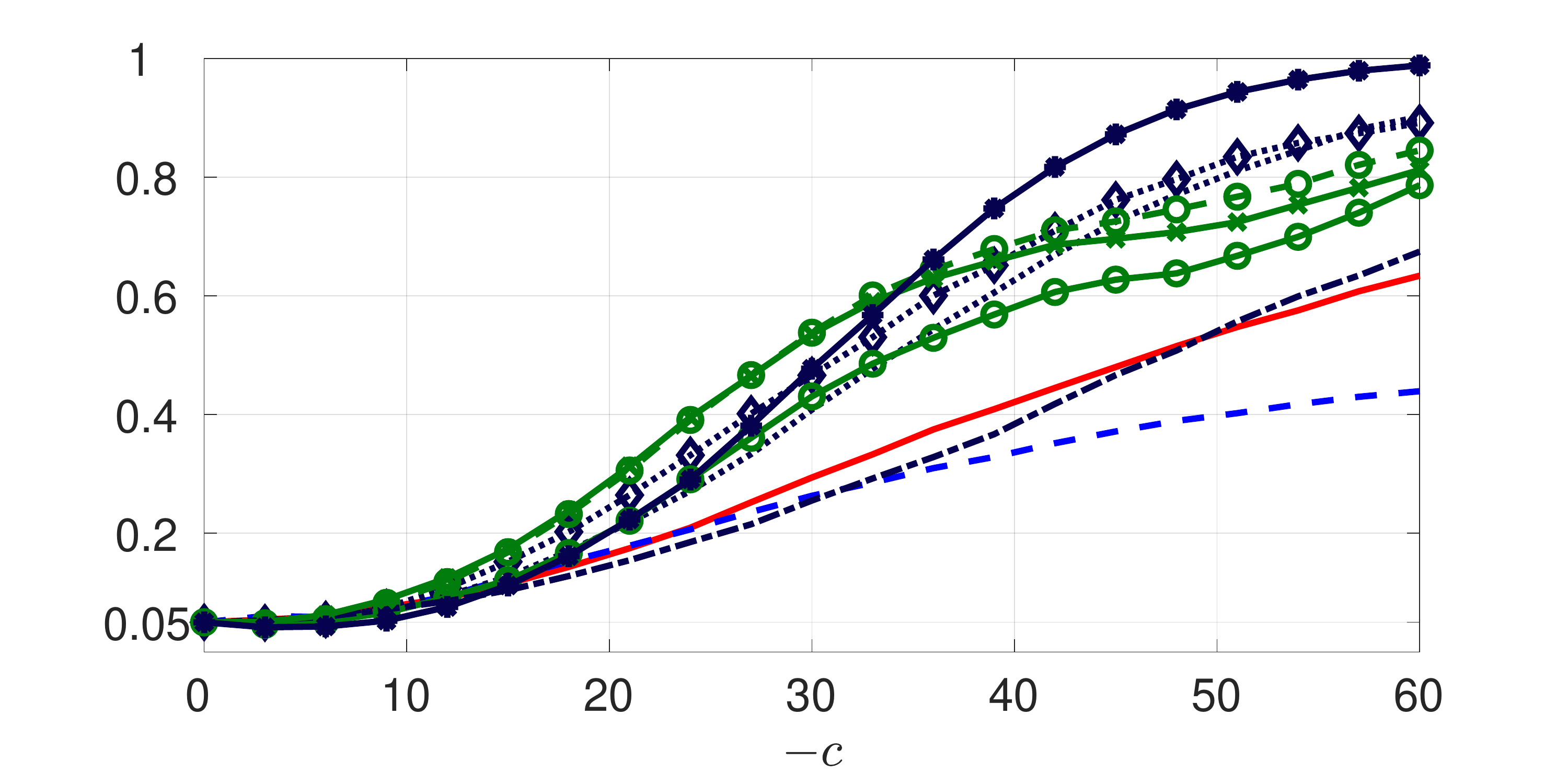}
	\end{subfigure}\begin{subfigure}{0.25\textheight}
		\centering
		\caption*{GARCH, $a_1=0.05,a_2=0.93$}
		\vspace{-1ex}
		\includegraphics[trim={2cm 0.2cm 2cm 0.5cm},width=0.98\textwidth,clip]{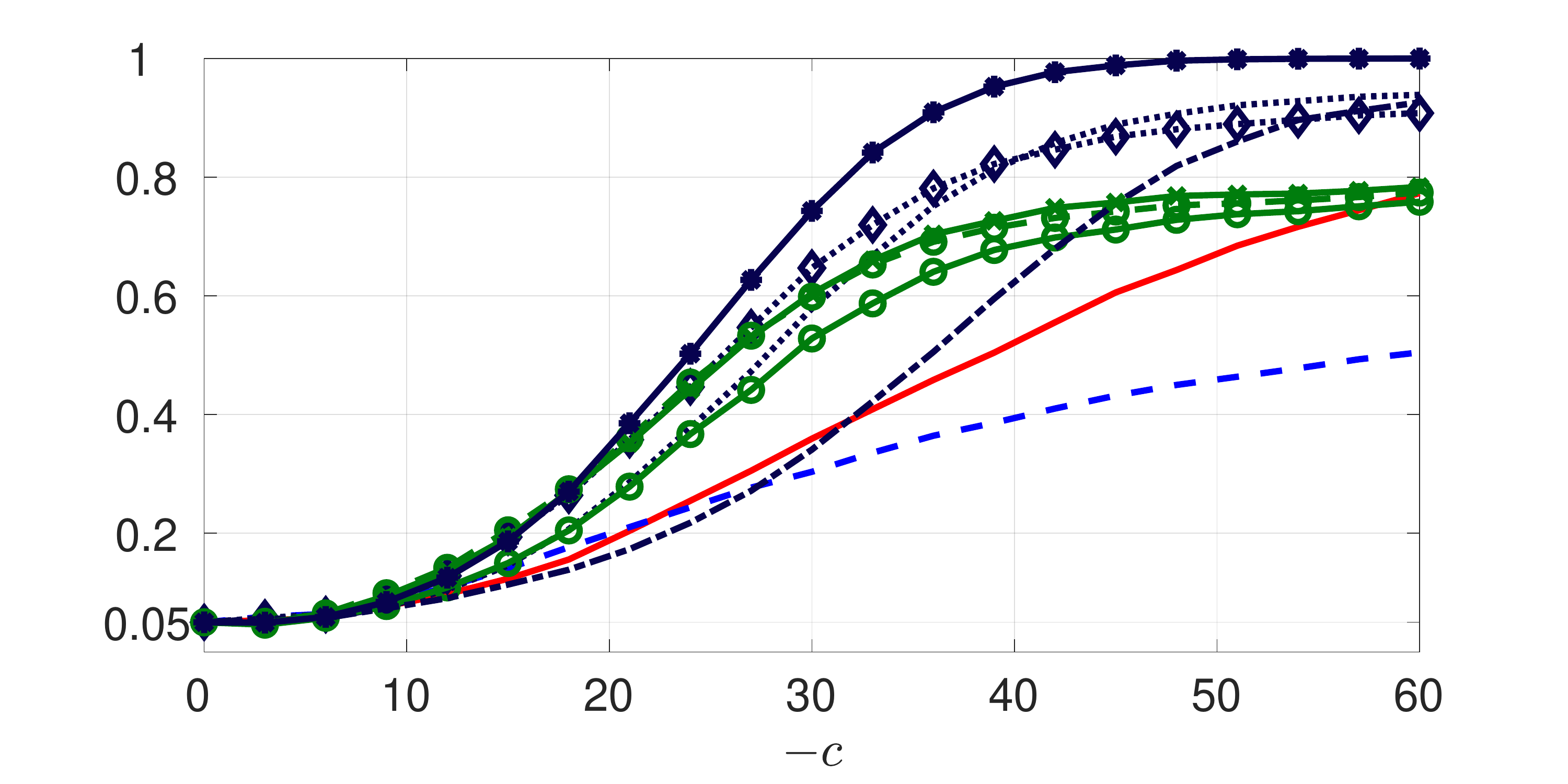}
	\end{subfigure}\begin{subfigure}{0.25\textheight}
		\centering
		\caption*{GARCH, $a_1=0.01,a_2=0.98$}
		\vspace{-1ex}
		\includegraphics[trim={2cm 0.2cm 2cm 0.5cm},width=0.98\textwidth,clip]{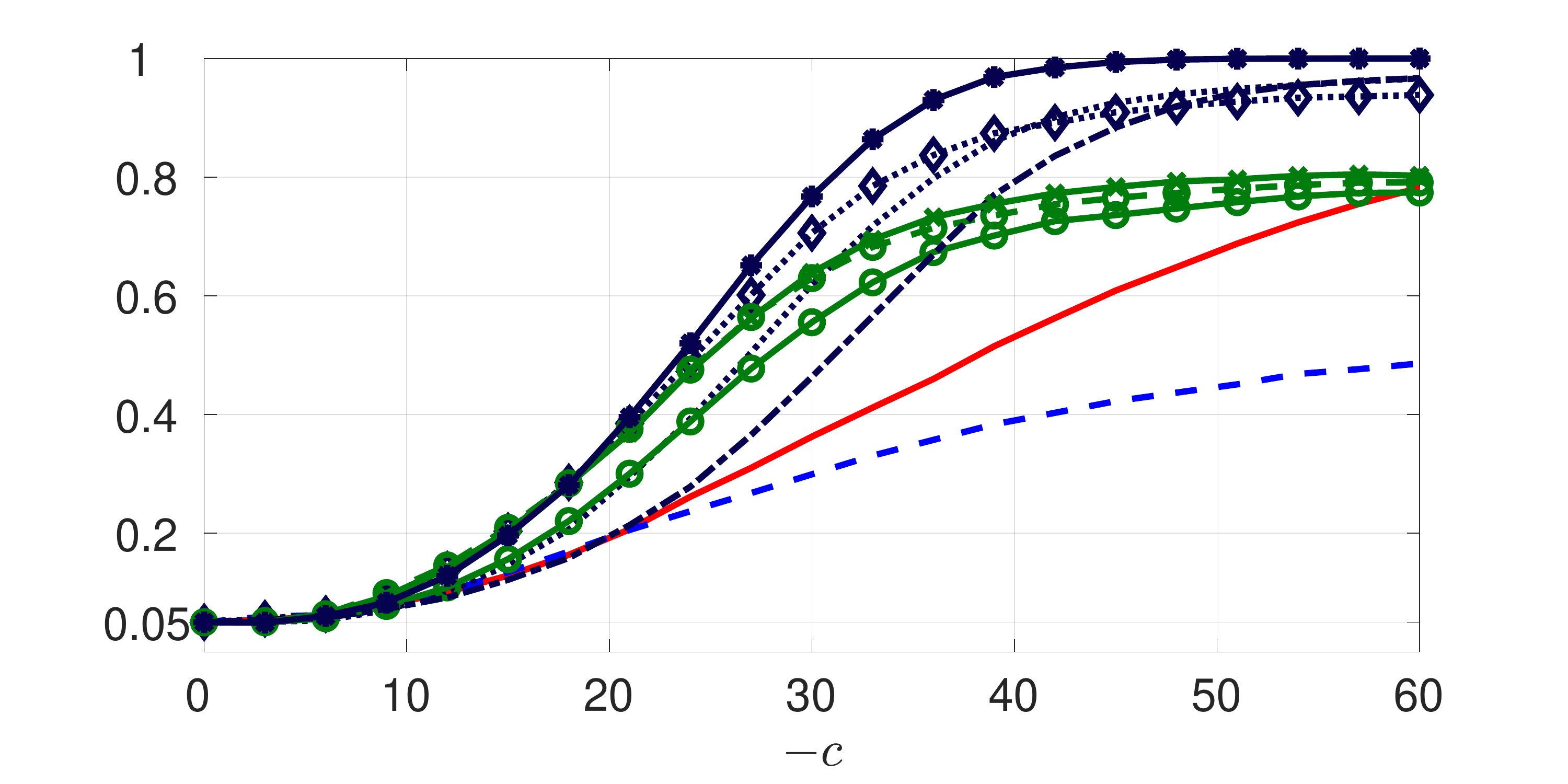}
	\end{subfigure}
	
\end{center}
\vspace{-3ex}
\caption{Size-corrected power of the tests at the nominal $5\%$ level for $\text{$\text{H}_0$:}\ \rho = 1$ under the alternative $\rho = 1+c/T$ in case D2, for $T=250$ and $R^2=0.4$\\{\footnotesize Note: See note to Figure~\ref{fig:size_adjusted_power_m_1_T_100_deter_2_R2_04_AIC}.}}
\label{fig:size_adjusted_power_m_1_T_250_deter_2_R2_04_AIC}
%\end{figure}
\end{sidewaysfigure}
%\end{landscape}

%\begin{landscape}
\begin{sidewaysfigure}[!ht]
%\begin{figure}[!ht]
\begin{center}
	\begin{subfigure}{0.25\textheight}
		\centering
		\caption*{IID}
		\vspace{-1ex}
		\includegraphics[trim={2cm 0.2cm 2cm 0.5cm},width=0.98\textwidth,clip]{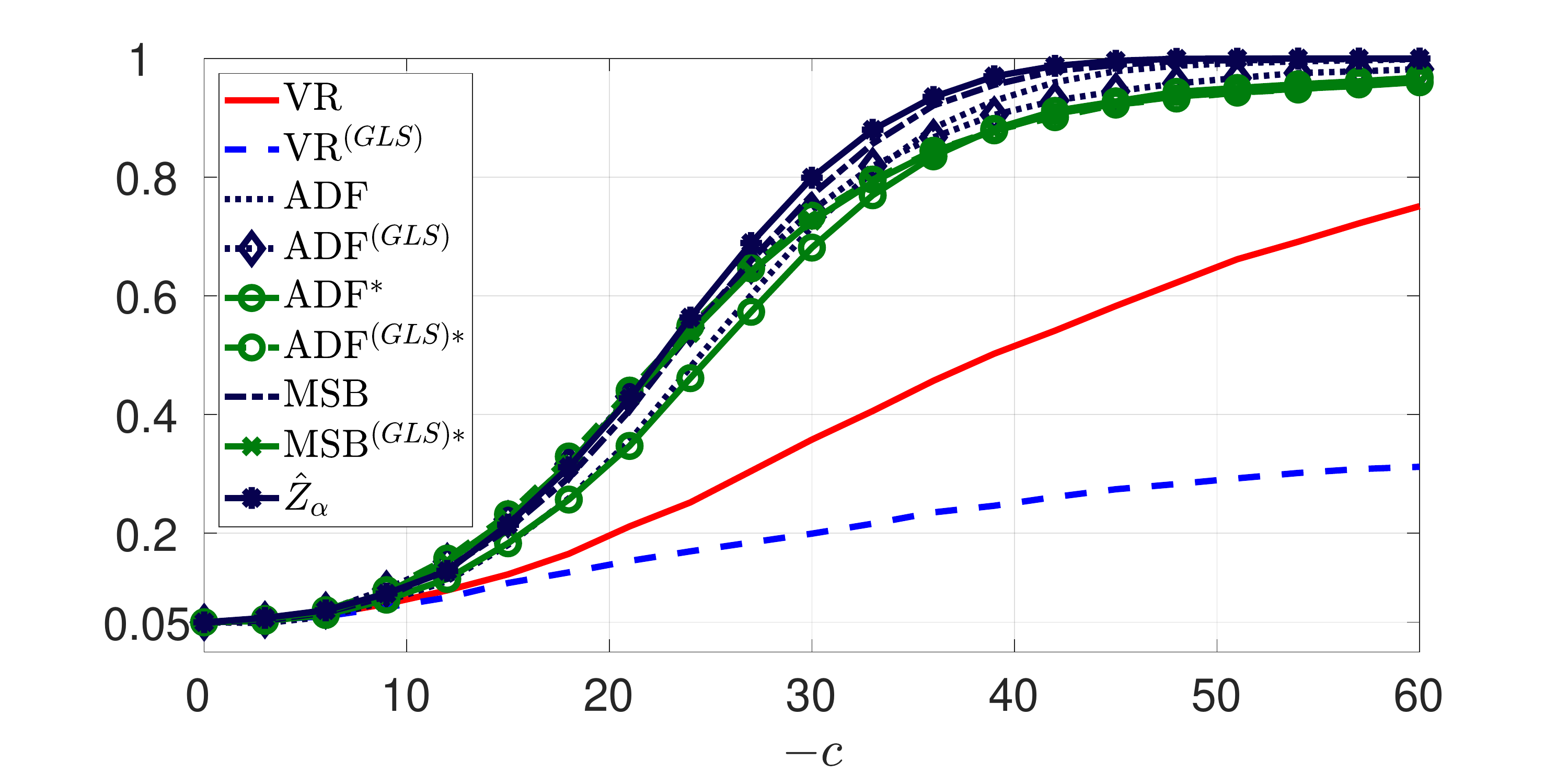}
	\end{subfigure}\begin{subfigure}{0.25\textheight}
		\centering
		\caption*{AR, $\phi=0.3$}
		\vspace{-1ex}
		\includegraphics[trim={2cm 0.2cm 2cm 0.5cm},width=0.98\textwidth,clip]{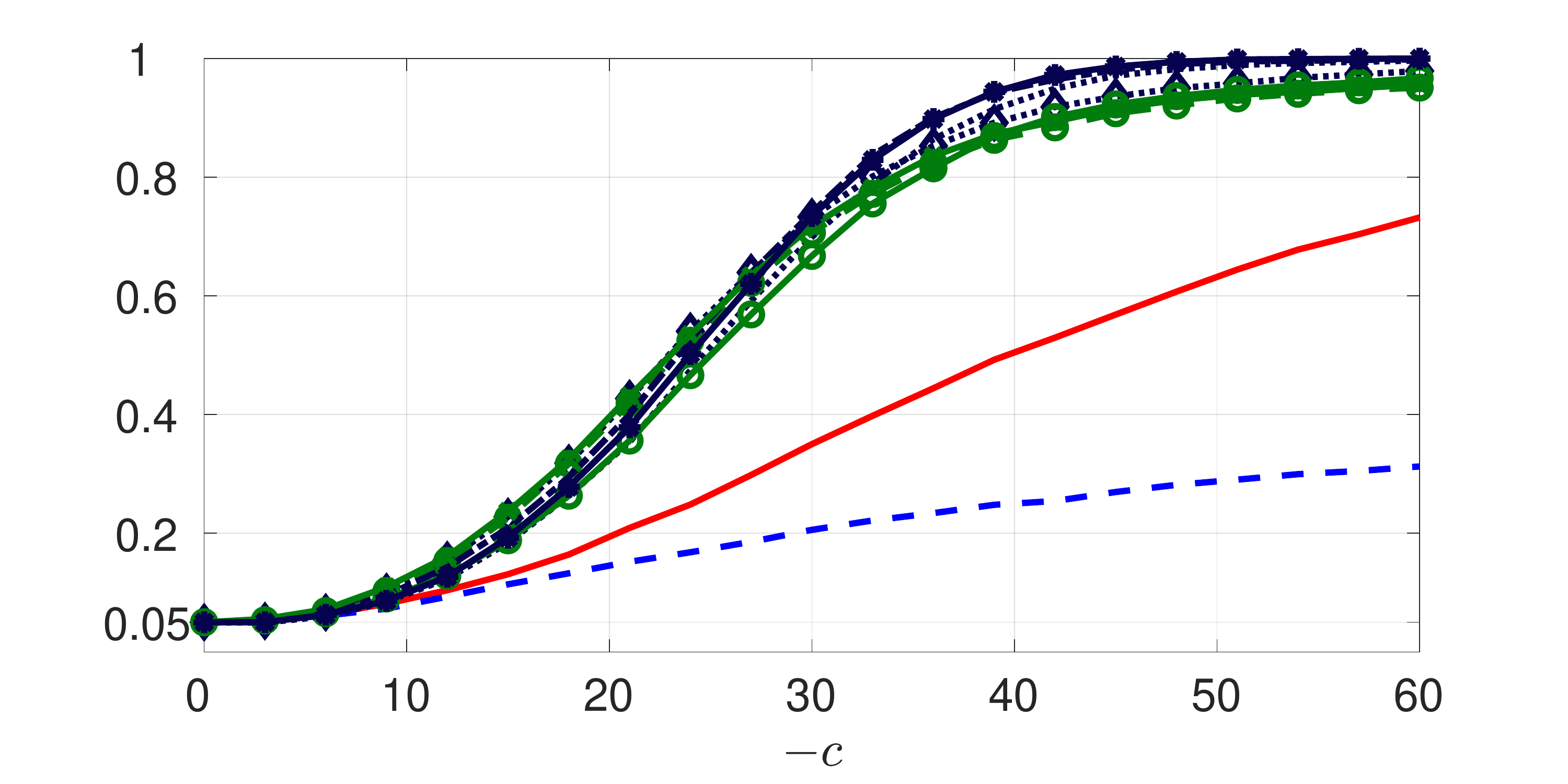}
	\end{subfigure}\begin{subfigure}{0.25\textheight}
		\centering
		\caption*{AR, $\phi=0.6$}
		\vspace{-1ex}
		\includegraphics[trim={2cm 0.2cm 2cm 0.5cm},width=0.98\textwidth,clip]{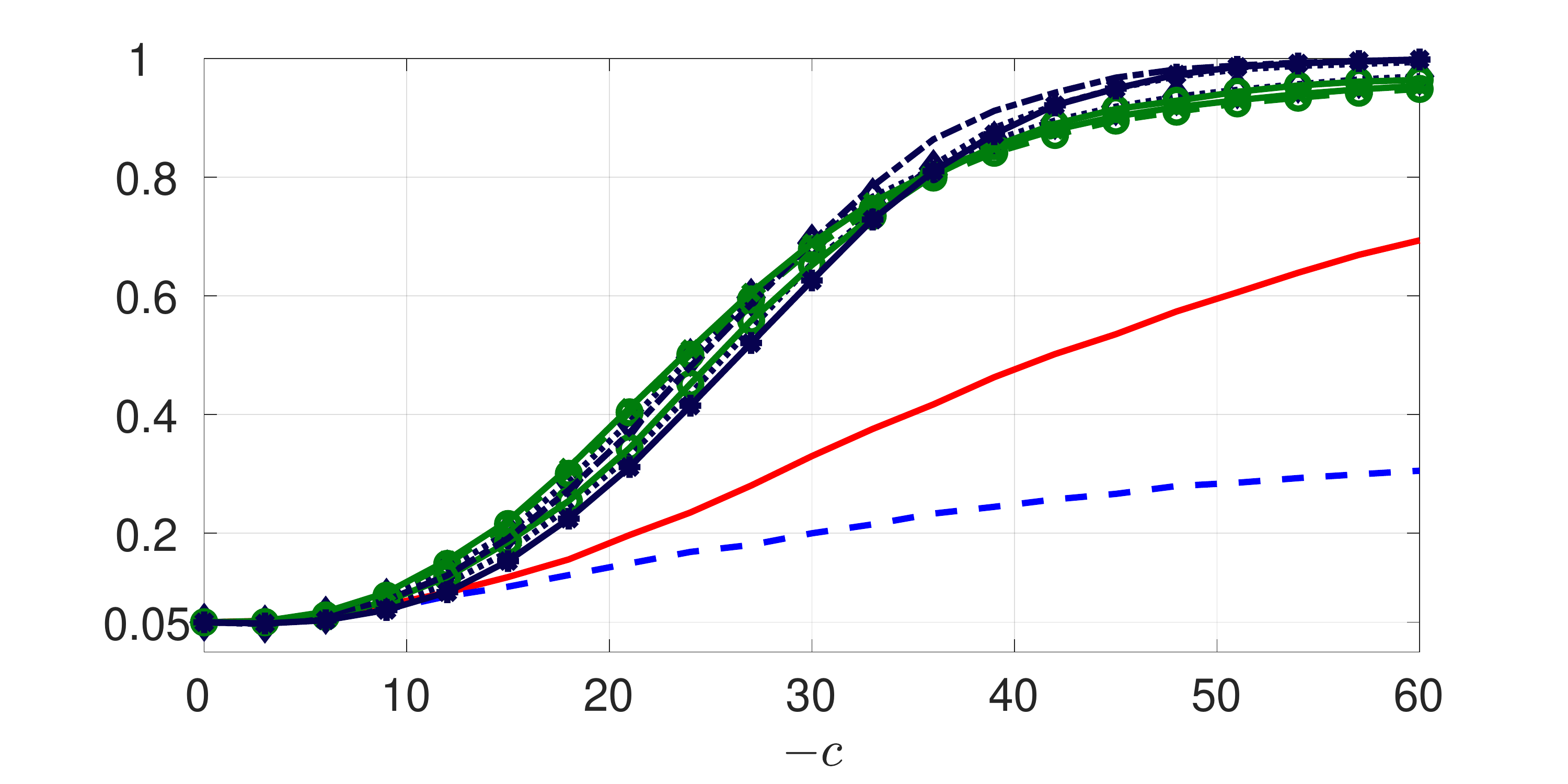}
	\end{subfigure}\begin{subfigure}{0.25\textheight}
		\centering
		\caption*{AR, $\phi=0.9$}
		\vspace{-1ex}
		\includegraphics[trim={2cm 0.2cm 2cm 0.5cm},width=0.98\textwidth,clip]{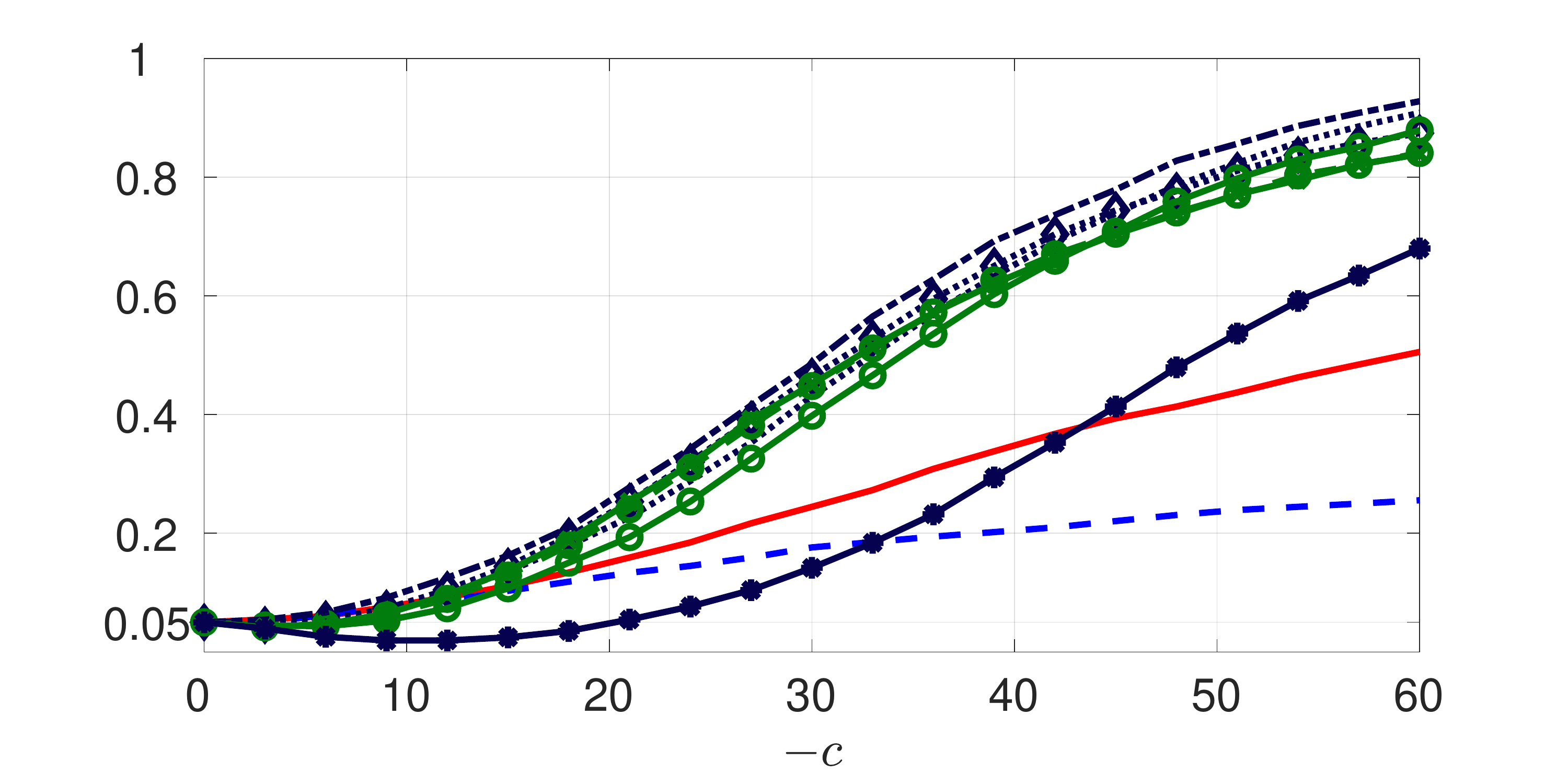}
	\end{subfigure}
	
	\vspace{1ex}
	
	\begin{subfigure}{0.25\textheight}
		\centering
		\caption*{MA, $\theta=0.3$}
		\vspace{-1ex}
		\includegraphics[trim={2cm 0.2cm 2cm 0.5cm},width=0.98\textwidth,clip]{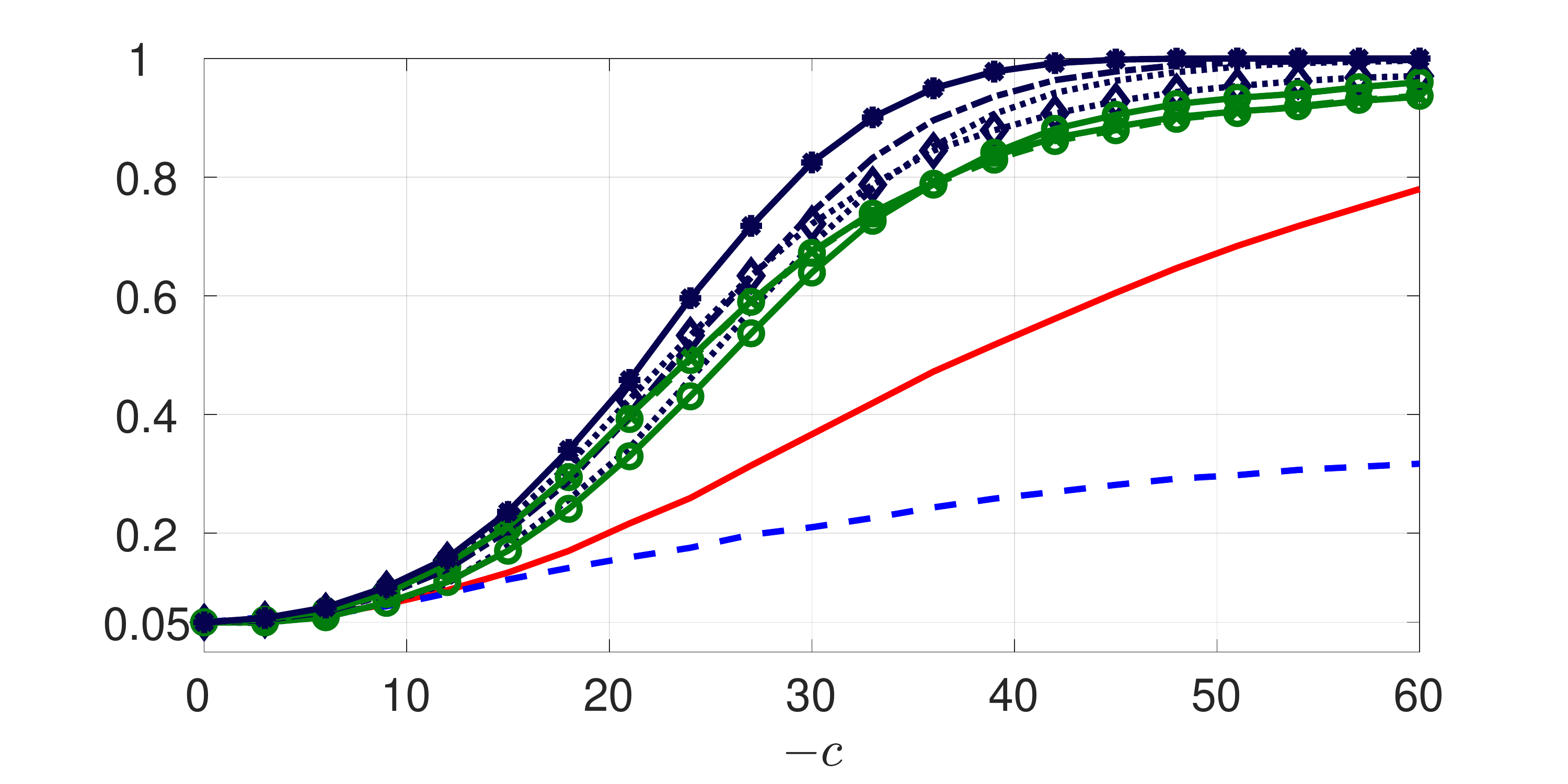}
	\end{subfigure}\begin{subfigure}{0.25\textheight}
		\centering
		\caption*{MA, $\theta=0.6$}
		\vspace{-1ex}
		\includegraphics[trim={2cm 0.2cm 2cm 0.5cm},width=0.98\textwidth,clip]{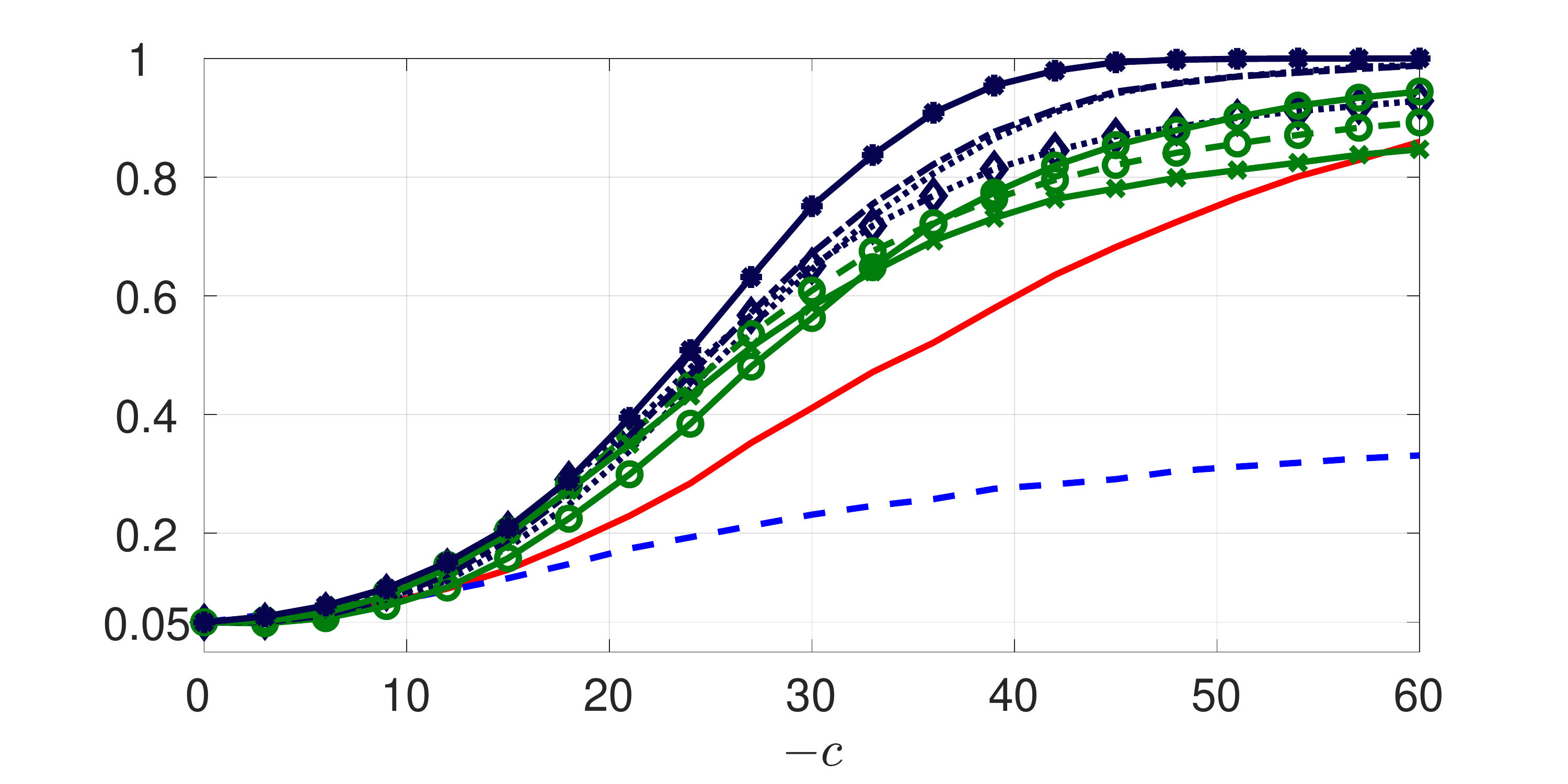}
	\end{subfigure}\begin{subfigure}{0.25\textheight}
		\centering
		\caption*{MA, $\theta=0.9$}
		\vspace{-1ex}
		\includegraphics[trim={2cm 0.2cm 2cm 0.5cm},width=0.98\textwidth,clip]{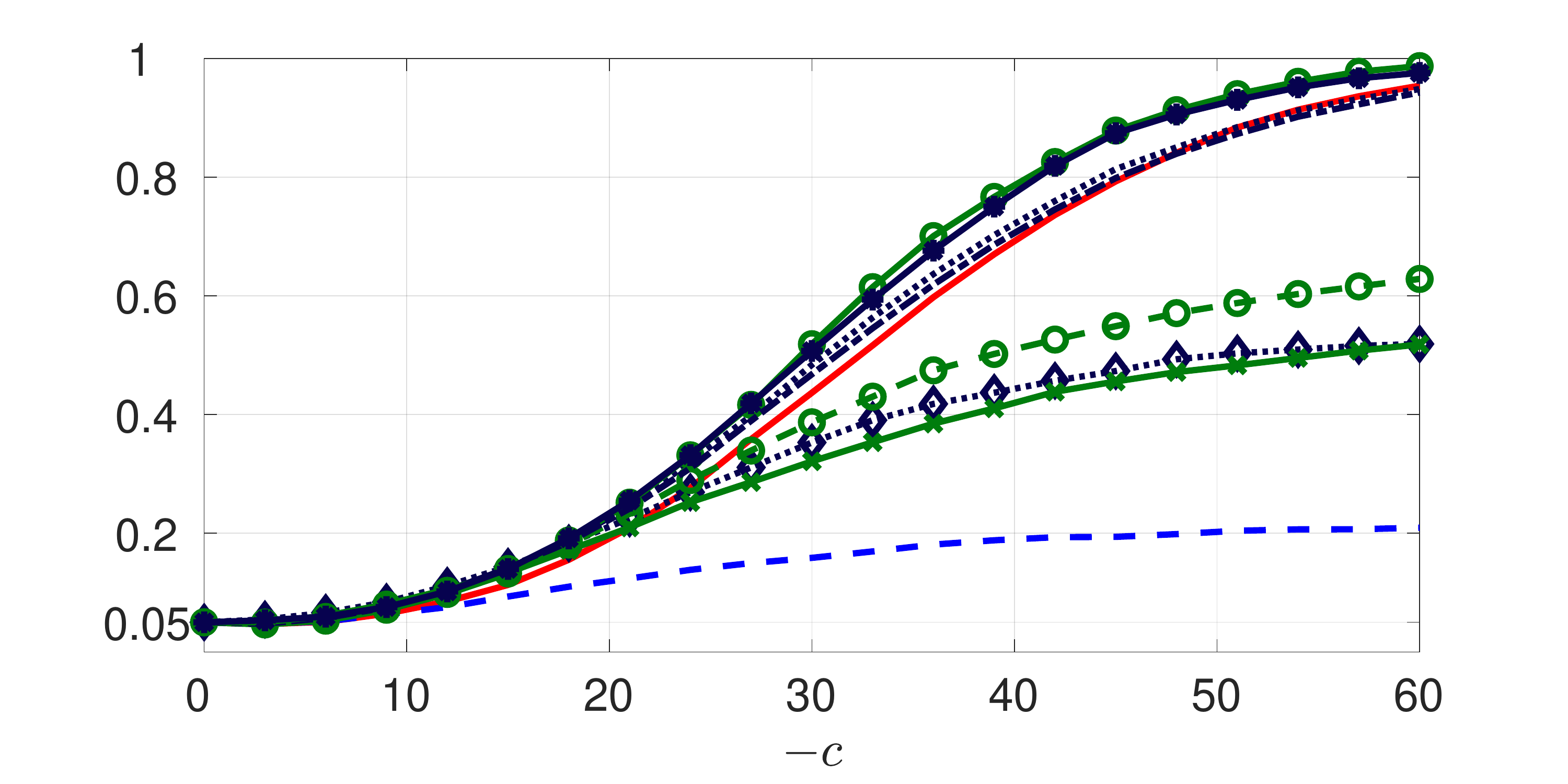}
	\end{subfigure}\begin{subfigure}{0.25\textheight}
		\centering
		\caption*{ARMA, $\phi=0.3,\theta=0.6$}
		\vspace{-1ex}
		\includegraphics[trim={2cm 0.2cm 2cm 0.5cm},width=0.98\textwidth,clip]{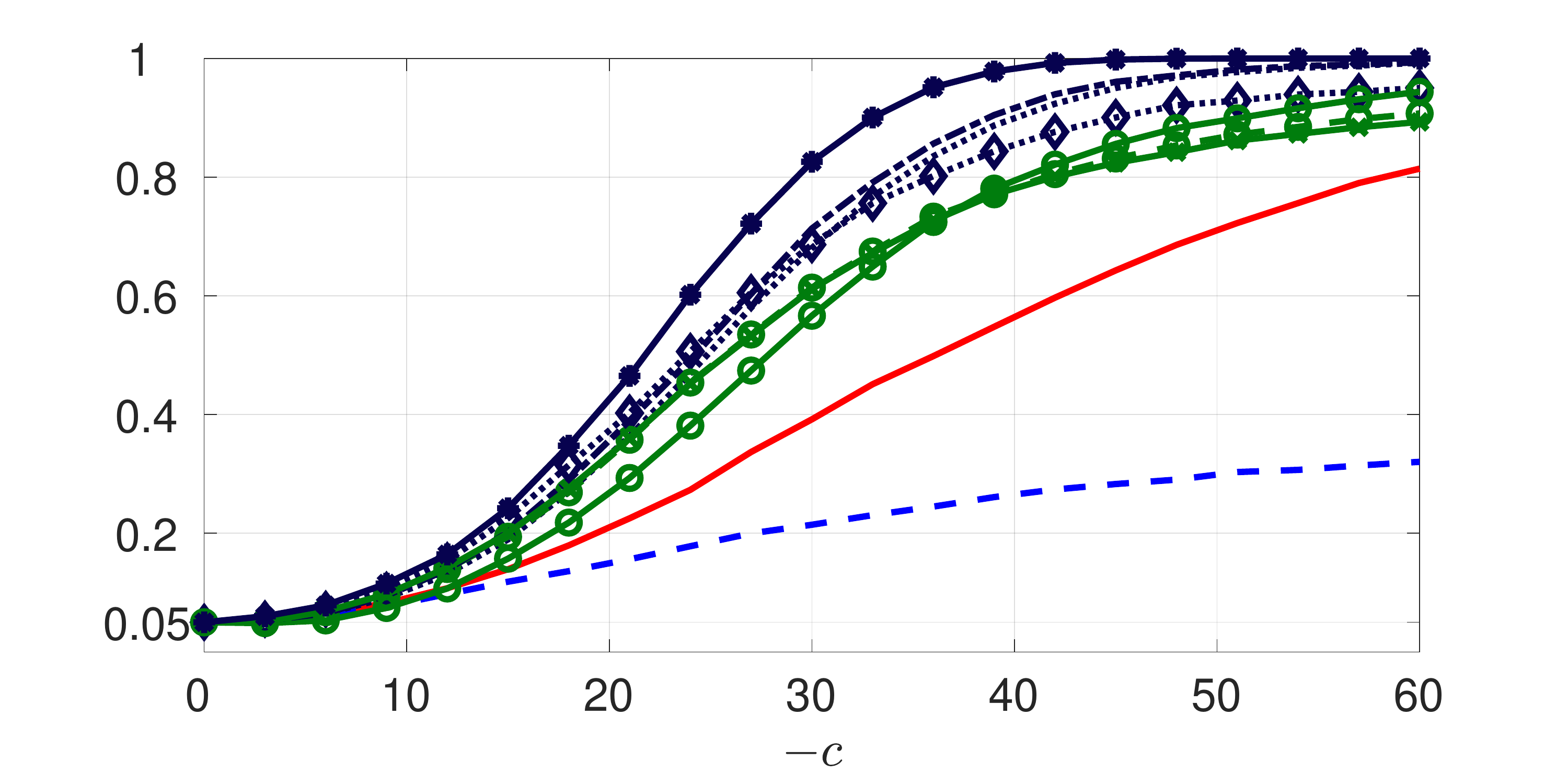}
	\end{subfigure}
	
	\vspace{1ex}
	
	\begin{subfigure}{0.25\textheight}
		\centering
		\caption*{ARMA, $\phi=0.3,\theta=0.3$}
		\vspace{-1ex}
		\includegraphics[trim={2cm 0.2cm 2cm 0.5cm},width=0.98\textwidth,clip]{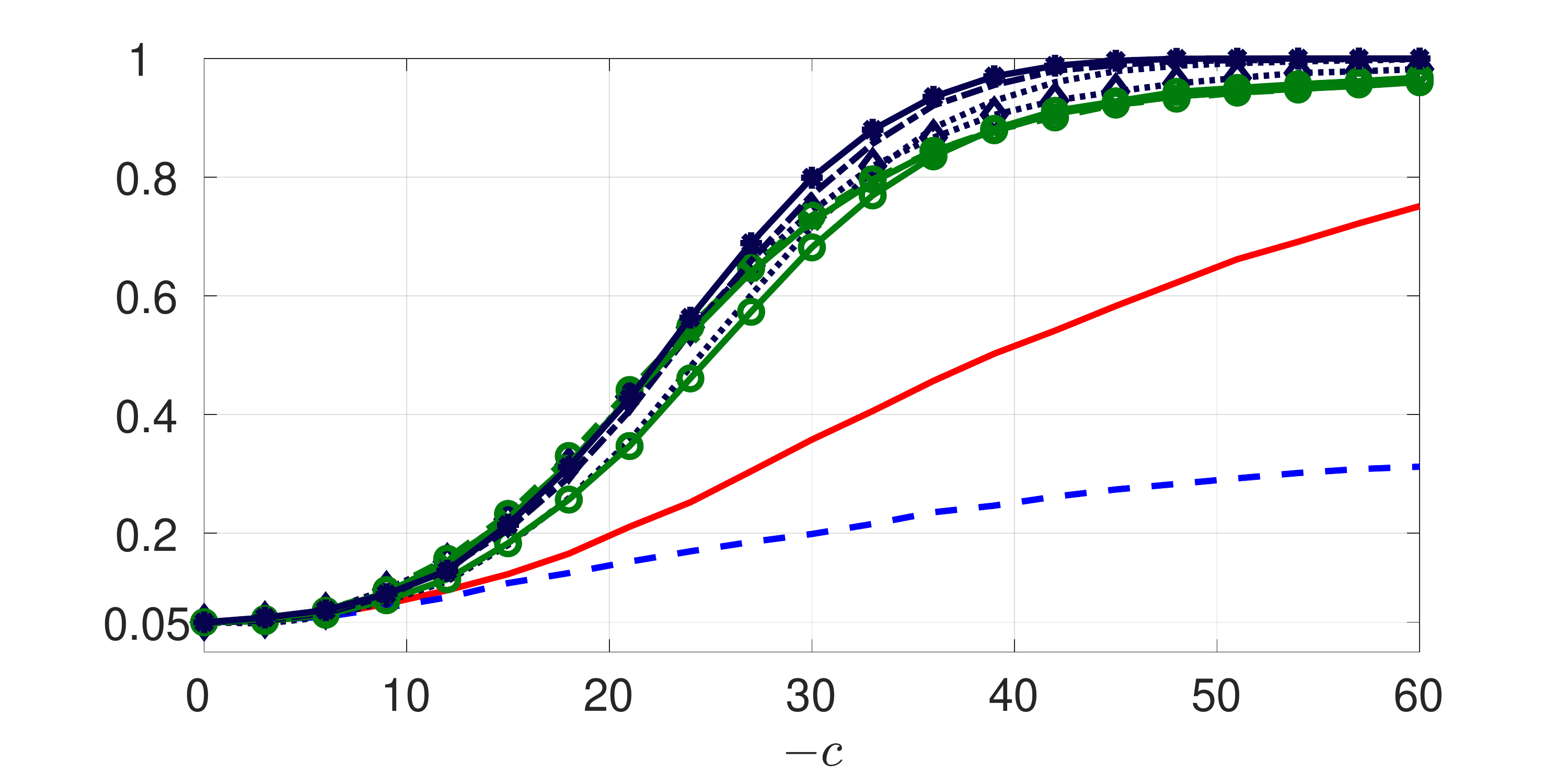}
	\end{subfigure}\begin{subfigure}{0.25\textheight}
		\centering
		\caption*{ARMA, $\phi=0.6,\theta=0.3$}
		\vspace{-1ex}
		\includegraphics[trim={2cm 0.2cm 2cm 0.5cm},width=0.98\textwidth,clip]{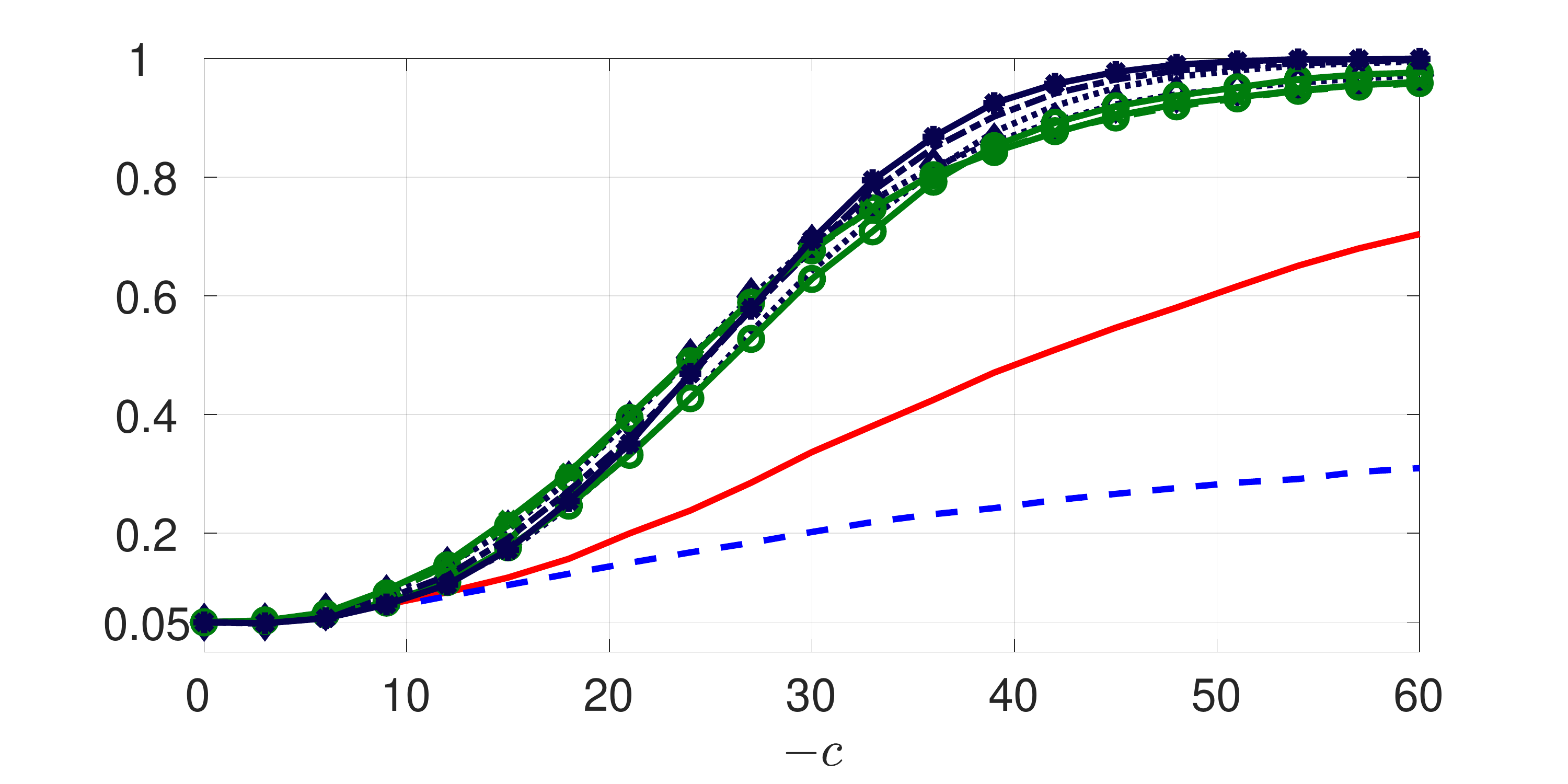}
	\end{subfigure}\begin{subfigure}{0.25\textheight}
		\centering
		\caption*{GARCH, $a_1=0.05,a_2=0.93$}
		\vspace{-1ex}
		\includegraphics[trim={2cm 0.2cm 2cm 0.5cm},width=0.98\textwidth,clip]{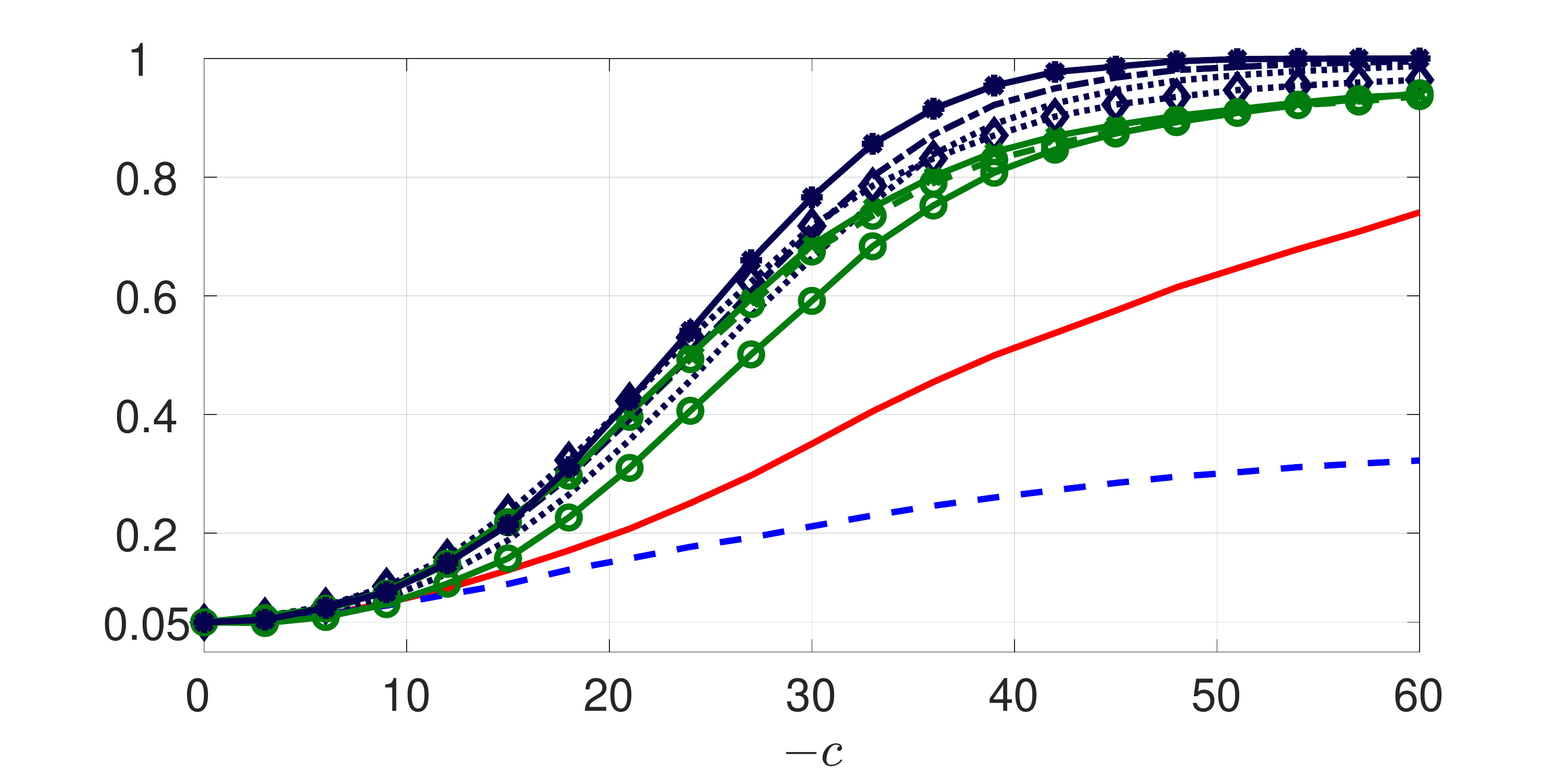}
	\end{subfigure}\begin{subfigure}{0.25\textheight}
		\centering
		\caption*{GARCH, $a_1=0.01,a_2=0.98$}
		\vspace{-1ex}
		\includegraphics[trim={2cm 0.2cm 2cm 0.5cm},width=0.98\textwidth,clip]{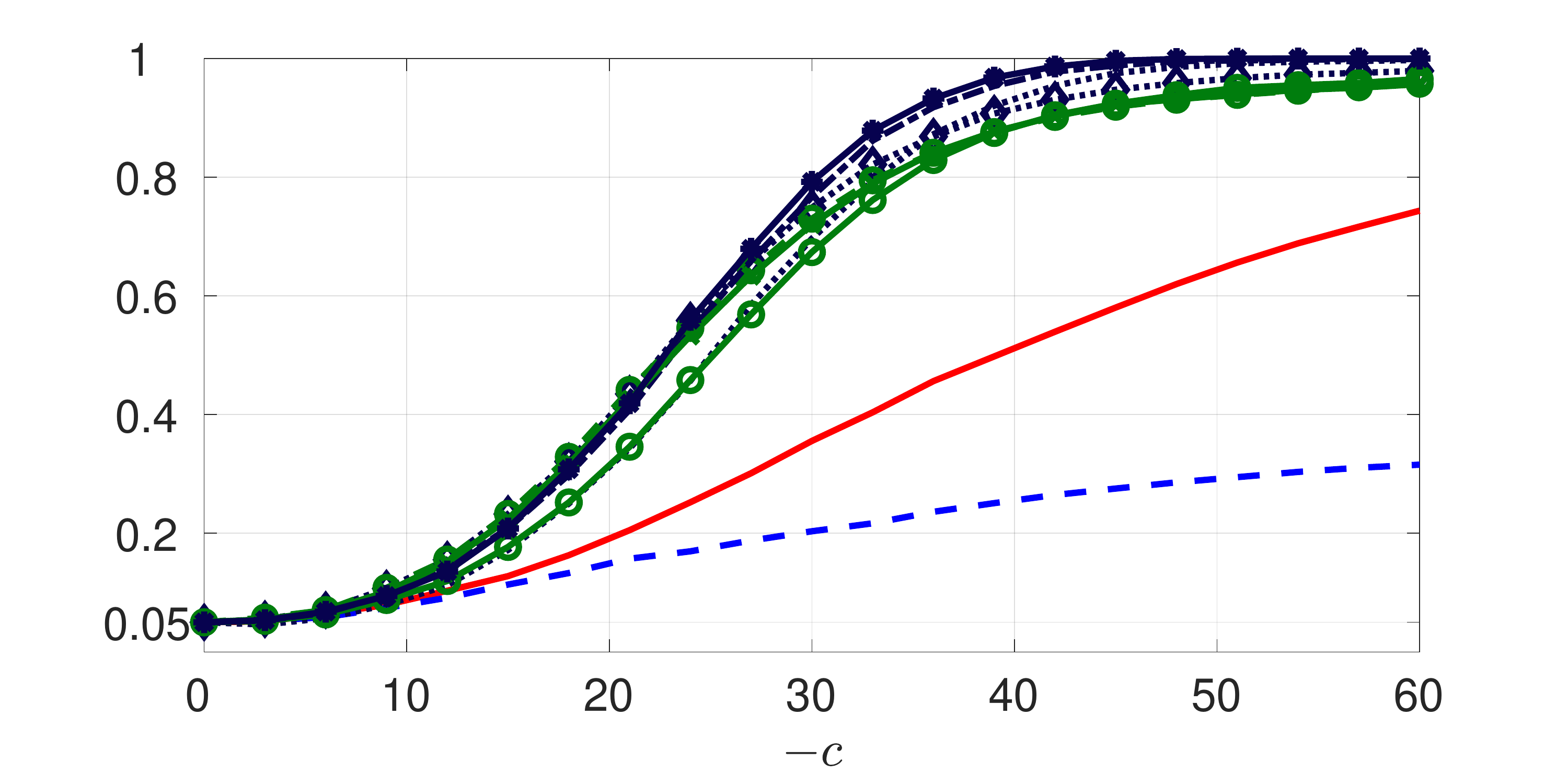}
	\end{subfigure}
	
\end{center}
\vspace{-3ex}
\caption{Size-corrected power of the tests at the nominal $5\%$ level for $\text{$\text{H}_0$:}\ \rho = 1$ under the alternative $\rho = 1+c/T$ in case D2, for $T=1{,}000$ and $R^2=0.4$\\{\footnotesize Note: See note to Figure~\ref{fig:size_adjusted_power_m_1_T_100_deter_2_R2_04_AIC}.}}
\label{fig:size_adjusted_power_m_1_T_1000_deter_2_R2_04_AIC}
%\end{figure}
\end{sidewaysfigure}
%\end{landscape}

\newpage
\clearpage

\section{Proofs}\label{app:mainproofs}
The proofs frequently use the fact that in cases D1 and D2 the deterministic component $d_t$ fulfills 
\begin{align}
	\underset{T\rightarrow \infty}{\text{lim}}T^{1/2} A_D d_{\floor{rT}} = D(r), \quad 0\leq r \leq 1,
\end{align}
with $\int_0^1 D(r) D(r)'dr>0$, where $A_D = T^{-1/2}$ and $D(r)=1$ in case D1 and $A_D = \text{diag}\left(T^{-1/2},T^{-3/2}\right)$ and $D(r)=[1,r]'$ in case D2. The (transposed) potentially multivariate detrended stochastic process $\widetilde P(r)$, introduced in Section~\ref{sec:Th-Detrending}, can thus be written more generally as
\begin{align}
	\widetilde P(r)' = P(r)' - D(r)'\left(\int_0^1 D(s)D(s)'ds\right)^{-1}\int_0^1 D(s)P(s)'ds.
\end{align}

{\bf Proof of Proposition~\ref{prop:VR_H0}.} Under the null hypothesis of no cointegration ($\rho=1$), it holds that $u_t = u_0 + \sum_{s=1}^t \xi_s$. In cases D1 and D2, the regression errors in~\eqref{eq:ytilde} are given by $\widetilde u_t = \sum_{s=1}^t \xi_s - d_t'\left(\sum_{s=1}^T d_s d_s'\right)^{-1}\sum_{s=1}^T d_s \sum_{l=1}^s \xi_l$. Under Assumption~\ref{ass:FCLTw}, we thus obtain
\begin{align}
	T^{-1/2}\widetilde u_{\floor{rT}} &= T^{-1/2}\sum_{t=1}^{\floor{rT}} \xi_t - d_{\floor{rT}}'\left(\sum_{t=1}^T d_t d_t'\right)^{-1}\sum_{t=1}^T d_t T^{-1/2}\sum_{s=1}^t \xi_s\notag\\
	&=T^{-1/2}\sum_{t=1}^{\floor{rT}} \xi_t - \left(T^{1/2}A_Dd_{\floor{rT}}\right)'\left(T^{-1}\sum_{t=1}^T \left(T^{1/2}A_Dd_t\right) \left(T^{1/2}A_Dd_t\right)'\right)^{-1}\notag\\
	&\hspace{3.5cm} \times T^{-1}\sum_{t=1}^T \left(T^{1/2}A_Dd_t\right) T^{-1/2}\sum_{s=1}^t \xi_s\notag\\
	& \overset{w}{\longrightarrow} B_{\xi}(r) - D(r)'\left(\int_0^1 D(s)D(s)'ds\right)^{-1}\int_0^1 D(s)B_{\xi}(s)ds\notag \\
	&= \widetilde B_{\xi}(r).
\end{align}
By construction, it holds that $B_{\xi}(r) = \Omega_{\xi \cdot v}^{1/2} W_{\xi \cdot v}(r) + \Omega_{\xi v}(\Omega_{vv}^{-1/2})'W_v(r)$, which implies $\widetilde B_{\xi}(r) = \Omega_{\xi \cdot v}^{1/2} \widetilde W_{\xi \cdot v}(r) + \Omega_{\xi v}(\Omega_{vv}^{-1/2})'\widetilde W_v(r)$. Analogously, it follows from~\eqref{eq:FCLTw} and the fact that OLS detrending annihilates $x_0$ (and $\mu t$) in $x_t$ that 
\begin{align}
	T^{-1/2}\widetilde x_{\floor{rT}}' &= T^{-1/2}x_{\floor{rT}}' - d_{\floor{rT}}'\left(\sum_{t=1}^T d_t d_t'\right)^{-1}\sum_{t=1}^T d_t T^{-1/2}x_t'\notag\\
	&= T^{-1/2}\left(\sum_{t=1}^{\floor{rT}}v_t\right)' - d_{\floor{rT}}'\left(\sum_{t=1}^T d_t d_t'\right)^{-1}\sum_{t=1}^T d_t T^{-1/2}\left(\sum_{s=1}^t v_s\right)'\notag\\
	& \overset{w}{\longrightarrow} B_v(r)' - D(r)'\left(\int_0^1 D(s)D(s)'ds\right)^{-1}\int_0^1 D(s)B_v(s)'ds\notag \\
	&= \widetilde B_v(r)',
\end{align}
where it follows from $B_v(r) = \Omega_{vv}^{1/2} W_v(r)$ that $\widetilde B_v(r) = \Omega_{vv}^{1/2} \widetilde W_v(r)$. In case D0, it holds that, using the notation $\widetilde P(r) = P(r)$,
\begin{align}
	T^{-1/2}\widetilde u_{\floor{rT}} &= T^{-1/2} u_0 + T^{-1/2} u_{\floor{rT}} = T^{-1/2}\sum_{t=1}^{\floor{rT}} \xi_t + o_\mP(1)\notag \\ 
	&\overset{w}{\longrightarrow} \widetilde B_{\xi}(r) = \Omega_{\xi \cdot v}^{1/2} \widetilde W_{\xi \cdot v}(r) + \Omega_{\xi v}(\Omega_{vv}^{-1/2})'\widetilde W_v(r),
\end{align}
and
\begin{align}
	T^{-1/2}\widetilde x_{\floor{rT}}' &= T^{-1/2}x_{\floor{rT}}' \overset{w}{\longrightarrow} \widetilde B_v(r)' = \Omega_{vv}^{1/2} \widetilde W_v(r)'.
\end{align}
In cases D0, D1, and D2, it thus holds for the OLS residuals $\widehat u_t$ defined in~\eqref{eq:uhat} that
\begin{align}
	T^{-1/2} \widehat u_{\floor{rT}} &= T^{-1/2}\widetilde u_{\floor{rT}} - T^{-1/2}\widetilde x_{\floor{rT}}'\left(T^{-1}\sum_{t=1}^T T^{-1/2}\widetilde x_t T^{-1/2}\widetilde x_t'\right)^{-1}T^{-1}\sum_{t=1}^T T^{-1/2}\widetilde x_t T^{-1/2}\widetilde u_t \notag\\
	& \overset{w}{\longrightarrow} \widetilde B_{\xi}(r) - \widetilde B_v(r)'\left(\int_0^1 \widetilde B_v(s)\widetilde B_v(s)'ds\right)^{-1}\int_0^1 \widetilde B_v(s)\widetilde B_{\xi}(s)ds\notag\\
	&= \Omega_{\xi \cdot v}^{1/2} \widetilde W_{\xi\cdot v}^+(r),
\end{align}
with $\widetilde W_{\xi\cdot v}^+(r)$ as defined in the main text. For the denominator of the variance ratio test statistic it directly follows that 
\begin{align}\label{eq:denom}
	T^{-2}\sum_{t=1}^T \widehat{u}_t^2 = T^{-1}\sum_{t=1}^T (T^{-1/2}\widehat{u}_t)^2 \overset{w}{\longrightarrow} \Omega_{\xi \cdot v} \int_0^1 \left(\widetilde W_{\xi\cdot v}^+(r)\right)^2dr.
\end{align}
Analogously,
\begin{align}
	T^{-3/2}\sum_{t=1}^{\floor{rT}} \widehat u_t = T^{-1}\sum_{t=1}^{\floor{rT}} T^{-1/2}\widehat u_t\overset{w}{\longrightarrow} \Omega_{\xi \cdot v}^{1/2} \int_0^r \widetilde W_{\xi\cdot v}^+(s)ds.
\end{align}
For the numerator of the variance ratio test statistic we thus obtain
\begin{align}\label{eq:num}
	\widehat{\eta}_T = T^{-1} \sum_{t=1}^T \left(T^{-3/2}\sum_{s=1}^t \widehat{u}_s \right)^2 \overset{w}{\longrightarrow} \Omega_{\xi \cdot v} \int_0^1 \left( \int_0^r \widetilde W_{\xi\cdot v}^+(s)ds \right)^2 dr,
\end{align}
Since the vector of numerator and denominator of the variance ratio test statistic can be expressed as a continuous functional of $T^{-1/2} \widehat u_{\floor{rT}}$ up to an error of $o_\mP(1)$, the weak convergence results in~\eqref{eq:denom} and~\eqref{eq:num} hold jointly \citeaffixed[Proof of Lemma~1]{Ph87}{cf., \eg,}. The limiting null distribution of the variance ratio test statistic stated in the proposition thus follows from the continuous mapping theorem and the fact that the scalar long-run variance parameter $\Omega_{\xi \cdot v}>0$ in~\eqref{eq:denom} and~\eqref{eq:num} cancels out. \hfill$\square$\\

{\bf Proof of Proposition~\ref{prop:VR_H1}.} Under the alternative of cointegration ($\vert\rho\vert<1$), it holds that $\widehat u_t = \widetilde u_t - \widetilde x_t'(\widehat \beta - \beta)$, where $T(\widehat \beta - \beta) = O_\mP(1)$, see, \eg, \citeasnoun{PhHa90}. The proof works similarly under all three deterministic specifications. We consider the case D0, where $\widetilde u_t=u_t$ and $\widetilde x_t=x_t$. For notational brevity, define $u_t^o \coloneqq \sum_{j=0}^{t-1}\rho^j \xi_{t-j}$, such that $u_t = \rho^t u_0 + u_t^o$. For the denominator of the variance ratio test statistic it holds that
\begin{align}\label{eq:aux0}
	T^{-1} \sum_{t=1}^T \widehat u_t^2 &= T^{-1} \sum_{t=1}^T \left(\rho^t u_0 + u_t^o -  x_t'(\widehat \beta - \beta) \right)^2 \notag\\
	&= u_0  T^{-1} \sum_{t=1}^T \left(\rho^2\right)^t + 2u_0 T^{-1} \sum_{t=1}^T \rho^t  \left(u_t^o -  x_t'(\widehat \beta - \beta) \right) \notag\\
	&\quad + T^{-1} \sum_{t=1}^T \left(u_t^o - x_t'(\widehat \beta - \beta) \right)^2.
\end{align}
Since $\sum_{t=0}^\infty \left(\rho^2\right)^t$ is a geometric series and $u_0=O_\mP(1)$ it follows that the first term in~\eqref{eq:aux0} is $o_\mP(1)$. Next, note that
\begin{align}
	\big\vert T^{-1} \sum_{t=1}^T \rho^t \left(u_t^o - x_t'(\widehat \beta - \beta) \right)\big\vert \leq T^{-1} \sum_{t=1}^T \big\vert \rho^t u_t^o \big\vert + T^{-1} \sum_{t=1}^T \big\vert \rho^t \left(  x_t'(\widehat \beta - \beta) \right)\big\vert.
\end{align}
It follows from Markov's inequality, stationarity of $u_t^o$ and the fact that $\sum_{t=0}^\infty \vert\rho\vert^t$ is a geometric series that $T^{-1} \sum_{t=1}^T \vert \rho^t u_t^o \vert=o_\mP(1)$. From
\begin{align}
	T^{-1} \sum_{t=1}^T \big\vert \rho^t \left(x_t'(\widehat \beta - \beta) \right)\big\vert \leq T^{-3/2} \sum_{t=1}^T \big\vert \frac{x_t}{\sqrt{T}} \big\vert \big\vert T(\widehat \beta - \beta)\big\vert = O_\mP(T^{-1/2})
\end{align}
it thus follows that also the second term in~\eqref{eq:aux0} is $o_\mP(1)$. Therefore,
\begin{align}
	T^{-1} \sum_{t=1}^T \widehat u_t^2 &= T^{-1} \sum_{t=1}^T \left(u_t^o - x_t'(\widehat \beta - \beta) \right)^2 + o_\mP(1)\notag\\
	&=T^{-1} \sum_{t=1}^T (u_t^o)^2 - 2T^{-1} \left(T^{-1} \sum_{t=1}^T  u_t^o x_t\right)'T\left(\widehat \beta-\beta\right) \notag\\
	&\quad + T^{-1} T\left(\widehat \beta-\beta\right)'\left(T^{-1} \sum_{t=1}^T \frac{x_t}{\sqrt{T}}\frac{x_t'}{\sqrt{T}}\right) T\left(\widehat \beta-\beta\right) + o_\mP(1) \notag\\
	& = T^{-1} \sum_{t=1}^T \left(u_t^o\right)^2 + o_\mP(1),
\end{align}
since $T^{-1} \sum_{t=1}^T \frac{x_t}{\sqrt{T}}\frac{x_t'}{\sqrt{T}}=O_\mP(1)$ by Assumption~\ref{ass:FCLTw} in combination with the continuous mapping theorem and $T^{-1} \sum_{t=1}^T  u_t^o x_t=O_\mP(1)$ \citeaffixed{PhHa90}{see, \eg,}. By Assumption~\ref{ass:FCLTw}, $\{\xi_t\}_{t\in\mZ}$ is strictly stationary and ergodic and $\mE\left(\xi_t^2\right)<\infty$. This implies that $\{u_t^o\}_{t\in\mZ}$ and thus also $\{\left(u_t^o\right)^2\}_{t\in\mZ}$ are strictly stationary and ergodic \cite[Theorem~3.35]{Wh01} and that $\mE\left(\left(u_t^o\right)^2\right)<\infty$. It thus follows from the law of large numbers for strictly stationary and ergodic time series \cite[Theorem~3.34]{Wh01} that 
\begin{align}
	T^{-1} \sum_{t=1}^T \widehat u_t^2 = T^{-1} \sum_{t=1}^T \left(u_t^o\right)^2 + o_\mP(1)\overset{p}{\longrightarrow} \mE(\left(u_1^o\right)^2),
\end{align}
where $0<\mE(\left(u_1^o\right)^2)<\infty$.

Turning to the numerator of the variance ratio test statistic, we note that
\begin{align}\label{eq:aux1}
	T^{-1/2} \sum_{t=1}^{\floor{rT}} \widehat u_t = T^{-1/2} \sum_{t=1}^{\floor{rT}} u_t^o - \left(T^{-1} \sum_{t=1}^{\floor{rT}} \frac{x_t}{\sqrt{T}}\right)'T\left(\widehat \beta-\beta\right) + o_\mP(1),
\end{align}
since $\sum_{t=0}^\infty \rho^t$ is a geometric series. The Beveridge-Nelson decomposition \cite{PhSo92} yields $T^{-1/2} \sum_{t=1}^{\floor{rT}} u_t^o \overset{w}{\longrightarrow} (1-\rho)^{-1} B_\xi(r)$. The second term in~\eqref{eq:aux1} is also $O_\mP(1)$, with a limit that is different from $(1-\rho)^{-1} B_\xi(r)$. Hence, $T^{-1/2} \sum_{t=1}^{\floor{rT}} \widehat u_t=O_{\mP}(1)$, which implies that $T^{-2} \sum_{t=1}^T \left(\sum_{s=1}^t \widehat{u}_s \right)^2 = O_\mP(1)$. In total, we thus have
\begin{align}
	\text{VR}= T^{-1} \frac{T^{-2} \sum_{t=1}^T \left(\sum_{s=1}^t \widehat{u}_s \right)^2}{T^{-1} \sum_{t=1}^T \widehat{u}_t^2} = O_\mP(T^{-1}),
\end{align}
as stated in the proposition. \hfill$\square$\\

{\bf Proof of Proposition~\ref{prop:VR_Local}.} Under the local alternative $\rho=\rho_T = 1+c/T$, with $c\leq0$, it holds that $T^{-1/2}u_{\floor{rT}} \overset{w}{\longrightarrow} \Omega_{\xi \cdot v}^{1/2}J_{\xi\cdot v}^c(r)$, $0\leq r \leq 1$, with $J_{\xi\cdot v}^c(r)$ as defined in the main text \citeaffixed[Lemma~5.1]{PeRo16}{cf., \eg,}. In cases D1 and D2, it follows that $T^{-1/2}\widetilde u_{\floor{rT}} \overset{w}{\longrightarrow} \Omega_{\xi \cdot v}^{1/2} \widetilde J_{\xi \cdot v}^c(r)$, $0\leq r \leq 1$, where 
\begin{align}
	\widetilde J_{\xi \cdot v}^c(r) = J_{\xi \cdot v}^c(r) - D(r)'\left(\int_0^1 D(s)D(s)'ds\right)^{-1}\int_0^1 D(s)J_{\xi \cdot v}^c(s)ds.
\end{align}
Analogously, in case D0, it holds that, using the notation $\widetilde P(r) = P(r)$, $T^{-1/2}\widetilde u_{\floor{rT}} \overset{w}{\longrightarrow} \Omega_{\xi \cdot v}^{1/2}\widetilde J_{\xi\cdot v}^c(r)=\Omega_{\xi \cdot v}^{1/2}J_{\xi\cdot v}^c(r)$, $0\leq r \leq 1$. The rest of the proof is similar to the proof of Proposition~\ref{prop:VR_H0} and therefore omitted. \hfill $\square$\\

{\bf Proof of Remark~\ref{rem:GLSLAP}.} \citeasnoun[Lemma~5.3]{PeRo16} show that $\widetilde u_t^{(\GLS)}$, as defined in Remark~\ref{rem:GLSDetrending}, fulfills $T^{-1/2} \widetilde u_{\floor{rT}}^{(\GLS)} \overset{w}{\longrightarrow}\Omega_{\xi \cdot v}^{1/2}J_{\xi\cdot v}^{c,\GLS}(r)$, $0\leq r\leq 1$, where $J_{\xi\cdot v}^{c,\GLS}(r)$ is given by $J_{\xi\cdot v}^c(r)$ in case D1 and by \linebreak $J_{\xi\cdot v}^c(r) - \left(\lambda J_{\xi\cdot v}^c(1) + 3(1-\lambda)\int_0^1 s J_{\xi\cdot v}^c(s)ds\right)r$ in case D2, with $\lambda$ as defined in the main text. The rest of the proof uses similar arguments as the proof of Theorem~5.2 in \citeasnoun{PeRo16} and the proof of Proposition~\ref{prop:VR_H0} and is therefore omitted. \hfill $\square$

\section{Computation of the ADF, MSB and $\widehat Z_{\alpha}$ Tests}\label{app:Conv-Tests}

\subsection{Test Statistics}\label{app:Conv-Tests-Def}
Given the regression residuals $\widehat u_t$, $t=1,\ldots,T$, as defined in~\eqref{eq:uhat}, the ADF test statistic, the MSB test statistic, and the $\widehat Z_{\alpha}$ test statistic are defined as follows:\footnote{In case of GLS detrending, the test statistics are defined by replacing $\widehat u_t$ with $\widehat u_t^{(\GLS)}$.}

\begin{itemize}
	\item The ADF statistic is defined as the usual $t$-test statistic for testing $b_0=0$ in the auxiliary regression
	\begin{align}\label{eq:ADFaux}
		\Delta \widehat u_t = b_0 \widehat u_{t-1} + \sum_{j=1}^p \pi_j \Delta \widehat u_{t-j} + r_{tp},
	\end{align}
	$t=p+2,\ldots,T$. The lag parameter $p$ is determined by means of information criteria, compare the discussion in Section~\ref{app:Conv-Tests-IC}.
	
	\item Let $\widehat \pi_j$ denote the estimates of $\pi_j$ obtained by estimating~\eqref{eq:ADFaux} with OLS and let $\widehat r_{tp}$ denote the corresponding residuals. The MSB statistic is then defined as 
	\begin{align}
		\text{MSB} \coloneqq \left(\frac{T^{-2}\sum_{t=1}^T \widehat{u}_t^2}{\widehat s^2}\right)^{1/2},
	\end{align}
	where $\widehat s^2 \coloneqq \widehat s_{rp}^2/(1-\widehat\pi(1))^2$, with $\widehat s_{rp}^2 \coloneqq T^{-1} \sum_{t=p+2}^T \widehat r_{tp}^2$ and $\widehat \pi(1)\coloneqq \sum_{j=1}^p \widehat \pi_j$.
	
	\item To define the $\widehat Z_\alpha$ statistic, consider the auxiliary regression
	\begin{align}
		\widehat u_t = \alpha \widehat u_{t-1} + k_t
	\end{align}
	$t=2,\ldots,T$. Let $\widehat \alpha$ and $\widehat k_t$ denote the OLS estimate of $\alpha$ and the corresponding OLS residuals, respectively. Define $s_k^2 \coloneqq (T-1)^{-1} \sum_{t=2}^T \widehat k_t^2$ and
	\begin{align}
		s_{Tb}^2 \coloneqq s_k^2 + 2(T-1)^{-1}\sum_{h=1}^{b_T} \mathcal{K}\left(\frac{h}{b_T}\right)\sum_{t=h+2}^T \widehat k_t \widehat k_{t-h},
	\end{align}
	where the kernel function $\mathcal{K}\left(\cdot\right)$ and the bandwidth parameter $b_T$ fulfill some technical assumptions, see, \eg, \citeasnoun{An91}, \citeasnoun{NeWe94} and \citeasnoun{Ja02} for details. The $\widehat Z_\alpha$ statistic is then defined as
	\begin{align}
		\widehat Z_\alpha \coloneqq (T-1)(\widehat \alpha-1)- \frac{1}{2}\left(s_{Tb}^2-s_k^2\right)\left((T-1)^{-2}\sum_{t=2}^T \widehat u_{t-1}^2\right)^{-1}.
	\end{align}
\end{itemize}

The three tests are left-tailed tests, rejecting the null hypothesis of no cointegration if the realization of the statistic is smaller than the corresponding critical value. Asymptotically valid critical values for the ADF and $\widehat Z_\alpha$ statistics in cases D0, D1, and D2 are tabulated in \citeasnoun{PhOu90}, whereas for the MSB statistic we use (unreported) critical values based on own simulations.\footnote{In case of GLS detrending, critical values for the ADF and MSB statistics are tabulated in \citeasnoun{PeRo16}.}

\subsection{Information Criteria}\label{app:Conv-Tests-IC}
Implementing the ADF and MSB tests requires the specification of the lag parameter $0\leq p \leq p_\text{max}$ in the auxiliary regression~\eqref{eq:ADFaux}. This is typically achieved by means of information criteria evaluated on exactly the same period $t=p_\text{max}+2,\ldots,T$ for each choice of $p$ \cite[p.\,56]{KiLu17}. The AIC and the BIC are defined as
\begin{align}
	\text{AIC}(p)\coloneqq \log(\widehat s_{rp_\text{max}}^2) + \frac{2p}{T}
\end{align}
and
\begin{align}
	\text{BIC}(p)\coloneqq \log(\widehat s_{rp_\text{max}}^2) + \frac{p\log(T)}{T},
\end{align}
respectively, where $\widehat s_{rp_\text{max}}^2 = T^{-1} \sum_{t=p_\text{max}+2}^T \widehat r_{tp}^2$, with $\widehat r_{tp}$ denoting the OLS residuals in~\eqref{eq:ADFaux}.\footnote{We follow \citeasnoun[p.\,56]{KiLu17} and use $\widehat s_{rp_\text{max}}^2$ rather than $(T-p)^{-1}T\widehat s_{rp_\text{max}}^2$.} \citeasnoun{NgPe01} and \citeasnoun{PeQu07} propose a modified AIC (MAIC) criterion. Applied to the regression residuals $\widehat u_t$ -- in the presence of deterministic components always based on OLS detrended data, even in case the test statistic is constructed using GLS detrended data -- the MAIC becomes
\begin{align}
	\text{MAIC}(p)\coloneqq \log((T-p_\text{max})^{-1}T\widehat s_{rp_\text{max}}^2) + \frac{2(p+\tau_T(p))}{T-p_\text{max}},
\end{align}
where $\tau_T(p)\coloneqq \left((T-p_\text{max})^{-1}T\widehat s_{rp_\text{max}}^2\right)^{-1} \widehat b_0^2 \sum_{t=p_\text{max}+2}^T \widehat u_{t-1}^2$, with $\widehat b_0$ denoting the OLS estimate of $b_0$ in~\eqref{eq:ADFaux}. A similarly modified version of the BIC is then given by
\begin{align}
	\text{MBIC}(p)\coloneqq \log((T-p_\text{max})^{-1}T\widehat s_{rp_\text{max}}^2) + \frac{\log(T-p_\text{max})(p+\tau_T(p))}{T-p_\text{max}}.
\end{align}

\end{appendices}
\end{document}